\documentclass[acmsmall, screen]{acmart} %
\usepackage[capitalise,noabbrev,nameinlink]{cleveref}

\setcopyright{none}
\acmPrice{}
\acmDOI{}
\copyrightyear{}
\acmConference[Conference acronym 'XX]{Make sure to enter the correct conference title from your rights confirmation emai}{June 03--05, 2018}{Woodstock, NY}

\makeatletter
\patchcmd{\@verbatim}
  {\verbatim@font}
  {\verbatim@font\fontsize{8.5pt}{11pt}\selectfont}
  {}{}
\makeatother
\usepackage{newunicodechar}
\newunicodechar{Π}{$\Pi$}
\newunicodechar{λ}{$\lambda$}
\newunicodechar{﹫}{$@$}
\newunicodechar{ℕ}{$\mathbb{N}$}

\usepackage[T1]{fontenc}

\usepackage{zi4}
\usepackage{hyperref}
\usepackage{color}

\AtEndPreamble{%
\theoremstyle{acmdefinition}
\newtheorem{notation}[theorem]{Notation}
\newtheorem{remark}[theorem]{Remark}}

\AtEndPreamble{%
\theoremstyle{acmdefinition}
}

\usepackage{todonotes} %

\usepackage{mathpartir}
\usepackage{float} %
\usepackage{mathtools}

\newcommand\tte{\texttt{e}}
\newcommand\ttf{\texttt{f}}
\newcommand\ttA{\texttt{A}}

\newcommand\tti{\texttt{i}}
\newcommand\ttB{\texttt{B}}
\newcommand\ttt{\texttt{t}}
\newcommand\ttl{\texttt{l}}
\newcommand\ttu{\texttt{u}}
\newcommand\ttp{\texttt{p}}
\newcommand\ttq{\texttt{q}}
\newcommand\ttP{\texttt{P}}
\newcommand\tta{\texttt{a}}
\newcommand\ttn{\texttt{n}}
\newcommand\ttx{\texttt{x}}
\newcommand\ttm{\texttt{m}}
\newcommand\ttb{\texttt{b}}

\newcommand{\red}{\longrightarrow}
\newcommand{\invred}{\longleftarrow}
\newcommand\redd{\mathrel{{\red}{}^*}}
\newcommand\iredd{\mathrel{{}^*{\invred}}}

\newcommand\Bt{\textbf{\textup{t}}}
\newcommand\Bu{\textbf{\textup{u}}}
\newcommand\Bv{\textbf{\textup{v}}}
\newcommand\Bs{\textbf{\textup{s}}}

\usepackage{xcolor}
\definecolor{gray2}{RGB}{155, 155, 155}
\newcommand\mkgray[1]{\text{{\footnotesize {\color{gray2} \hspace{0.25em}$#1$}}}} %
\definecolor{objectblue}{RGB}{0,30,160}
\definecolor{objectred}{RGB}{220,100,0} %
\newcommand{\stext}[1]{\text{ \footnotesize #1}}
\newcommand\sep{\mathrel{\text{\textbf{;}}}}

\newcommand\nEqPii{{\color{objectblue}\textup{Eq}\Pi_i}}
\newcommand\nW{{\color{objectblue}\textup{W}}}
\newcommand\nSup{{\color{objectblue}\textup{sup}}}
\newcommand\nTm{{\color{objectblue}\textup{Tm}}}
\newcommand\nTy{{\color{objectblue}\textup{Ty}}}

\newcommand\nPrf{{\color{objectblue} \textup{Prf}}}
\newcommand\nProp{{\color{objectblue} \textup{Prop}}}

\newcommand\nForall{{\color{objectblue} \forall}}
\newcommand\nForallUP{{\color{objectblue} \rotatebox[origin=c]{180}{\textup{A}}}}
\newcommand\nForallIntro{{\color{objectblue} \forall_\text{i}}}
\newcommand\nZero{{\color{objectblue} 0}}
\newcommand\nSucc{{\color{objectblue} \textup{S}}}
\newcommand\nNat{{\color{objectblue} \textup{Nat}}}
\newcommand\nU{{\color{objectblue} \textup{U}}}
\newcommand\nLvl{{\color{objectblue} \textup{Lvl}}}
\newcommand\nUp{{\color{objectblue} \Uparrow}}

\newcommand\npi{{\color{objectblue} \pi}}
\newcommand\nuu{{\color{objectblue} \textup{u}}}
\newcommand\nNil{{\color{objectblue} \textup{nil}}}
\newcommand\nCons{{\color{objectblue} \textup{cons}}}
\newcommand\nBigLam{{\color{objectblue} \Lambda}}
\newcommand\nVec{{\color{objectblue} \textup{Vec}}}
\newcommand\nPi{{\color{objectblue} \Pi}}
\newcommand\nSigma{{\color{objectblue} \Sigma}}
\newcommand\nPair{{\color{objectblue} \textup{pair}}}
\newcommand\nrefl{{\color{objectblue} \textup{refl}}}

\newcommand\ncode{{\color{objectblue} \textup{c}}}
\newcommand\nEq{{\color{objectblue} \textup{Eq}}}
\newcommand\nLam{{\color{objectblue} \lambda}}
\newcommand\nEx{{\color{objectblue} \textup{Ex}}}
\newcommand\nErr{{\color{objectblue} \textup{err}}}
\newcommand\nRaise{{\color{objectblue} \textup{raise}}}
\newcommand\nBool{{\color{objectblue} \mathbb{B}}}
\newcommand\nTrue{{\color{objectblue} \textup{true}}}
\newcommand\nFalse{{\color{objectblue} \textup{false}}}

\newcommand\nEqUpiel{{\color{objectred}\textup{EqU}^{\pi,\pi}_{e1}}}
\newcommand\nEqUpier{{\color{objectred}\textup{EqU}^{\pi,\pi}_{e2}}}
\newcommand\nCast{{\color{objectred}\textup{cast}}}
\newcommand\nEqPie{{\color{objectred}\textup{Eq}\Pi_e}}
\newcommand\nWRec{{\color{objectred}\textup{rec}_{\textup{W}}}}
\newcommand\nBoolRec{{\color{objectred}\textup{rec}_\mathbb{B}}}
\newcommand\nForallElim{{\color{objectred} \forall_\text{e}}}

\newcommand\nProjLeft{{\color{objectred} \textup{proj}_1}}
\newcommand\nProjRight{{\color{objectred} \textup{proj}_2}}
\newcommand\nJ{{\color{objectred} \textup{J}}}
\newcommand\nInst{{\color{objectred} \textup{inst}}}
\newcommand\nApp{{\color{objectred} \textbf{@}}}
\newcommand\nEl{{\color{objectred} \textup{El}}}
\newcommand\nVecRec{{\color{objectred} \textup{rec}_\textup{Vec}}}

\newcommand\nCatchBool{{\color{objectred} \textup{catch}_\mathbb{B}}}

 \newenvironment{smalldisplay}
  {\par\nopagebreak\small\noindent\ignorespaces\vspace{-0.5em}}
  {}

\begin{document}

\title{Generic bidirectional typing for dependent type theories}
\author{Thiago Felicissimo}
\email{thiago.felicissimo@inria.fr}
\affiliation{
  \institution{Université Paris-Saclay, INRIA, LMF, ENS Paris-Saclay}
  \city{Gif-sur-Yvette}
  \country{France}}
\begin{abstract}
  Bidirectional typing is a discipline in which the typing judgment is decomposed explicitly into inference and checking modes, allowing to control the flow of type information in typing rules and to specify algorithmically how they should be used. Bidirectional typing has been fruitfully studied and bidirectional systems have been developed for many type theories. However, the formal development of bidirectional typing has until now been kept confined to specific theories, with general guidelines remaining informal.  In this work, we give a generic account of bidirectional typing for a general class of dependent type theories. This is done by first giving a general definition of type theories (or equivalently, a logical framework), for which we define declarative and bidirectional type systems. We then show, in a theory-independent fashion, that the two systems are equivalent. Finally, we establish the decidability of bidirectional typing for normalizing theories, yielding a generic type-checking algorithm that has been implemented in a prototype and used in practice with many theories.
\end{abstract}


\maketitle

\section{Introduction}

When defining the syntax of programming languages and type theories, many choices are available.
The approach of seeing such systems as algebraic theories\footnote{Or, equivalently, as theories encoded in a \textit{logical framework}~\cite{harper2021equational,dedukti,harper93jacm,pfenning2001logical}.}~\cite{sterling2019algebraic,CARTMELL1986209,thorsten-ambrus,gratzer2023syntax} leads one to consider \textit{fully-annotated} syntaxes in which every type argument is spelled out explicitly: an application is written as $t @_{A,x.B} u$, a dependent pair as $\langle t,u\rangle_{A, x.B}$, cons as $t ::_A l$, etc. On the one hand, this choice of syntax is very natural because all the premises of typing rules get recorded in the syntax, as illustrated by the typing rule for $\langle t, u \rangle_{A,x.B}$:
\begin{mathpar}
  \inferrule
  {\Gamma \vdash A~\textsf{type}\\\Gamma, x:A \vdash B~\textsf{type}\\
   \Gamma \vdash t : A\\ \Gamma \vdash u : B[t/x]}
  {\Gamma\vdash \langle t, u\rangle_{A,x.B} : \Sigma x :A.B}
\end{mathpar}

On the other hand, the verbosity of such type annotations unfortunately destroys any hope of practical usability. Fully-annotated terms are not only much slower to type-check and reduce, but it is also not reasonable to ask users of type theories to write all such annotations.
Because of this, the type theories used in practice  omit the majority of these annotations, so one writes $t ~u$ for application, $\langle t, u\rangle$ for a dependent pair, $t:: l$ for cons, etc. This unannotated syntax is so common that many might not even realize that an omission is being~made.

\newcommand\unknown{\hspace{0.2em}?}

The omission of type arguments has nevertheless a cost: because knowing them is still important when typing terms, it becomes unclear how to do this algorithmically, even when the definitional equality of the theory is decidable. For instance, if we omit the annotations in $\langle t,u\rangle_{A, x.B}$, then when building a typing derivation for this term one has to guess the values of $A$ and $B$:
\begin{mathpar}
  \inferrule
  {\Gamma \vdash \unknown~\textsf{type}\\\Gamma, x: \unknown \vdash \unknown~\textsf{type}\\
   \Gamma \vdash t : \unknown\\ \Gamma \vdash u : \unknown}
  {\Gamma\vdash \langle t, u\rangle : \Sigma x :\unknown.\unknown}
\end{mathpar}

\paragraph*{Bidirectional typing}

A solution to this problem is provided by \textit{bidirectional typing}~\cite{mcbride,coquand1996algorithm,dunfield2021bidirectional,10.1145/964001.964025,pierce2000local}, a typing discipline in which the declarative typing judgment $\Gamma\vdash t:A$ is decomposed explicitly into inference $\Gamma \vdash t \Rightarrow A$, where $\Gamma$ and $t$ are inputs and $A$ is an output, and checking $\Gamma\vdash t \Leftarrow A$, where $\Gamma$, $t$ and $A$ are all inputs. The important point is that, by using these new judgments to control the flow of type information in typing rules, one can specify algorithmically how these rules should be used. For instance, the following rule clarifies how one should type $\langle t, u\rangle$: the types $A$ and $B$ are not to be guessed, but instead recovered from the type $C$, which should be given as~input.
\begin{mathpar}
  \inferrule
  {C \red^* \Sigma x:A. B \\ \Gamma \vdash t \Leftarrow A\\ \Gamma \vdash u \Leftarrow B[t/x]}
  {\Gamma\vdash \langle t, u\rangle \Leftarrow C}
\end{mathpar}

In general, whenever a term starts by a \textit{constructor} (that is, an \textit{introduction form}), bidirectional typing allows the recovery of annotations by asking its type to be given as input. Dually to how constructors can be type-checked, bidirectional typing supports type-inference of \textit{destructors} (that is, of \textit{elimination forms}) by recovering the missing arguments from the type of its first argument, which gets inferred. This can be illustrated with the bidirectional typing rule for application:
\begin{mathpar}
  \inferrule
  {\Gamma \vdash t \Rightarrow C \\ C \red^* \Pi x:A. B \\ \Gamma \vdash u \Leftarrow A}
  {\Gamma\vdash t~u \Rightarrow B[u/x]}
\end{mathpar}

We therefore see that bidirectional typing is the natural companion for an unannotated syntax, as it allows to algorithmically explain how the missing information can be retrieved.

Bidirectional typing has been fruitfully studied and bidirectional systems have been developed for many type theories \cite{coquand1996algorithm,lennonbertrand:LIPIcs.ITP.2021.24,gratzer2019implementing,abel2011partial,norell:thesis,abel2005untyped}. However, the formal development of bidirectional typing has until now remained confined to specific theories, with general guidelines remaining informal. One can then naturally wonder if it would be possible to define a framework in which bidirectional typing could be  studied generically, putting its general theory in solid ground. This is exactly the goal of~this~paper.

\paragraph*{Our contribution}

We contribute a generic account of bidirectional typing for a general class of dependent type theories.

In order to study bidirectional typing generically, our first contribution is to give a general definition of type theories (or equivalently, a logical framework). Our proposal is much inspired by previous frameworks such as Uemura's SOGATs~\cite{uemura2021abstract} and Haselwarter and Bauer's FTTs~\cite{haselwarter2021finitary}, yet differs from them in significant ways, in particular by allowing for the usual unannotated syntaxes most often used in practice --- after all, bidirectional typing is only interesting for explaining why annotations can be omitted ---, and being more geared towards implementation by only supporting definitional equalities defined by rewrite rules. %

Our notion of theory then allows for two kinds of entries: rewrite rules and \textit{schematic typing rules}, for specifying the theory's typing relation. One important feature of our proposal is the separation of (term-level) schematic rules as either \textit{constructor} or \textit{destructor} rules, motivated by the differences of their roles with regards to bidirectionality.
The schematic rules can then be instantiated into actual typing rules, yielding the main type system associated with the theory, which we refer to as \textit{declarative} in order to distinguish it from the \textit{bidirectional} system. We then establish that, for each of the theories, its associated declarative system satisfies desirable properties, like weakening and substitution. These properties are not only good sanity checks, but also essential for when establishing the correctness of the bidirectional type system subsequently.

To formulate our bidirectional system we address the well-known problem that some unannotated terms cannot be algorithmically typed --- a limitation that is not at all specific to bidirectional typing~\cite{dowek1993undecidability}. This is typically the case for redices like $(\lambda x.t)u$: while the bidirectional rule for application requires the type of its first argument to be inferred, the rule for abstraction requires its type to be given an input.
To rule out this issue, we define our bidirectional system over a syntax of \textit{inferable} and \textit{checkable} terms, in which an \textit{ascription} $t :: T$ must be inserted when a destructor meets a constructor --- similarly to bidirectional typing \textit{à~la}~McBride~\cite{mcbride}.

We then prove our main results, showing the correctness of the  bidirectional type  system with respect to the declarative one.
We first establish \textit{soundness}, ensuring that any term typed by the bidirectional system is also typable by the declarative one when forgetting about ascriptions.
Dually, we show \textit{annotability}~\cite{dunfield2021bidirectional}, ensuring that any (declaratively) typed term can be sufficiently annotated with ascriptions to get a term typable by the bidirectional system.
Finally, we also show that the bidirectional system is decidable for strongly-normalizing theories, allowing it to be used for typing terms algorithmically. We emphasize that we do not show theses results for a specific  type theory but instead for a whole general class of theories. %

\paragraph*{The implementation}

The framework we develop is not only of theoretical interest, but also has important practical applications.
The decidability result for our bidirectional system allowed us to implement it in the theory-independent type-checker BiTTs, publicly available at
\begin{center}
  \url{https://doi.org/10.5281/zenodo.10996395}
\end{center}
This has allowed our system to be used in practice with multiple theories, from variants of Martin-L\"of Type Theory, to Higher-Order Logic and Observational Type Theory.

Moreover, because our framework allows for defining theories with less annotations in the syntax, we expect it to allow for better performances when compared with other theory-independent type-checkers such as Dedukti, in which the lack of support for omitting annotations can have an important impact on type-checking times~\cite[Section~9]{felicissimo:LIPIcs.FSCD.2022.25}.

Finally, our implementation can also be used to prototype with new type theories without having to implement a new bidirectional algorithm from scratch, an important commodity given the large number of new theories that are proposed each year.

\paragraph*{Plan of the paper}
We start in \cref{sec:theories} by formulating our general definition of type theories, before moving to \cref{sec:declarative-type-sytem} in which we specify the declarative system of the theories and show its expected properties. \cref{sec:bidirectional-type-system} is then dedicated to the definition of the bidirectional type system and its proofs of correctness with respect to the declarative system and of decidability. We then give in \cref{sec:more-examples} more examples of theories covered by our framework, and discuss its implementation in  \cref{implementation}. We finish by discussing related work in \cref{sec:related-work} before concluding with \cref{sec:conclusion}.

\paragraph*{Related version} A preliminary version of our work had first appeared in the proceedings of the 33rd European Symposium on Programming~\cite{felicissimo-esop}. This version contains multiple improvements, among which are the following:

\begin{itemize}
  \item We have addressed the main deficiency of our preliminary work, which was its lack of support for indexed types, such as vectors and, most importantly, the equality type: our notion of theory has been updated in order to support such types.
  \item
  We have extended the bidirectional syntax by adding support for ascriptions, allowing to turn a checkable term~$t$ into an inferable one $t::T$. Accordingly, our notion of completeness has been updated to annotability, as discussed in the introduction. Our previous completeness result, here called \textit{ascription-free completeness}, is now shown as a simple corollary of annotability.

  \item Our treatment of matching modulo rewriting has been simplified, in particular by removing its reliance on the maximal-outermost strategy. This brings the theory closer to our implementation, where we use a reduction strategy based on call-by-value.

  \item The section on examples of theories covered by our framework has been considerably extended, better illustrating the generality of our approach.

  \item Finally, due the space constraints, the proofs of our results could only be sketched in our preliminary version, whereas this extended version provides detailed proofs.

\end{itemize}

\section{A general definition of bidirectional type theories}\label{sec:theories}

In this section, we give a general definition of type theories (or equivalently, a logical framework) for which we will give declarative and bidirectional type systems in later sections.
We start the section by defining the raw syntax of our theories. Then, after defining patterns and substitution, we give one of the central definitions of our work: the one of theory. We finish the section by describing the rewrite judgment used to specify the definitional equality of our type theories.

\subsection{Raw Syntax}%
\label{subsec:label}

\newcommand\Ctx{\textup{\textsf{Ctx}}}
\newcommand\MCtx{\textup{\textsf{MCtx}}}
\newcommand\Tm{\textup{\textsf{Tm}}}
\newcommand\sort{\textup{\textsf{sort}}}
\newcommand\Scope{\textup{\textsf{Scope}}}
\newcommand\Sig{\textup{\textsf{Sig}}}
\newcommand\MScope{\textup{\textsf{MScope}}}
\newcommand\Expr{\textup{\textsf{Expr}}}
\newcommand\Sub{\textup{\textsf{Sub}}}
\newcommand\MSub{\textup{\textsf{MSub}}}
\newcommand\id{\textup{\textsf{id}}}

\newcommand{\condition}[1]{\text{{\footnotesize $#1$}}}

\subsubsection{Intrinsically-scoped syntax} When defining the syntax of type theory, one most usually defines a set $\Tm$ containing all terms. However, it is sometimes useful to distinguish terms $t\in\Tm$ according to the variables that can appear in $t$ --- for instance, when defining the application of a substitution, it is useful to know that it is defined for all variables in $t$. Therefore, we elect an \textit{intrinsically-scoped} presentation of syntax, in which we define instead a \textit{family} of terms, indexed by \textit{scopes} specifying which variables can appear.

\subsubsection{Scopes and signatures} The basic ingredients of our raw expressions are actually not only \textit{variables} $x,y,\dots$, but also \textit{metavariables} $\texttt{x},\texttt{y},\dots$ and \textit{operations} $o,\dots$, which are specified respectively by \textit{(variable) scopes}, \textit{metavariable scopes} and \textit{signatures}.
\begin{align*}
  \fbox{\Scope} \ni & &\gamma, \delta  &::= \cdot \mid \gamma,x\\
  \fbox{\MScope} \ni & & \theta, \xi  &::=  \cdot \mid \theta, \texttt{x} \{\delta\}  & &\\
  \fbox{\Sig} \ni & & \Sigma &::= \cdot\mid \Sigma, o (\theta)
\end{align*}

A scope $\gamma$ is simply a list of variables, whereas a metavariable scope $\theta$ is a list of metavariables accompanied by a variable scope $\delta$, explaining the arguments each metavariable expects. For instance, $\texttt{x}\{x,y\}\in\theta$ specifies a metavariable $\texttt{x}$ taking two arguments, each one named by the variables $x$ or $y$.   We have an obvious operation of concatenation $\gamma.\delta$  and $\theta.\xi$, for variable scopes and metavariable scopes respectively. A signature $\Sigma$ is then a list of operations accompanied by a metavariable scope explaining its arguments: for instance, $o(\texttt{x}\{\cdot\}, \texttt{y}\{x\})\in\Sigma$ specifies an~operation~$o$ taking two arguments and binding one variable in its second argument. We allow ourselves to abbreviate $\texttt{x}\{\cdot\}$ as $\texttt{x}$ and $o(\cdot)$ as $o$ when convenient. Finally, we suppose that operations $o$ are partitioned between \textit{constructors} $c$, whose names we write in {\color{objectblue} blue}, and \textit{destructors} $d$, whose names we write in {\color{objectred} orange}. %

\begin{example}
  The following signature \ref{ex:sig} defines the raw syntax of a minimalistic Martin-L\"of Type Theory (MLTT) with only dependent functions.
  \begin{align*}\label{ex:sig}
    \tag{$\Sigma_{\lambda\Pi}$} &\nTy,~\nTm(\texttt{A}),~\nPi (\texttt{A}, \texttt{B}\{x\}),~\nLam(\texttt{t}\{x\}),~\nApp(\texttt{t},\texttt{u})
  \end{align*}
  The operations are the ones we would expect, except  for $\nTy$ and $\nTm$ whose role will become~clear~later.
\end{example}

\begin{remark}
  When working with de Bruijn indices, a scope $\gamma$ becomes just a natural number and a metavariable scope $\theta$ becomes just a list of natural numbers. In this setting, our definition of signatures corresponds exactly to the well-known \textit{binding-signatures}~\cite{general-church,fiore1999abstract}, where $o(\texttt{x}_1\{x^1_1,\dots,x^1_{k_1}\},\dots,\texttt{x}_n\{x^n_1,\dots,x^n_{k_n}\})$ is represented by $(k_1,\dots,k_n)$.
\end{remark}

\subsubsection{Terms and substitutions} Given a fixed signature $\Sigma$, we define \textit{terms}, \textit{(variable) substitutions} and \textit{metavariable substitutions} by the following grammars. Note that, as explained before, we do not define a set of terms but instead a family $\Tm~\theta~\gamma$ in which the indices $\theta$ and $\gamma$ explain which metavariables and variables are in scope. %
\begin{align*}
  \boxed{\Tm~\theta~\gamma} \ni & &t, u, T, U ::=
  & \mid x & &\stext{if $ x \in \gamma$}\\
  &&&\mid \texttt{x}\{\vec{t} \mkgray{\in \Sub~\theta~\gamma~\delta}\}& &\stext{if $\texttt{x} \{\delta\}  \in \theta$}\\
  &&&\mid o(\Bt\mkgray{\in\MSub~\theta~\gamma~\xi})& &\stext{if $o (\xi)  \in \Sigma$}\\
  \boxed{\Sub~\theta~\gamma~\delta} \ni & & \vec{t}, \vec{u}, \vec{s}, \vec{v}::=
  &\mid \varepsilon & & \stext{if $\delta = \cdot$}\\
  &&& \mid \vec{u} \mkgray{\in \Sub~\theta~\gamma~\delta'}, t \mkgray{\in \Tm~\theta~\gamma} &&\stext{if $\delta = \delta', x$} \\
  \boxed{\MSub~\theta~\gamma~\xi} \ni & &\Bt,\Bu,\Bs,\Bv  ::=
  &\mid \varepsilon && \stext{if $\xi = \cdot$}\\
  &&& \mid \Bu \mkgray{\in \MSub~\theta~\gamma~\xi'}, \vec{x}_{\delta}.t \mkgray{\in \Tm~\theta~\gamma.\delta} &&\stext{if $ \xi = \xi', \texttt{x} \{\delta\}$}
\end{align*}

A term is either a variable $x$, a metavariable $\texttt{x}$ applied to a substitution~$\vec{t}$, or an operation $o$ applied to a metavariable substitution $\Bt$.  A (variable) substitution $\vec{t}\in\Sub~\theta~\gamma~\delta$ is then simply a list of terms, in which each term corresponds to one of the variables in $\delta$. Similarly, a metavariable substitution $\Bt\in\MSub~\theta~\gamma~\xi$ is also a list of terms, with the difference that each position $\texttt{x}\{\delta\}\in \xi$ extends the current scope $\gamma$ with the variables in $\delta$. We write this variable binding as $\vec{x}_\delta. t$, which can be seen in action in the cases of $\nLam$ and $\nPi$ in \cref{example-terms}. Finally, we allow ourselves to abbreviate $\texttt{x}\{\varepsilon\}$ as $\texttt{x}$ and  $o(\varepsilon)$ as $o$ when convenient.

\begin{example}\label{example-terms}
  The terms defined by the signature \ref{ex:sig} are given by the following grammar, where we omit the scope requirements for variables and metavariables.
  \begin{align*}
    t, u, A, B  ::= x \mid \texttt{x}\{\vec{t}\} \mid \nTy \mid \nTm(A) \mid \nLam(x.t) \mid \nPi(A,x.B)\mid\nApp(t,u)
  \end{align*}
\end{example}

Given a metavariable substitution $\Bt\in\MSub~\theta~\gamma~\xi$ and $\texttt{x}\{\delta\}\in\xi$, we write $\Bt_\texttt{x}\in \Tm~\theta~\gamma.\delta$ for the term in $\Bt$ at the position pointed by $\texttt{x}$. Similarly, given a substitution $\vec{t}\in\Sub~\theta~\gamma~\delta$ and $x\in\delta$, we write $t_x\in \Tm~\theta~\gamma$ for the term in $\vec{t}$ at the position pointed by $x$.

For each $\gamma$ and $\theta$ we have the \textit{identity substitutions} $\id_\gamma \in \Sub~(\cdot)~\gamma~\gamma$ and $\id_\theta\in\MSub~\theta~(\cdot)~\theta$, defined by $\id_{(\cdot)}:=\varepsilon$ and $\id_{\gamma,x}:= \id_\gamma,x$ and $\id_{\theta,\texttt{x}\{\gamma\}} := \id_\theta,\vec{x}_\gamma. \texttt{x}\{\id_\gamma\}$. Note that, while the identity variable substitution $\id_\gamma$ is just the list of variables from $\gamma$, the identity metavariable substitution $\id_\theta$ needs to "eta-expand" each metavariable $\texttt{x}\{\delta\}\in\theta$ to $\vec{x}_\delta.\texttt{x}\{\id_\delta\}$ in order for the result to be a valid metavariable substitution. Finally, we allow ourselves to omit the index of $\id_\gamma$ or $\id_\theta$ when it can be inferred from the context.

\subsubsection{Contexts} Given a fixed signature $\Sigma$, we define  \textit{(variable) contexts} and \textit{metavariable contexts} by the following grammars. These are defined simultaneously with two functions $|\_{}|$ computing their \textit{underlying scopes} $|\Gamma|\in\Scope$ and $|\Theta|\in\MScope$, given by $|\cdot|:=\cdot$ and $|\Gamma,x:T|:=|\Gamma|,x$ and $|\Theta,\texttt{x}\{\Delta\}:T|:=|\Theta|,\texttt{x}\{|\Delta|\}$.
\begin{align*}
  \boxed{\Ctx~\theta~\gamma} \ni & &\Gamma, \Delta ::=& ~ \cdot \mid \Gamma \mkgray{\in \Ctx~\theta~\gamma}, x : T\mkgray{\in \Tm~\theta~\gamma.|\Gamma|} \\
  \boxed{\MCtx~\theta} \ni & &\Theta, \Xi ::=&~ \cdot \mid \Theta \mkgray{\in \MCtx~\theta}, \texttt{x} \{\Gamma \mkgray{\in\Ctx~\theta.|\Theta|~(\cdot)}\}: T\mkgray{\in \Tm~\theta.|\Theta|~|\Gamma|}
\end{align*}

A context $\Gamma \in\Ctx~\gamma~\theta$ is either empty, or composed by a context $\Gamma' \in\Ctx~\gamma~\theta$ and a variable~$x$ with a term $T\in\Tm~\theta~\gamma.|\Gamma'|$. An important point to note is that the term $T$ does not live in scope $\gamma$, but in the extension of $\gamma$ with the underlying scope of $\Gamma'$. This means that if a context $x_1:T_1,\dots,x_k:T_k$ lives in some scope $\gamma$, then each term $T_i$ lives in $\gamma,x_1,\dots,x_{i-1}$ and thus also has access to the previously occurring variables. The case of a metavariable context $\Theta\in\MCtx~\theta$ is similar: we have either $\Theta$ empty or $\Theta = \Theta',\texttt{x}\{\Delta\}:T$, where $\Delta$ has access to metavariables in $\theta$ and $\Theta$, and $T$ has moreover access to the variables in $\Delta$.%

\begin{notation} We establish the following notations. %
  \begin{itemize}
    \item We write $e \in \Expr~\theta~\gamma$ for either $e \in \Tm~\theta~\gamma$ or $e \in \Sub~\theta~\gamma~\delta$ or $e \in \MSub~\theta~\gamma~\xi$ or $e \in \Ctx~\theta~\gamma$.
    \item We write $\Tm_\Sigma$, $\Sub_\Sigma$, $\MSub_\Sigma$, $\Ctx_\Sigma$ and $\MCtx_\Sigma$ when $\Sigma$ is not clear from the context.%
    \item We write $\Ctx~\theta$ for $\Ctx~\theta~(\cdot)$, $\Ctx$ for $\Ctx~(\cdot)~(\cdot)$ and $\MCtx$ for $\MCtx~(\cdot)$.
  \end{itemize}
\end{notation}

\begin{remark}
  We work with a nameful syntax, allowing us to implicitly weaken expressions: if $e \in \Expr~\theta~\gamma$ and $\theta$ is a subsequence of $\theta'$ and $\gamma$ is a subsequence of $\gamma'$ then we also have $e\in\Expr~\theta'~\gamma'$. Nevertheless, we expect that our proofs can be formally carried out using de Bruijn indices, by properly inserting weakenings whenever needed, and showing the associated lemmata.
\end{remark}

\subsubsection{Substitution application}\label{subsec:substitution}

We define in \cref{fig:var-subst} the application of a (variable or metavariable) substitution to an expression. Given a variable substitution $\vec{v}\in\Sub~\theta~\gamma_1~\gamma_2$ its application to an expression $e \in \Expr~\theta~\gamma_2$ gives $e[\vec{v}]\in\Expr~\theta~\gamma_1$, and given a metavariable substitution $\Bv\in\MSub~\theta_1~\delta~\theta_2$ its application to an expression $e\in\Expr~\theta_2~\gamma$ gives $e[\Bv]\in\Expr~\theta_1~\delta.\gamma$.%

The main case of the definition is when we substitute $\Bv\in\MSub~\theta_1~\delta~\theta_2$ in the term $\texttt{x}\{\vec{t}\} \in \Tm~\theta_2~\gamma$, where $\texttt{x}\{\gamma_\texttt{x}\}\in\theta_2$. By first recursively substituting $\Bv$ in $\vec{t}\in\Sub~\theta_2~\gamma~\gamma_\texttt{x}$ we get $\vec{t}[\Bv] \in \Sub~\theta_1~\delta.\gamma~\gamma_\texttt{x}$. We moreover have $\Bv_\texttt{x}\in\Tm~\theta_1~\delta.\gamma_\texttt{x}$, so by substituting the variables in $\gamma_\texttt{x}$ by $\vec{t}[\Bv]$ and the ones in $\delta$ by themselves we get $\Bv_\texttt{x}[\id, \vec{t}[\Bv]]\in\Tm~\theta_1~\delta.\gamma$ as the final result.

\begin{figure}
  \hspace{-0.8em}\begin{minipage}{0.5\linewidth}{\small
  \begin{align*}
  &\_{}[\_{}] : \Tm~\theta~\gamma_{2} \to \Sub~\theta~\gamma_{1}~\gamma_{2} \to \Tm~\theta~\gamma_{1}\\
  &x[\vec{v}] :=v_{x}\\
  &\texttt{x}\{\vec{t}\}[\vec{v}] := \texttt{x}\{\vec{t}[\vec{v}]\}\\
  &o(\Bt)[\vec{v}] := o(\Bt[\vec{v}])\\\\
  &\_{}[\_{}] : \Sub~\theta~\gamma_{2}~\delta \to \Sub~\theta~\gamma_{1}~\gamma_{2} \to \Sub~\theta~\gamma_{1}~\delta\\
  &\varepsilon[\vec{v}] := \varepsilon\\
  &(\vec{t}, u)[\vec{v}] := \vec{t}[\vec{v}], u[\vec{v}]\\\\
  &\_{}[\_{}] : \MSub~\theta~\gamma_{2}~\xi \to \Sub~\theta~\gamma_{1}~\gamma_{2} \to \MSub~\theta~\gamma_{1}~\xi\\
  &\varepsilon[\vec{v}] := \varepsilon\\
  &(\Bt, \vec{x}_{\delta}. u)[\vec{v}] := \Bt[\vec{v}], \vec{x}. u[\vec{v}, \id]\\\\
  &\_{}[\_{}] : \Ctx~\theta~\gamma_{2} \to \Sub~\theta~\gamma_{1}~\gamma_{2} \to \Ctx~\theta~\gamma_{1}\\
  &(\cdot)[\vec{v}] := \cdot\\
    &(\Gamma, x : T)[\vec{v}] := \Gamma[\vec{v}], x : T[\vec{v}, \id]%
\end{align*}}
\end{minipage}
\begin{minipage}{0.45\linewidth}{\small
  \begin{align*}
    &\_{}[\_{}] : \Tm~\theta_{2}~\gamma \to \MSub~\theta_{1}~\delta~\theta_{2} \to \Tm~\theta_{1}~\delta.\gamma\\
    &x[\Bv] :=x\\
    &\texttt{x}\{\vec{t}\}[\Bv] := \Bv_{\texttt{x}}[\id,\vec{t}[\Bv]]\\
    &o(\Bt)[\Bv] := o(\Bt[\Bv])\\\\
    &\_{}[\_{}] : \Sub~\theta_{2}~\gamma~\gamma_{0} \to \MSub~\theta_{1}~\delta~\theta_{2} \to \Sub~\theta_{1}~\delta.\gamma~\gamma_{0}\\
    &\varepsilon[\Bv]:=\varepsilon\\
    &(\vec{t},u)[\Bv] := \vec{t}[\Bv], u[\Bv]\\\\
    &\_{}[\_{}] : \MSub~\theta_{2}~\gamma~\xi \to \MSub~\theta_{1}~\delta~\theta_{2} \to \MSub~\theta_{1}~\delta.\gamma~\xi\\
    &\varepsilon[\Bv]:=\varepsilon\\
    &(\Bt,\vec{x}.u)[\Bv] := \Bt[\Bv], \vec{x}.u[\Bv]\\\\
    &\_{}[\_{}] : \Ctx~\theta_{2}~\gamma \to \MSub~\theta_{1}~\delta~\theta_{2} \to \Ctx~\theta_{1}~\delta.\gamma\\
    &(\cdot)[\Bv] := \cdot\\
    &(\Gamma, x :T)[\Bv] := \Gamma[\Bv], x : T[\Bv]%
    \end{align*}}
\end{minipage}
\caption{Application of a variable or metavariable substitution}
\label{fig:var-subst}
\end{figure}

\begin{example}
  If $t \in \Tm~(\cdot)~(\gamma, x)$ and $u \in \Tm~(\cdot)~\gamma$, then by applying $x. t, u \in \MSub~(\cdot)~\gamma~(\texttt{t}\{x\}, \texttt{u})$ to  $\nApp(\nLam(x.\texttt{t}\{x\}),\texttt{u}) \in \Tm~(\texttt{t}\{x\}, \texttt{u})~(\cdot)$ we get the term $\nApp(\nLam(x.t), u) \in \Tm~(\cdot)~\gamma$.
\end{example}

\begin{remark}
Compared to frameworks derived from contextual modal type theory~\cite{nanevski2008contextual}, our metavariable substitutions are not required to be closed and can introduce new variables in the scope of the resulting term. For instance, in the previous example, while the term $\nApp(\nLam(x.\texttt{t}\{x\}),\texttt{u})$ lives in an empty variable scope, the application of the metavariable substitution yields $\nApp(\nLam(x.t), u)$, which lives in the scope~$\gamma$. Therefore, a metavariable $\texttt{t}\{x_1,\dots,x_k\}$ should not be seen as a placeholder for a term containing only $x_1,\dots,x_k$, but instead for a term in \textit{any} scope extended by $x_1,\dots,x_k$.

\end{remark}

Substitution application satisfies the following basic laws.%

\begin{proposition}[Unit laws for $\id$]\label{unit-subst}
We have $e[\id_{\gamma}] = e$ and $e[\id_{\theta}]=e$ for all $e \in \Expr~\theta~\gamma$, and $\id_{\delta}[\vec{t}]=\vec{t}$ for all $\vec{t}\in \Sub~\theta~\gamma~\delta$, and $\id_{\theta}[\Bv]=\Bv$ for all $\Bv \in \MSub~\xi~\gamma~\theta$.
\end{proposition}
\begin{proof}
  We first show $e[\id_{\gamma}]=e$ by induction on $e$ and $\id_{\delta}[\vec{t}]=\vec{t}$ by induction on $\delta$. We then show $e[\id_{\theta}]=e$ by induction on $e$ and $\id_{\theta}[\Bv]=\Bv$ by induction on $\theta$.
\end{proof}

The following two properties are shown simultaneously.

\begin{proposition}[Commutation lemmas]\label{commutation-subst}
We have $e[\Bu][\vec{v}, \id_{\delta}]=e[\Bu[\vec{v}]]\in \Expr~\theta_{1}~\gamma_{1}.\delta$ for all $e \in \Expr~\theta_{2}~\delta$ and $\Bu \in \MSub~\theta_{1}~\gamma_{2}~\theta_{2}$ and $\vec{v}\in \Sub~\theta_{1}~\gamma_{1}~\gamma_{2}$. We have $e[\vec{u}][\Bv] = e[\Bv][\id_{\delta}, \vec{u}[\Bv]] \in \Expr~\theta_{1}~\delta.\gamma_{1}$ for all $e \in \Expr~\theta_{2}~\gamma_{2}$ and $\vec{u}\in \Sub~\theta_{2}~\gamma_{1}~\gamma_{2}$ and $\Bv \in \MSub~\theta_{1}~\delta~\theta_{2}$.
\end{proposition}

\begin{proposition}[Associativity of substitution]\label{assoc-subst}
  Let $e \in \Expr~\theta_{3}~\gamma_{3}$. For all $\vec{v} \in \Sub~\theta_{3}~\gamma_{2}~\gamma_{3}$ and $\vec{u} \in \Sub~\theta_{3}~\gamma_{1}~\gamma_{2}$ we have $e[\vec{v}][\vec{u}]=e[\vec{v}[\vec{u}]]$, and for all $\Bv \in \MSub~\theta_{2}~\gamma_{3}~\theta_{3}$ and $\Bu \in \MSub~\theta_{1}~\gamma_{3}~\theta_{2}$ we have $e[\Bv][\Bu]= e[\Bv[\Bu]]$.
\end{proposition}
\begin{proof}
  All results are shown by induction on $e$, on the following order: first $e[\vec{v}][\vec{u}]=e[\vec{v}[\vec{u}]]$, then $e[\Bu][\vec{v}, \id_{\delta}]=e[\Bu[\vec{v}]]$, then $e[\vec{u}][\Bv] = e[\Bv][\id_{\delta}, \vec{u}[\Bv]]$, then $e[\Bv][\Bu]= e[\Bv[\Bu]]$.
\end{proof}

\subsubsection{Patterns}
\newcommand\Pat{\textsf{\textup{P}}}
\newcommand\pat{\textsf{\textup{P}}}

We finish this subsection by isolating the \textit{pattern} fragment of the syntax, which will be used in the next subsection to define the theories. Given a signature $\Sigma$, \textit{term patterns} and \textit{metavariables substitution patterns} are defined by the following grammars.
\begin{align*}
  \boxed{\Tm^\pat~\theta~\gamma} \ni & &t, u, v, s ::= &\mid \texttt{x}\{\id\}& &\stext{if $ \theta = \texttt{x} \{\gamma\}$}\\
  &&&\mid c(\Bt\mkgray{\in\MSub^{\pat}~\theta~\gamma~\xi})& &\stext{if $ c(\xi) \in \Sigma$}\\
  \boxed{\MSub^\pat~\theta~\gamma~\xi} \ni & &\Bt,\Bu,\Bs,\Bv  ::= &\mid \epsilon && \stext{if $\xi = \cdot$  and $\theta = \cdot$}\\
  &&& \mid \Bt \mkgray{\in \MSub^{\pat}~\theta_{1}~\gamma~\xi'}, \vec{x}_{\delta}.t \mkgray{\in \Tm^{\pat}~\theta_{2}~\gamma.\delta} &&\stext{if $\xi = \xi', \texttt{x}\{\delta\}$ and $\theta=\theta_{1}.\theta_{2}$}
\end{align*}

Compared with the regular syntax, the pattern condition imposes that each metavariable $\texttt{x}\{\delta\}\in\theta$ must occur precisely once, and moreover applied to all variables occurring in the scope of its occurrence --- in particular, imposing it to be equal to $\delta$. These restrictions ensure that, differently from regular terms, patterns support decidable and unitary matching~\cite{miller1991logic}. However, as we will see in \cref{matching-modulo}, we will not only need syntactic matching but also \textit{matching modulo rewriting}, leading us also to additionally impose that only constructors may appear in patterns.
\begin{example}
  In \ref{ex:sig}, we can build the pattern $
    \nTm(\nPi(\texttt{A}, x.\texttt{B}\{x\})) \in \Tm^\Pat~(\texttt{A},~\texttt{B}\{x\})~(\cdot)$.
\end{example}
Note that we have inclusions $\Tm^\Pat~\theta~\gamma\subset \Tm~\theta~\gamma$ and $\MSub^\Pat~\theta~\gamma~\xi\subset\MSub~\theta~\gamma~\xi$, which we use to implicitly coerce patterns into regular expressions when needed.

\subsection{Theories}\label{subsec:theories}

\newcommand\Thy{\textsf{\textup{Thy}}}

We now come to a central definition in our work, that of a \textit{theory} $\mathbb{T}\in\textsf{Thy}$. We define inductively how a theory is built, simultaneously with its underlying signature~$|\mathbb{T}|\in\Sig$ --- technically, our definition is by small induction-recursion~\cite{dybjer2000general}. Assuming that a theory $\mathbb{T}$ is given we show how to extend it with two types of entries: \textit{schematic typing rules} and \textit{rewrite rules}. We start with the first, which come in three kinds: sort, constructor and destructor rules.

\subsubsection{Sort rules}
In our framework, a \textit{sort} $T$ is a term that can appear in the second position of the typing judgment $t : T$.\footnote{We purposely avoid calling them "types" to prevent a name clash with the types of the theories we define. Still, we allow ourselves to say "$t$ is typed by sort $T$" to mean $t:T$.}
Like in algebraic presentations of type theory~\cite{CARTMELL1986209,uemura2021abstract,sterling2019algebraic}, the sorts of a theory correspond to its \textit{judgment forms}.
For instance, vanilla Martin-L\"of Type Theory features two judgment forms:  $A~\textsf{type}$, for asserting that $A$ is a type, and $t : A$ for asserting that $t$ is a term of type $A$.
In our framework, these are materialized by the following sort rules.
\begin{mathpar}
  \inferrule
  {  }
  {\nTy~\sort}
  \and
  \inferrule
  {\texttt{A}:\nTy}
  {\nTm(\texttt{A})~\sort}
\end{mathpar}

Formally, a sort rule is actually of the form\[
c (\Xi \mkgray{\in \MCtx_{|\mathbb{T}|}}) ~ \sort
\]and the previously shown rules are just an informal notation for $\nTy(\cdot) ~ \sort$ and $\nTm(\texttt{A}:\nTy) ~ \sort$. More precisely, our notation represents metavariables $\texttt{x}\{\Gamma\}: T \in \Xi$ as premises $ \Gamma \vdash \texttt{x} : T$, or just $\texttt{x}:T$ when $\Gamma$ is empty. In the following, we will make use of such informal representations in order to enhance readability of schematic rules.

\subsubsection{Constructor rules}

As discussed in the introduction, the way in which missing arguments can be recovered in bidirectional typing is deeply linked with whether a symbol is a constructor or a destructor: whereas constructors support type-checking, destructors support type-inference. In order to capture this distinction, we classify our term-level schematic rules as either constructor or destructor rules, and we start  by describing constructor rules.

In order to specify the distinction between the arguments that are missing or not from the syntax, we split them in two metavariable contexts: $\Xi_o$ containing the omitted arguments (sometimes also referred to as \textit{erased}), and $\Xi_c$ containing the constructor arguments present in the syntax. Then, to ensure that the arguments in $\Xi_o$ can be recovered from the sort given as input, our first idea is to ask the sort of the rule $T$ to be a pattern over the metavariables of $\Xi_o$, leading to rules of the form
\begin{align*}\label{problem}
  c(\Xi_o\mkgray{\in\MCtx_{|\mathbb{T}|}}\sep\Xi_c\mkgray{\in\MCtx_{|\mathbb{T}|}~|\Xi_o|}) : T \mkgray{\in\Tm^\pat_{|\mathbb{T}|}~|\Xi_o|~(\cdot)}  \tag*{$(*)$}
\end{align*}

Note that, whereas $\Xi_o$ is closed, $\Xi_c$ is allowed to depend on the underlying scope of $\Xi_o$, so the sorts and contexts of the non-omitted arguments can depend on the missing information.
Two examples of rules fitting this definition are the ones for $\nPi$ and $\nLam$ --- note however that the one for $\nPi$ is slightly degenerate, given that we have $\Xi_{o}=\cdot$ and thus no erased premises.
\begin{mathpar}
  \inferrule
  { \vdash\texttt{A}:\nTy\\ x:\nTm(\texttt{A}) \vdash \texttt{B}:\nTy }
  {\vdash \nPi(\texttt{A},x.\texttt{B}\{x\}):\nTy}
  \and
  \inferrule
  {\vdash\texttt{A}:\nTy\\ x:\nTm(\texttt{A}) \vdash \texttt{B}:\nTy\\\\
    x:\nTm(\texttt{A}) \vdash \texttt{t}:\nTm(\texttt{B}\{x\})}
  {\vdash\nLam(x.\texttt{t}\{x\}):\nTm(\nPi(\texttt{A},x.\texttt{B}\{x\}))}
\end{mathpar}

Once again, these are just notations for the formal rules $\nPi(\cdot\sep\texttt{A}:\nTy, \texttt{B}\{x : \nTm(\texttt{A})\} : \nTy) : \nTy$ and $ \nLam(\texttt{A}:\nTy, \texttt{B}\{x : \nTm(\texttt{A})\} : \nTy \sep \texttt{t}\{x : \nTm(\texttt{A})\} : \nTm(\texttt{B}\{x\}) ) : \nTm(\nPi(\texttt{A},x.\texttt{B}\{x\}))$.

In the rule for $\nLam$, $\Xi_o$ stores the \textit{parameters} of the dependent function type. This is actually not specific to this one example, but in general all non-indexed type constructors fit \ref{problem} when taking $\Xi_o$ to be its parameters. The situation is however trickier  when trying to handle \textit{indexed inductive types}. Consider for instance the constructors $\nrefl$ for equality and $\nCons$ for vectors.
\begin{mathpar}
  \inferrule
  { \ttA:\nTy\\ \tta : \nTm(\ttA)  }
  {\nrefl : \nTm(\nEq(\ttA, \tta, \tta))}
  \and
  \inferrule
  {\ttA:\nTy\\ \ttn:\nTm(\nNat)\\
  \ttt:\nTm(\ttA)\\
  \ttl:\nTm(\nVec(\ttA,\ttn))}
  {\nCons(\ttn,\ttt, \ttl) :
  \nTm(\nVec(\ttA,\nSucc(\ttn)))}
\end{mathpar}

In the case of $\nrefl$, the  metavariable $\tta$ occurs \textit{non-linearly} in its sort $\nTm(\nEq(\ttA, \tta, \tta))$, which is therefore not a pattern in our sense. In the case of $\nCons$ its sort actually \textit{is} a pattern, however in order for the rule to fit \ref{problem} we would have to omit the size of the vector $\texttt{n}$. This in principe can appear to be a good thing, after all the goal of bidirectional typing is precisely to remove annotations that are not needed. However when writing the reduction rules associated with the eliminator for vectors we realize that $\texttt{n}$ actually \textit{is} needed. Indeed, the argument $\texttt{n}$ is \textit{computationally-relevant}, meaning that the result of a computation might depend on it, so it  cannot be erased.

In order to account for such rules we propose to split the metavariables occurring in the sort $T$ of the rule between \textit{parameters} $\Xi_p$ and \textit{indices} $\Xi_i$. Whereas the parameters correspond to the erased arguments of the rule, indices correspond to terms depending on $\Xi_p$ and $\Xi_c$. More precisely, the metavariables $\Xi_i$ are instantiated by an \textit{index metavariable substitution} $\Bv_i$ living in the underlying scopes of $\Xi_p$ and $\Xi_c$. This now yields the definitive form for our constructor rules:
\[
  c (\Xi_p\mkgray{\in \MCtx_{|\mathbb{T}|}}\sep \Xi_c\mkgray{\in \MCtx_{|\mathbb{T}|}~|\Xi_p|}\sep
  \Bv_i \mkgray{\in \MSub_{|\mathbb{T}|}~|\Xi_p.\Xi_c|~(\cdot)~|\Xi_i|} ~/~ \Xi_i\mkgray{\in \MCtx_{|\mathbb{T}|}~|\Xi_p|}): T \mkgray{\in \Tm_{|\mathbb{T}|}^{\Pat}~|\Xi_p.\Xi_i|~(\cdot)}
\]

The previously shown rules for $\nPi$ and $\nLam$ still fit this scheme by taking $\Xi_i=\cdot$ and $\Bv_i = \varepsilon$, however now the rules for $\nrefl$ and $\nCons$ can also be defined:
\begin{mathpar}
  \inferrule
  { \ttA:\nTy\\ \tta : \nTm(\ttA) \\ \ttb \mapsto \tta : \nTm(\ttA) }
  {\nrefl : \nTm(\nEq(\ttA, \tta, \ttb))}
  \and
  \inferrule
  {\ttA:\nTy\\ \ttn:\nTm(\nNat)\\
  \ttt:\nTm(\ttA)\\\\
  \ttl:\nTm(\nVec(\ttA,\ttn))\\
  \ttm \mapsto \nSucc(\ttn) : \nTm(\nNat)
  }
  {\nCons(\ttn,\ttt, \ttl) :
  \nTm(\nVec(\ttA,\ttm))}
\end{mathpar}

Here, we update our informal notation by  writing the premise $\Gamma\vdash \texttt{x} \mapsto t : T$ for when the index substitution $\Bv_i$ instantiates the metavariable $\ttx\{\Gamma\} : T\in \Xi_i$ with the term $t$. These rules can thus be "parsed" into the formal notation as $\nrefl(\ttA:\nTy,\tta:\nTm(\ttA)\sep\cdot\sep\tta /\ttb : \nTm(\ttA)) : \nTm(\nEq(\ttA,\tta,\ttb))$ and $\nCons(\ttA:\nTy\sep \ttn : \nTm(\nNat), \ttt : \nTm(\ttA), \ttl : \nTm(\nVec(\ttA, \ttn))\sep\nSucc(\ttn) /\ttm : \nTm(\nNat)) : \nTm(\nVec(\ttA,\ttm))$.

As we will see in \cref{sec:bidirectional-type-system}, from the sort given as input we will now not only recover the parameters but also the expected values for the indices. Once we have type-checked the non-erased arguments, we will then have to check that, by replacing the parameters and non-erased arguments in $\Bv_i$, we get something convertible to what we found before. For instance, in the case of the the rule for $\nCons$, we will need to check that the value of $\ttm$ is convertible to successor of the value of $\ttn$.

\begin{remark}
  Whenever a constructor has no erased arguments (like is the case for $\nPi$), we do not actually need its sort to be given as input, and if we wanted we could also bidirectionally type it in mode infer. Nevertheless, we will prefer to value uniformity by imposing that constructors should \textit{always} be bidirectionally typed in mode check.%
\end{remark}

\subsubsection{Destructor rules}

Whereas constructors support type-checking, destructors support \textit{type-inference}. Therefore, instead of recovering the omitted arguments from the sort of the rule, they are retrieved by inferring the sort of a specific argument we call the \textit{principal argument}. The arguments of a destructor rule are thus separated between erased arguments $\Xi_{pi}$, the principal argument $\texttt{x}:T$ and the other non-omitted destructor arguments $\Xi_d$, where the sort of the principal argument should be a pattern over the metavariables of $\Xi_{pi}$. We therefore have rules of the following form:
\begin{align*}
  d ( \Xi_{pi} \mkgray{\in \MCtx_{|\mathbb{T}|}} \sep
    \texttt{x} : T\mkgray{\in \Tm_{|\mathbb{T}|}^{\Pat}~|\Xi_{pi}|~(\cdot)} \sep
    \Xi_d \mkgray{\in \MCtx_{|\mathbb{T}|}~|\Xi_{pi}, \texttt{x}:T|})
  : U\mkgray{ \in \Tm_{|\mathbb{T}|}~|\Xi_{pi}. (\texttt{x}:T).\Xi_d|~(\cdot) }
\end{align*}

The prime example of a destructor rule is the one for functional application:
\begin{mathpar}
  \inferrule
  {\texttt{A}:\nTy\\ x:\nTm(\texttt{A}) \vdash \texttt{B}:\nTy\\
    \texttt{t}:\nTm(\nPi(\texttt{A},x.\texttt{B}\{x\}))\\
    \texttt{u}:\nTm(\texttt{A})}
  {\nApp(\texttt{t},\texttt{u}):\nTm(\texttt{B}\{\texttt{u}\})}
\end{mathpar}
which is just an informal notation for \[
  \nApp(\texttt{A}:\nTy, \texttt{B}\{x : \nTm(\texttt{A})\} : \nTy \sep
    \texttt{t} : \nTm(\nPi(\texttt{A},x.\texttt{B}\{x\})) \sep
    \texttt{u}:\nTm(\texttt{A}) ) : \nTm(\texttt{B}\{\texttt{u}\})\]

Note that, in destructor rules, the principal argument is always the first non-erased argument. Because all the other non-erased arguments are typed in check mode, it is important that the erased arguments are known when typing them, given that their sorts or contexts might depend on the erased arguments. For instance, if we reorder the rule for application by placing $\ttu : \nTm(\ttA)$ before $\texttt{t}:\nTm(\nPi(\texttt{A},x.\texttt{B}\{x\}))$, then how could we type-check $\ttu$ against $\nTm(\ttA)$ if $\ttA$ is still not known? A similar reasoning justifies why the context corresponding to the principal argument is always empty. Indeed, if we had principal arguments of the form $\ttx\{\Gamma\}:T$ instead of $\ttx:T$, we would have to extend the current context by $\Gamma$ before inferring $\ttx$. However, $\Gamma$ may make reference to the erased arguments which are still not known at this point.

\subsubsection{Rewrite rules}

Finally, the last type of rules are rewrite rules, which specify the definitional equality (also known as conversion) of the theory. As suggested by our constructor/destructor separation of symbols, the left-hand sides of rewrite rules are to be headed by destructors. The right-hand side of the rule is then a term containing only the metavariables introduced by the left-hand side. The rewrite rules are hence of the form, where $\xi$ is such that $d(\xi) \in |\mathbb{T}|$. \[
  \theta\Vdash d(\Bt\mkgray{\in\MSub^\pat~\theta~\xi})\longmapsto r \mkgray{\in\Tm~\theta~(\cdot)}
\]

We shall also ask for a supplementary condition: in order to extend a theory~$\mathbb{T}$ with a rule $\theta \Vdash d(\Bt) \longmapsto r$, there can be no rule $\theta' \Vdash d(\Bt') \longmapsto r'$  in $\mathbb{T}$ such that we have $\Bt[\Bv]=\Bt'[\Bv']$ for some $\Bv$ and $\Bv'$.  As discussed in the next subsection, this will ensure that the rewrite system of a theory is confluent by construction.

The prototypical example of a rewrite rule is the computation rule for functions: the $\beta$-rule from the $\lambda$-calculus.
\begin{mathpar}
\texttt{t}\{x\},\texttt{u}\Vdash\nApp(\nLam(x.\texttt{t}\{x\}),\texttt{u})\longmapsto\texttt{t}\{\texttt{u}\}
\end{mathpar}

In the following we allow ourselves to omit the rule metavariable scope as it can be easily reconstructed by inspecting the rewrite rule's left hand side.

\begin{figure}
  \begin{align*}
    \fbox{\Thy}\ni \mathbb{T} ::= & \mid \cdot\\
    &\mid\mathbb{T},~c (\Xi \mkgray{\in \MCtx_{|\mathbb{T}|}}) ~\sort\\
    &\mid\mathbb{T},~c (\Xi_p\mkgray{\in \MCtx_{|\mathbb{T}|}}\sep \Xi_c\mkgray{\in \MCtx_{|\mathbb{T}|}~|\Xi_p|}\sep \Bv_i \mkgray{\in \MSub_{|\mathbb{T}|}~|\Xi_p.\Xi_c|~(\cdot)~|\Xi_i|} ~/~ \Xi_i\mkgray{\in \MCtx_{|\mathbb{T}|}~|\Xi_p|})\\
    &\hspace{4em}: T \mkgray{\in \Tm_{|\mathbb{T}|}^{\Pat}~|\Xi_p.\Xi_i|~(\cdot)}\\
    &\mid\mathbb{T},~d (\Xi_{pi}\mkgray{\in \MCtx_{|\mathbb{T}|}}\sep \ttx : T \mkgray{\in \Tm_{|\mathbb{T}|}^{\Pat}~|\Xi_{pi}|~(\cdot)}\sep\Xi_d\mkgray{\in \MCtx_{|\mathbb{T}|}~|\Xi_{pi},\texttt{x}:T|}) \\
    &\hspace{4em}: U\mkgray{\in \Tm_{|\mathbb{T}|}~|\Xi_{pi}.(\texttt{x}:T).\Xi_d|~(\cdot)}\\
    &\mid\mathbb{T},~ \theta\Vdash d(\Bt\mkgray{\in\MSub^\pat~\theta~\xi})\longmapsto r \mkgray{\in\Tm~\theta~(\cdot)}\\
    &\hspace{1.5em} \stext{if $d(\xi)\in|\mathbb{T}|$ and, for all $(\theta' \Vdash d(\Bt') \longmapsto r')\in\mathbb{T}$ and $\Bu, \Bu'$, we have $\Bt[\Bu] \neq \Bt'[\Bu']$}
  \end{align*}
  \caption{Definition of theories}
  \label{theories}
\end{figure}
\subsubsection{Underlying signature} As explained in the beginning of the section, theories $\mathbb{T} \in \Thy$ are specified mutually with their underlying signatures $|\mathbb{T}|\in\Sig$. We now give the definition of underlying signature, by the following clauses. As we can see, in both constructor and destructor rules the metavariable contexts of erased premises are indeed omitted from the syntax. Moreover, rewrite rules are simply ignored when computing the underlying signature.
\begin{align*}
  &|\_{}| : \Thy  \to \Sig &&|\mathbb{T},c(\Xi_{p} \sep \Xi_{c}\sep \Bv_i/\Xi_i):T| := |\mathbb{T}|, c(|\Xi_{c}|)\\
  &|\cdot| := \cdot &&|\mathbb{T},d(\Xi_{pi}\sep\texttt{x}:T\sep\Xi_{d}):U| := |\mathbb{T}|, d(\ttx,|\Xi_{d}|)\\
  &|\mathbb{T},c(\Xi)~\sort| := |\mathbb{T}|, c(|\Xi|)&&|\mathbb{T}, \theta\Vdash d(\Bt) \longmapsto r| := |\mathbb{T}|
  \end{align*}

  This concludes the definition of our theories, which for ease of reference we recapitulate  in \cref{theories}.

\begin{example}
  By putting together some of the rules seen in this subsection, we get the following theory \ref{ex:theory} defining a basic version of MLTT with only dependent functions.
  \begin{align*}\label{ex:theory}
    \tag{$\mathbb{T}_{\lambda\Pi}$} &\nTy(\cdot) ~ \sort, \hspace{0.7em} \nTm(\texttt{A}:\nTy) ~ \sort, \hspace{0.7em}\nPi(\cdot \sep \texttt{A}:\nTy,~ \texttt{B}\{x : \nTm(\texttt{A})\} : \nTy \sep \varepsilon/(\cdot)) : \nTy,\\
    &\nLam(\texttt{A}:\nTy,~ \texttt{B}\{x : \nTm(\texttt{A})\} : \nTy \sep
      \texttt{t}\{x : \nTm(\texttt{A})\} : \nTm(\texttt{B}\{x\}) \sep \varepsilon/(\cdot)) : \nTm(\nPi(\texttt{A},x.\texttt{B}\{x\})),\\
    &\nApp(\texttt{A}:\nTy,~ \texttt{B}\{x : \nTm(\texttt{A})\} : \nTy \sep
      \texttt{t} : \nTm(\nPi(\texttt{A},x.\texttt{B}\{x\})) \sep
      \texttt{u}:\nTm(\texttt{A}) ) : \nTm(\texttt{B}\{\texttt{u}\}),\\
    &\nApp(\nLam(x.\texttt{t}\{x\}), \texttt{u}) \longmapsto \texttt{t}\{\texttt{u}\}
  \end{align*}
  When computing its underlying signature $|\mathbb{T}_{\lambda\Pi}|$ we get the signature \ref{ex:sig}.
\end{example}

\subsection{Rewriting}

The rewrite rules of a theory $\mathbb{T}$ are used to define the \textit{rewriting relation} $e \red e'$ in \cref{reduction}. This definition is done simultaneously with the proof that reduction preserves underlying scopes: we have $|\Gamma| = |\Gamma'|$ whenever $\Gamma \red \Gamma'$, by an easy induction on $\Gamma$. The relations $\redd$ and $\equiv$ are then defined as usual, respectively as the  reflexive-transitive and reflexive-symmetric-transitive closures of $\red$. The relation $\equiv$ is called the \textit{definitional equality} (or \textit{conversion}) of the theory. One of the key properties of rewriting is its stability under substitution:

\begin{figure}\raggedright

  \begin{minipage}{0.52\linewidth}
    \boxed{t \red u\quad(
      t\in \Tm~\theta~\gamma;~
      u\in \Tm~\theta~\gamma)}
  \begin{mathpar}
    \inferrule
    {\vec{t} \red \vec{t}'}
    {\texttt{x}\{\vec{t}\} \red \texttt{x}\{\vec{t}'\}}
    \and
    \inferrule
    {\Bv \red \Bv'}
    {o(\Bv) \red o(\Bv')}
    \and
    \inferrule
    {(\theta_0 \Vdash d(\Bu) \longmapsto r)\in\mathbb{T}\\\Bt \in \MSub~\theta~\gamma~\theta_0}
    {d(\Bu[\Bt]) \red r[\Bt]}
  \end{mathpar}
  \vspace{0.5em}
  \boxed{\vec{t} \red \vec{u}\quad(
      \vec{t}\in \Sub~\theta~\gamma~\delta;~
      \vec{u}\in \Sub~\theta~\gamma~\delta)}
  \begin{mathpar}
    \inferrule
    {\vec{t}\red\vec{t}'}
    {\vec{t},u\red \vec{t}',u}
    \and
    \inferrule
    {u\red u'}
    {\vec{t},u\red \vec{t},u'}
  \end{mathpar}
  \end{minipage}
  \begin{minipage}{0.47\linewidth}
  \boxed{\Bv \red \Bu\quad(
      \Bv\in \MSub~\theta~\gamma~\xi;~
      \Bu\in \MSub~\theta~\gamma~\xi)}
  \begin{mathpar}
    \inferrule
    {\Bv\red\Bv'}
    {\Bv,\vec{x}.u\red \Bv',\vec{x}.u}
    \and
    \inferrule
    {u\red u'}
    {\Bv,\vec{x}.u\red \Bt,\vec{x}.u'}
  \end{mathpar}

  \vspace{0.5em}

  \boxed{\Gamma \red \Delta\quad(
      \Gamma\in \Ctx~\theta~\gamma;~
      \Delta\in \Ctx~\theta~\gamma)}
  \begin{mathpar}
    \inferrule
    {\Gamma \red \Gamma'}
    {\Gamma,x:T \red \Gamma', x:T}
    \and
    \inferrule
    {T \red T'}
    {\Gamma,x:T \red \Gamma, x:T'}
  \end{mathpar}\end{minipage}
  \caption{Rewriting relation defined by theory $\mathbb{T}$}
  \label{reduction}
  \end{figure}

\begin{proposition}[Stability of rewriting under substitution]
  Let $e \in \Expr~\theta~\gamma$ with $e \red^{*}e'$. If $\vec{v}\in\Sub~\theta~\delta~\gamma$ and $\vec{v} \red^{*} \vec{v}'$ then $e[\vec{v}]\red^{*} e'[\vec{v}']$. If $\Bv\in\MSub~\xi~\gamma~\theta$ and $\Bv \red^{*} \Bv'$ then $e[\Bv]\red^{*} e'[\Bv']$.
  \end{proposition}
  \begin{proof}
    We first show that $\vec{v}\red^{*}\vec{v}'$ implies $e[\vec{v}]\red^{*} e[\vec{v}']$, by induction on $e$, and then that $e \red e'$ implies $e[\vec{v}] \red e'[\vec{v}]$, by induction on $e \red e'$. By iterating these two statements, we then conclude that $e \red^* e'$ and $\vec{v} \red^* \vec{v}'$ imply $e[\vec{v}]\red^*e'[\vec{v}']$.

    Then, we show that $\Bv\red^{*}\Bv'$ implies $e[\Bv]\red^{*} e[\Bv']$, by induction on $e$, and then that $e \red e'$ implies $e[\Bv] \red e'[\Bv]$, by induction on $e \red e'$. By iterating these two statements, we then conclude that $e \red^* e'$ and $\Bv \red^* \Bv'$ imply $e[\Bv]\red^*e'[\Bv']$.
  \end{proof}

  This implies in particular that conversion is stable under substitution.

  \begin{corollary}[Stability of conversion under substitution]\label{stability-conversion-subst}
  Suppose $e \equiv e'$. We have $e[\vec{v}]\equiv e'[\vec{v}']$ for all $\vec{v}\equiv \vec{v}'$ and $e[\Bv]\equiv e'[\Bv']$ for all $\Bv\equiv \Bv'$.
  \end{corollary}

  \subsubsection{Confluence}

  Recall that when defining theories we asked that no two different left-hand sides should unify. Because this is the only way two rule can overlap, this means that there are no \textit{critical pairs}~\cite{bezem2003term}. Therefore, because rules are also all \textit{left-linear}, it follows that the rewrite system of any theory is \textit{orthogonal}, hence confluent by~\cite[Theorem~6.11]{mayr1998higher}.
\begin{proposition}[Confluence]\label{confluence}
  If $e'\iredd e \redd e''$ then there is some $e'''$ such that $e' \redd e''' \iredd e''$. In particular, if $e \equiv e'$ then we have $e \redd e'' \iredd e'$ for some~$e''$.
  \end{proposition}

  One of the main consequences of confluence is that patterns are injective modulo  conversion:

  \begin{proposition}[Injectivity of patterns]\label{pat-inj}
  If $t \in \Tm^{\Pat}~\theta~\gamma$ and $t[\Bv] \equiv t[\Bv']$ for some $\Bv \in \MSub~\theta'~\delta~\theta$ and $\Bv' \in \MSub~\theta'~\delta~\theta$  then $\Bv \equiv \Bv'$.
  \end{proposition}
  \begin{proof}
  We need to show simultaneously a similar result for metavariable substitution patterns: if $\Bt \in \MSub^\Pat~\theta~\gamma~\xi$ and $\Bt[\Bv]\equiv\Bt[\Bv']$ then $\Bv\equiv\Bv'$. The proof is done by induction on the pattern, using confluence for the case $c(\Bt[\Bv]) \equiv c(\Bt[\Bv'])$ to deduce $\Bt[\Bv]\equiv\Bt[\Bv']$.
  \end{proof}

\section{Declarative type system}%
\label{sec:declarative-type-sytem}

In the previous section we have seen that theories are specified by rewrite rules and schematic typing rules. In this section, we shall see how such schematic rules can be instantiated into concrete typing rules, defining the declarative type system of the corresponding theory.
This system is then used to define the \textit{valid theories}, a refinement of our notion of theory in which typing information is also taken into account. We then conclude the section by showing that the declarative system satisfies desirable properties, such as weakening and substitution.

\begin{figure}\raggedright
  \begin{minipage}{0.45\linewidth}
    \fbox{$ \Theta\vdash \quad (
      \Theta \in \MCtx)$}

    \begin{mathpar}
      \inferrule[EmptyMCtx]
      { }
      {\cdot \vdash}
      \and
      \inferrule[ExtMCtx]
      {\Theta;\Gamma \vdash T~\sort}
      {\Theta, \texttt{x} \{\Gamma\} : T \vdash }
    \end{mathpar}
  \end{minipage}
  \begin{minipage}{0.45\linewidth}
    \fbox{$ \Theta;\Gamma\vdash \quad (
      \Theta \in \MCtx ;~
      \Gamma \in \Ctx~|\Theta|)$}

    \begin{mathpar}
      \inferrule[EmptyCtx]
      {\Theta \vdash }
      {\Theta;\cdot \vdash}
      \and
      \inferrule[ExtCtx]
      {\Theta;\Gamma \vdash T~\sort}
      {\Theta; \Gamma, x : T \vdash }
    \end{mathpar}
  \end{minipage}

  \vspace{0.5em}

  \fbox{$ \Theta;\Gamma  \vdash T~\sort \quad (
    \Theta \in \MCtx;~
    \Gamma \in \Ctx~|\Theta|;~
    T \in \Tm~|\Theta|~|\Gamma|)$}
    \begin{mathpar}
      c(\Xi)~\sort\in\mathbb{T}
      \inferrule[Sort]
      { \Theta;\Gamma \vdash \Bt : \Xi}
      {\Theta;\Gamma  \vdash c(\Bt)~\sort}
    \end{mathpar}

  \vspace{0.5em}

  \fbox{$ \Theta;\Gamma  \vdash t:T \quad (
    \Theta \in \MCtx;~
    \Gamma \in \Ctx~|\Theta|;~
    T \in \Tm~|\Theta|~|\Gamma|;~
    t \in \Tm~|\Theta|~|\Gamma|)$}
\begin{mathpar}
  x:T\in\Gamma~
  \inferrule[Var]
  {\Theta; \Gamma \vdash}
  {\Theta;\Gamma  \vdash x:T}
  \and
  \texttt{x} \{\Delta\} :T \in \Theta~
  \inferrule[MVar]
  { \Theta;\Gamma \vdash \vec{t}: \Delta}
  {\Theta;\Gamma  \vdash \texttt{x}\{\vec{t}\} : T[\vec{t}]}
  \and
  c (\Xi_p\sep  \Xi_c\sep \Bv_i/\Xi_i) : T \in \mathbb{T}~
  \inferrule[Cons]
  { \Theta;\Gamma \vdash \Bt_p, \Bt:\Xi_p.\Xi_c\\
    \Theta;\Gamma \vdash T[\Bt_p, \Bv_i[\Bt_p,\Bt]]~\sort}
  {\Theta;\Gamma  \vdash c(\Bt) : T[\Bt_p, \Bv_i[\Bt_p,\Bt]]}
  \and
  d (\Xi_{pi} \sep \texttt{x}:T \sep \Xi_d) : U \in \mathbb{T}~
  \inferrule[Dest]
  {\Theta;\Gamma \vdash \Bt_{pi}, \Bt : \Xi_{pi}.(\texttt{x}:T).\Xi_d}
  {\Theta;\Gamma  \vdash d(\Bt) : U[\Bt_{pi},\Bt]}
  \and
  T\equiv U~
  \inferrule[Conv]
  {\Theta;\Gamma \vdash t : T\\ \Theta;\Gamma \vdash U~\sort}
  {\Theta;\Gamma \vdash t : U}
\end{mathpar}

\vspace{0.5em}

\fbox{$ \Theta;\Gamma  \vdash \vec{t}: \Delta \quad
  (\Theta \in \MCtx;~
  \Gamma \in \Ctx~|\Theta|;~
  \Delta \in \Ctx~|\Theta|;~
  \vec{t} \in \Sub~|\Theta|~|\Gamma|~|\Delta|)$}
\begin{mathpar}
  \inferrule[EmptySub]
  {\Theta;\Gamma \vdash}
  {\Theta;\Gamma  \vdash \varepsilon : (\cdot)}
  \and
  \inferrule[ExtSub]
  {\Theta;\Gamma \vdash \vec{t} : \Delta \\ \Theta;\Gamma \vdash t : T[\vec{t}]}
  {\Theta;\Gamma \vdash \vec{t},  t : (\Delta , x : T)}
\end{mathpar}

\vspace{0.5em}

\fbox{$ \Theta;\Gamma  \vdash \Bt: \Xi \quad
  (\Theta \in \MCtx;~
  \Gamma \in \Ctx~|\Theta|;~
  \Xi \in \MCtx;~
  \Bt \in \MSub~|\Theta|~|\Gamma|~|\Xi|)$}
\begin{mathpar}
  \inferrule[EmptyMSub]
  {\Theta;\Gamma \vdash}
  {\Theta;\Gamma  \vdash \varepsilon : (\cdot)}
  \and
  \inferrule[ExtMSub]
  {\Theta;\Gamma \vdash \Bt : \Xi \\ \Theta;\Gamma.\Delta[\Bt] \vdash t : T[\Bt]}
  {\Theta;\Gamma \vdash \Bt, \vec{x}_\Delta.  t : (\Xi , \texttt{x}\{\Delta\} : T)}
\end{mathpar}
\caption{Declarative typing rules}
\label{fig:declarative-typing}
\end{figure}

\subsection{Declarative typing rules}

Given a fixed theory $\mathbb{T}$, the declarative type system is defined by the rules in \cref{fig:declarative-typing}. The system is split in 6 judgments:
\begin{itemize}
  \item $\Theta\vdash$ : Well-typedness of metavariable context $\Theta$.
  \item $\Theta;\Gamma\vdash$ : Well-typedness of variable context $\Gamma$ under metavariable context~$\Theta$.
  \item $\Theta;\Gamma \vdash T~\sort$ : Well-typedness of sort $T$ under contexts $\Theta;\Gamma$.
  \item $\Theta;\Gamma\vdash t : T$ : Typing of a term $t$ by $T$ under context $\Theta;\Gamma$.
  \item $\Theta;\Gamma \vdash \vec{t}:\Delta$ : Typing of a variable substitution $\vec{t}$ by $\Delta$ under context $\Theta;\Gamma$.
  \item $\Theta;\Gamma \vdash \Bt:\Xi$ : Typing of a metavariable substitution $\Bt$ by $\Xi$ under context~$\Theta;\Gamma$.
\end{itemize}

The most important rules are the ones which instantiate schematic typing rules: \textsc{Cons}, \textsc{Dest} and \textsc{Sort}. For instance, in order to use \textsc{Dest} to type $d(\Bt)$ a metavariable substitution $\Bt_{pi}$ not stored in the syntax must be "guessed", and then we must show that $\Bt_{pi},\Bt$ is typed by $\Xi_{pi}.(\texttt{x}:T).\Xi_d$. The rules for typing a metavariable substitution can then be applied, which has the effect of unfolding the judgment $\Bt_{pi},\Bt : \Xi_{pi}.(\texttt{x}:T).\Xi_d$ into regular term typing judgments.  At the end of this unfolding process, the resulting derivation has basically the same shape as the schematic typing rule for $d$, and it can be understood as its instantiation. Let us look at a concrete example of this.
\begin{example}
  Suppose we want to show that $\nApp(t,u)$ is well-typed in the theory~\ref{ex:theory}. Because $\nApp$ is a destructor symbol with schematic rule\[
    \nApp(\texttt{A}:\nTy, \texttt{B}\{x : \nTm(\texttt{A})\} : \nTy \sep
    \texttt{t} : \nTm(\nPi(\texttt{A},x.\texttt{B}\{x\})) \sep
    \texttt{u}:\nTm(\texttt{A})  ) : \nTm(\texttt{B}\{\texttt{u}\})
  \]by guessing some $A$ and $B$ we can start the derivation with rule \textsc{Dest}, giving
  \begin{mathpar}
    \inferrule
    {\Theta;\Gamma \vdash A, x.B, t, u : (\texttt{A}:\nTy,~\texttt{B}\{x : \nTm(\texttt{A})\} : \nTy,~\texttt{t} : \nTm(\nPi(\texttt{A},x.\texttt{B}\{x\})),~
    \texttt{u}:\nTm(\texttt{A}))}
    {\Theta;\Gamma \vdash \nApp(t,u):\nTm(\texttt{B}\{\texttt{u}\})[A,x.B,t,u]}
  \end{mathpar}
  If we note that $\nTm(\texttt{B}\{\texttt{u}\})[A,x.B,t,u] = \nTm(B[\id_\Gamma,u])$, and we continue by applying the rules defining the judgment $\Theta;\Gamma\vdash \Bt:\Xi$, we get
  \begin{mathpar}
    \inferrule
    {\Theta;\Gamma \vdash\\ \Theta;\Gamma \vdash A: \nTy\\ \Theta;\Gamma,x:\nTm(A) \vdash B:\nTy\\\\ \Theta;\Gamma \vdash t : \nTm(\nPi(A,x.B))\\ \Theta;\Gamma \vdash u : \nTm(A)}
    {\Theta;\Gamma \vdash \nApp(t,u):\nTm(B[\id_\Gamma,u])}
  \end{mathpar}
  which can be understood as the instantiation of the schematic rule for $\nApp$. Note that sometimes the first three premises of this rule are omitted, but this is only justified because they are admissible from the other ones, a result we could also show here by applying results from \cref{subsec:metaproperties}.

\end{example}

\begin{remark}\label{verbose-cons-remark}
  In the rule for \textsc{Cons} it might seem odd that we also ask the sort to be well-typed, whereas this hypothesis is not needed in rules \textsc{Dest}, \textsc{Meta} and \textsc{Var}.
  The reason is that in the proof of \cref{annotability} we will need to apply the induction hypothesis to the sort $T$ of the term $c(\Bt)$, and thus we need a derivation of $T~\sort$ smaller than the one of $c(\Bt):T$ we start with.
  Nevertheless, we will show in \cref{subsec:metaproperties} that this extra hypothesis is admissible, allowing us to use the economic version of the rule when building derivations.
  A similar technique is also employed by Harper et al~\cite{harper2005equivalence} and Abel et al~\cite{abelDecidabilityConversionType2018}.
\end{remark}

\newcommand\derives\triangleright

\begin{notation} We finish this subsection by establishing some notations.
  \begin{enumerate}
    \item We write $\Theta;\Gamma\vdash \mathcal{J}$ for any of the following: $\Theta;\Gamma\vdash$ or $\Theta;\Gamma\vdash T~\sort$ or $\Theta;\Gamma\vdash t:T$ or $\Theta;\Gamma\vdash \vec{t}:\Delta$ or $\Theta;\Gamma\vdash \Bt:\Xi$.
    \item We write $\mathbb{T}\derives \Theta;\Gamma\vdash\mathcal{J}$ when $\mathbb{T}$ is not clear from the context.
    \item We write $\Theta\vdash \mathcal{J}$ for $\Theta;\cdot\vdash\mathcal{J}$ and $\Gamma\vdash\mathcal{J}$ for $\cdot;\Gamma\vdash\mathcal{J}$.
  \end{enumerate}
\end{notation}

\subsection{Valid theories}\label{well-typed-theories}

\begin{figure}[b]
  \begin{mathpar}
    \inferrule
    {  }
    { \cdot \vdash  }
    \and
    \inferrule
    {\mathbb{T}\vdash\\ \mathbb{T} \derives \Xi \vdash}
    {\mathbb{T}, c(\Xi)~\sort\vdash }
    \and
    \inferrule
    {\mathbb{T}\vdash\\
     \mathbb{T} \derives \Xi_{pi}.(\texttt{x}:T). \Xi_d \vdash U~\sort}
    {\mathbb{T}, d(\Xi_{pi}\sep\texttt{x}:T\sep\Xi_d):U\vdash}
    \\
    \inferrule
    {\mathbb{T}\vdash \\ \mathbb{T} \derives \Xi_p. \Xi_c \vdash \id, \Bv_i : \Xi_p.\Xi_i\\ \mathbb{T} \derives \Xi_p.\Xi_i \vdash T~\sort}
    {\mathbb{T}, c(\Xi_p\sep \Xi_c\sep\Bv_i/\Xi_i):T\vdash }
    \and %
    \inferrule
    {\mathbb{T}\vdash}
    {\mathbb{T}, (\theta \Vdash d(\Bt) \longmapsto r)\vdash }
  \end{mathpar}
  \caption{Well-typed theories}\label{fig-well-typed-theories}
  \end{figure}

Our definition of theories given in \cref{subsec:theories} specifies the desired syntax but  imposes not typing constraints whatsoever, allowing for non-sensible and ill-behaved theories. But now that we have introduced typing rules, we can impose such constraints \textit{a posteriori} by defining the \textit{valid theories}.

Our first step to do this is to define the \textit{well-typed theories} with the judgment $\mathbb{T}\vdash$ specified by \cref{fig-well-typed-theories}.
The definition of $\mathbb{T}\vdash$ ensures that each time we extend a theory $\mathbb{T}$ with a schematic typing,  $\mathbb{T}$ can justify that the new rule is well-typed. For sort rules $c(\Xi)~\sort$ this amounts to ensuring that the metavariable context $\Xi$ is well-typed, whereas for destructor rules $d(\Xi_{pi}\sep\texttt{x}:T\sep \Xi_d): U$ this means ensuring that $U$ is a well-typed sort in metavariable context $\Xi_{pi}.(\texttt{x}:T). \Xi_d$ --- implying in particular that the metavariable context is also well-typed.
For typing a constructor $c(\Xi_p\sep\Xi_c\sep\Bv_i/\Xi_i):T$ we do not only ask $T$ to be a well-typed sort in $\Xi_p.\Xi_i$, but also $\Bv_i$ to be typed by $\Xi_i$. However, the judgment $\Xi_p.\Xi_c \vdash \Bv_i : \Xi_i$ would not be well-formed: in the definition of $\Theta;\Gamma\vdash\Bt:\Xi$ in \cref{fig:declarative-typing} we ask $\Xi\in\MCtx$, meaning that metavariable substitutions can only be typed by closed metavariable contexts. To eliminate the dangling metavariables of $\Xi_i$, we therefore prefix it with $\Xi_p$ and instead ask $\Xi_p.\Xi_c\vdash\id,\Bv_i : \Xi_p.\Xi_i$ to be derivable.

\begin{remark}
Note that the definition of $\mathbb{T}\vdash$ asks schematic typing rules to be typed incrementally, which excludes theories that rely on circularities --- for instance, when a rule depends on itself to be well-typed. Nevertheless, by a form of weakening for theories we can still deduce that all schematic typing rules of $\mathbb{T}$ can also be typed in $\mathbb{T}$ itself (for instance, if $c(\Xi)~\sort\in \mathbb{T}$ and $\mathbb{T}\vdash$ then $\mathbb{T}\derives\Xi\vdash$), a fact that we will often use without announcement in the proofs to come.
\end{remark}

The definition of well-typed theory ensures that the schematic rules are well-behaved, however it imposes no restriction whatsoever on rewrite rules. To remedy this, let us say that a rewrite rule $\theta\Vdash d(\Bt)\longmapsto r$ is \textit{type-preserving} if $\Gamma \vdash d(\Bt[\Bv]) : T$ implies $\Gamma\vdash r[\Bv] : T$ for all $\Gamma \in \Ctx$ and $T \in \Tm~(\cdot)~|\Gamma|$ and $\Bv \in \MSub~(\cdot)~|\Gamma|~\theta$. As we will see in the end of \cref{subsec:metaproperties}, this local form of subject reduction implies (global) subject reduction, allowing one to establish the latter more easily. We then say that a theory is \textit{valid} if it is well-typed and all its rewrite rules are type-preserving.

\begin{example}\label{ex:well-typed-theory}
  It is tedious but uncomplicated to see that the theory \ref{ex:theory} is valid. Checking its well-typedness is straightfoward, the most  interesting part is  verifying that the $\beta$-rule is type-preserving.  In order to do this, we start with a generic derivation of $\Gamma \vdash \nApp(\nLam(x.\ttt\{x\}), \ttu)[x.t,u] : T$, where $\Gamma$, $T$, and $x.t,u$ are any, and we must show $\Gamma\vdash \ttt\{\ttu\}[x.t,u]:T$. Note that we have $\nApp(\nLam(x.\ttt\{x\}), \ttu)[x.t,u] = \nApp(\nLam(x.t),u)$ and $\ttt\{\ttu\}[x.t,u] = t[\id,u]$, so we start by applying inversion of typing to $\Gamma \vdash \nApp(\nLam(x.t), u) : T$ to get
  \[
  \Gamma \vdash A : \nTy \hspace{2em}
  \Gamma,x:\nTm(A)\vdash B : \nTy \hspace{2em}
  \Gamma\vdash \nLam(x.t):\nTm(\nPi(A,x.B)) \hspace{2em}
  \Gamma\vdash u : \nTm(A)
  \]for some $A, B$ with $T \equiv \nTm(\ttB\{\ttu\})[A,x.B,t,u]=\nTm(B[\id,u])$. By inversion again, but this time on $\Gamma \vdash\nLam(x.t) : \nTm(\nPi(A,x.B))$, we have \[
  \Gamma \vdash A' : \nTy \hspace{2em}
  \Gamma,x:\nTm(A')\vdash B' : \nTy \hspace{2em}
  \Gamma,x:\nTm(A')\vdash t :\nTm(B')
  \]for some $A',B'$ with $\nTm(\nPi(A,x.B))\equiv \nTm(\nPi(A',x.B'))$, which by \cref{pat-inj} gives $A \equiv A'$ and $B \equiv B'$. Now we can conclude by applying some of the methatheorems we will~see~in~the next subsection, that crucially \textit{do not} rely on type-preservation (which would otherwise incur~a~circularity in our reasoning) but only on well-typedness of the theory, which we already have at this point.

  More precisely, from the above derivations we can show $\Gamma, x : \nTm(A) \vdash$ and $\Gamma, x: \nTm(A)\vdash \nTm(B)~\sort$, so starting from $\Gamma,x:\nTm(A')\vdash t:\nTm(B')$ we can applying \cref{conversion-in-context} and conversion to obtain $\Gamma,x:\nTm(A)\vdash t : \nTm(B)$. We  can also show $\Gamma \vdash \id, u : (\Gamma, x: \nTm(A))$, so by \cref{substitution-property} we then get $\Gamma \vdash t[\id,u]:\nTm(B[\id,u])$. Finally, by applying \cref{types-are-well-typed} to $\Gamma \vdash \nApp(\nLam(x.t),u):T$ we get $\Gamma\vdash T~\sort$, so by applying conversion with $\nTm(B[\id,u])\equiv T$ we conclude $\Gamma\vdash t[\id,u]: T$.
\end{example}

\begin{remark}
  Verifying that a theory is valid is often a tedious task, however we do not expect users of our framework to do such calculations by hand all the time. Indeed, our tool implements an automatic check that try to verify this condition automatically. The implemented criterion is incomplete, but works well in most common cases.

\end{remark}

\subsection{Metatheory}\label{subsec:metaproperties}

We now show some basic metaproperties satisfied by the declarative type system. Most of these properties hold even when the theory is not well-typed or valid, so such assumptions will be stated explicitly~when~needed.

  \begin{proposition}[Weakening]\label{weakening}
    Let us write $\Gamma \sqsubseteq \Delta$ if $\Gamma$ is a subsequence of $\Delta$, and $\Theta\sqsubseteq \Xi$ if $\Theta$ is a subsequence of $\Xi$. The following rules are admissible.
    \begin{mathpar}
      \Gamma \sqsubseteq \Delta
      \inferrule
      {\Theta;\Gamma \vdash \mathcal{J}\\ \Theta;\Delta\vdash}
      {\Theta;\Delta\vdash \mathcal{J}}
      \and
      \Theta \sqsubseteq \Xi
      \inferrule
      {\Theta;\Gamma\vdash \mathcal{J}\\ \Xi\vdash}
      {\Xi;\Gamma\vdash\mathcal{J}}
    \end{mathpar}
  \end{proposition}

  \begin{proof}
  In order for the induction to go through, we strengthen the first statement: instead we show that $\Theta;\Gamma.\Gamma'\vdash\mathcal{J}$ and $\Theta;\Delta\vdash$ and $\Gamma\sqsubseteq\Delta$ imply $\Theta;\Delta.\Gamma'\vdash\mathcal{J}$. The proof is then by induction on $\Theta;\Gamma.\Gamma' \vdash \mathcal{J}$ for the first statement, and on $\Theta;\Gamma \vdash \mathcal{J}$ for the second.
  \end{proof}

  Our theories also satisfy a substitution property, meaning that if $\Theta;\Gamma \vdash \mathcal{J}$ is derivable, then by applying any substitution typed by $\Gamma$ or metavariable substitution typed by  $\Theta$ the resulting judgment is still derivable. In order to state this property precisely, we first need to explain what it means to apply a substitution to a judgment $\mathcal{J}$. This is specified by the following table. Note that in the case $\mathcal{J}= \vec{t}: \Delta$, taking $\mathcal{J}[\Bv]_\Gamma := \vec{t}[\Bv] : \Delta[\Bv]$ would not in general yield a well-formed judgment, given that $\Bv$ might introduce dangling variables in $\Delta[\Bv]$. Therefore, we need to prefix~$\Delta[\Bv]$ with the context  $\Gamma$ of the substitution, and fill the positions with the identity $\id$ --- the same trick we used in \cref{well-typed-theories}. This is also the reason why the application of a metavariable substitution is annotated with $\Gamma$.

  \begin{table}[H]\centering
    \begin{tabular}{l|ll}
      $\vdash \mathcal{J}$ & $\vdash\mathcal{J}[\vec{v}]$  & $\vdash\mathcal{J}[\Bv]_\Gamma$ \\\hline
      $\vdash$  &$\vdash$          &$\vdash$   \\
      $\vdash T~\sort$ & $\vdash T[\vec{v}]~\sort $         & $\vdash T[\Bv]~\sort $      \\
      $\vdash t: T$          & $\vdash t[\vec{v}]:T[\vec{v}]$          & $\vdash t[\Bv]:T[\Bv] $\\
      $\vdash\vec{t}: \Delta$          & $\vdash \vec{t}[\vec{v}]: \Delta$          & $\vdash \id,\vec{t}[\Bv]: \Gamma.\Delta[\Bv]$\\
      $\vdash\Bt : \Xi$          & $\vdash \Bt[\vec{v}] : \Xi$          & $\vdash \Bt[\Bv] : \Xi$
    \end{tabular}
  \end{table}

  \begin{proposition}[Substitution property]\label{substitution-property} The following rules are admissible.
  \begin{mathpar}
    \inferrule
    {\Theta;\Gamma \vdash \vec{v}: \Delta\\\Theta;\Delta \vdash \mathcal{J}}
    {\Theta;\Gamma \vdash \mathcal{J}[\vec{v}]}
    \and
    \inferrule
    {\Xi;\Gamma \vdash \Bv: \Theta\\\Theta;\Delta \vdash \mathcal{J}}
    {\Theta;\Gamma.\Delta[\Bv] \vdash \mathcal{J}[\Bv]_\Gamma}
  \end{mathpar}
  \end{proposition}
  \begin{proof}
  For the proof to go through, we strengthen the first statement in the~following~way:
  \begin{mathpar}
    \inferrule
    {\Theta;\Gamma \vdash \vec{v}: \Delta\\\Theta;\Delta.\Gamma' \vdash \mathcal{J}}
    {\Theta;\Gamma.\Gamma'[\vec{v}] \vdash \mathcal{J}[\vec{v},\id]}
  \end{mathpar}

  Then both statements can be shown by induction, on $\Theta;\Delta.\Gamma' \vdash \mathcal{J}$ for the first statement, and $\Theta;\Delta\vdash \mathcal{J}$ for the second. Most cases follow directly from the induction hypothesis, the basic properties of substitution (\cref{unit-subst,commutation-subst,assoc-subst}) and, for the rule \textsc{Conv}, from the stability of  conversion under substitution (\cref{stability-conversion-subst}). We show the key cases, and illustrate the other ones by some representative cases.
  \begin{itemize}
    \item Case \textsc{Var} of the first statement.
    \begin{mathpar}
      x :T \in \Delta.\Gamma'~
      \inferrule
      {\Theta; \Delta.\Gamma' \vdash }
      {\Theta;\Delta.\Gamma'  \vdash x:T}
    \end{mathpar}
    We have either $x:T \in \Gamma'$ or $x:T\in\Delta$. In the first case, we apply the i.h. to get $\Theta;\Gamma.\Gamma'[\vec{v}]\vdash$ and then conclude with the variable rule. Otherwise, if $x:T\in\Delta$ then from $\Theta;\Gamma\vdash\vec{v}:\Delta$ we first get $\Theta;\Gamma\vdash v_x : T[\vec{v}]$. Then, by i.h. we have $\Theta;\Gamma.\Gamma'[\vec{v}]\vdash$, so we can apply \cref{weakening} to get $\Theta;\Gamma.\Gamma'[\vec{v}]\vdash v_x : T[\vec{v}]$. Because we have $T[\vec{v}]=T[\vec{v},\id]$ we are done.

    \item Case \textsc{Dest} of the first statement.
    \begin{mathpar}
      d (\Xi_{pi} \sep \texttt{x}:T \sep \Xi_d) : U \in \mathbb{T}~
      \inferrule
      {\Theta;\Delta.\Gamma' \vdash \Bt_{pi}, \Bu : \Xi_{pi}.(\texttt{x}:T).\Xi_d}
      {\Theta;\Delta.\Gamma'  \vdash d(\Bu) : U[\Bt_{pi},\Bu]}
    \end{mathpar}
    By i.h. we have $\Theta;\Gamma.\Gamma'[\vec{v}]\vdash (\Bt_{pi},\Bu)[\vec{v},\id] : \Xi_p.\Xi_i.(\texttt{x}:T).\Xi_d$, therefore we can derive $\Theta;\Gamma.\Gamma'[\vec{v}]  \vdash d(\Bu[\vec{v},\id]) : U[(\Bt_{pi},\Bu)[\vec{v},\id]]$. Because  $U[(\Bt_{pi},\Bu)[\vec{v},\id]]= U[\Bt_{pi},\Bu][\vec{v},\id]$, we are done.

    \item Case \textsc{ExtMSub} of the first statement.
    \begin{mathpar}
      \inferrule
      {\Theta;\Delta.\Gamma' \vdash \Bt : \Xi \\ \Theta;\Delta.\Gamma'.\Delta_\texttt{x}[\Bt] \vdash t : T[\Bt]}
      {\Theta;\Delta.\Gamma' \vdash \Bt, \vec{x}_\Delta.  t : (\Xi , \texttt{x}\{\Delta_\texttt{x}\} : T)}
    \end{mathpar}
    By i.h., $\Theta;\Gamma.\Gamma'[\vec{v}] \vdash \Bt[\vec{v},\id] : \Xi$ and $\Theta;\Gamma.(\Gamma'.\Delta_\texttt{x}[\Bt])[\vec{v}] \vdash t[\vec{v},\id] : T[\Bt][\vec{v},\id]$. We have  $(\Gamma'.\Delta_\texttt{x}[\Bt])[\vec{v}] = \Gamma'[\vec{v}].\Delta_\texttt{x}[\Bt[\vec{v},\id]]$ and $T[\Bt][\vec{v},\id]=T[\Bt[\vec{v},\id]]$, and so we also have\\ $\Theta;\Gamma.\Gamma'[\vec{v}].\Delta_\texttt{x}[\Bt[\vec{v},\id]]\vdash t[\vec{v},\id] : T[\Bt[\vec{v},\id]]$. Thus, we can build a derivation of $\Theta;\Gamma.\Gamma'[\vec{v}]\vdash \Bt[\vec{v},\id],\vec{x}_{\Delta_\texttt{x}}. t[\vec{v},\id] : (\Xi, \texttt{x}\{\Delta_\texttt{x}\} : T)$, and because $(\Bt,\vec{x}_{\Delta_\texttt{x}}.t)[\vec{v},\id]=\Bt[\vec{v},\id], \vec{x}_{\Delta_\texttt{x}}.t[\vec{v},\id]$ we are done.

    \item Case \textsc{MVar} of the second statement.
    \begin{mathpar}
      \texttt{x} \{\Delta'\} :T\in \Theta~
      \inferrule
      { \Theta;\Delta \vdash \vec{t}: \Delta'}
      {\Theta;\Delta  \vdash \texttt{x}\{\vec{t}\} : T[\vec{t}]}
    \end{mathpar}
    By i.h. we have $\Xi;\Gamma.\Delta[\Bv] \vdash \id,\vec{t}[\Bv]:\Gamma.\Delta'[\Bv]$. Moreover, from $\Xi;\Gamma\vdash\Bv:\Theta$ we can deduce $\Xi;\Gamma.\Delta'[\Bv]\vdash \Bv_\texttt{x}:T[\Bv]$, so by the substitution property for variable substitutions we get $\Xi;\Gamma.\Delta[\Bv]\vdash \Bv_\texttt{x}[\id,\vec{t}[\Bv]]:T[\Bv][\id,\vec{t}[\Bv]]$, and because $\texttt{x}\{\vec{t}\}[\Bv]=\Bv_\texttt{x}[\id,\vec{t}[\Bv]]$ and  $T[\Bv][\id,\vec{t}[\Bv]]~=~T[\vec{t}][\Bv]$ we are done

     \item Case \textsc{EmptySub} of the second statement.
    \begin{mathpar}
      \inferrule
      {\Theta;\Delta \vdash}
      {\Theta;\Delta  \vdash \varepsilon : (\cdot)}
    \end{mathpar}
    By i.h. we have $\Xi;\Gamma.\Delta[\Bv]\vdash$. We can show $\Xi;\Gamma.\Delta[\Bv]\vdash \id:\Gamma$ and so we are done.

    \item Case \textsc{ExtMSub} of the second statement.
    \begin{mathpar}
      \inferrule
      {\Theta;\Delta \vdash \Bt : \Xi \\ \Theta;\Delta.\Delta'[\Bt] \vdash t : T[\Bt]}
      {\Theta;\Delta \vdash \Bt, \vec{x}_\Delta.  t : (\Xi , \texttt{x}\{\Delta'\} : T)}
    \end{mathpar}
    By i.h. we have $\Xi;\Gamma.\Delta[\Bv]\vdash \Bt[\Bv]:\Xi$ and $\Xi;\Gamma.(\Delta.\Delta'[\Bt])[\Bv]\vdash t[\Bv]:T[\Bt][\Bv]$. We have $ (\Delta.\Delta'[\Bt])[\Bv] = \Delta[\Bv].\Delta'[\Bt[\Bv]]$ and $T[\Bt][\Bv]=T[\Bt[\Bv]]$, therefore $\Xi;\Gamma.\Delta[\Bv].\Delta'[\Bt[\Bv]]\vdash t[\Bv]:T[\Bt[\Bv]]$. We can thus conclude $\Xi;\Gamma.\Delta[\Bv]\vdash \Bt[\Bv],\vec{x}.t[\Bv]: (\Xi, \texttt{x}\{\Delta'\}:T)$.\qedhere
  \end{itemize}
  \end{proof}

  The declarative system of \cref{fig:declarative-typing} features a conversion rule allowing us to deduce $t : U$ from $t : T$ when $ T\equiv U$. This rule can actually be strengthen to contexts in the following manner:

  \begin{proposition}[Conversion in context]\label{conversion-in-context}
    The following rules are admissible.
    \begin{mathpar}
      \Delta \equiv \Delta'
      \inferrule
      {\Theta;\Gamma \vdash \vec{t}:\Delta\\\Theta;\Delta'\vdash}
      {\Theta;\Delta\vdash\vec{t}:\Delta'}
      \and
      \Gamma \equiv \Delta
      \inferrule
      {\Theta;\Gamma \vdash \mathcal{J}\\\Theta;\Delta\vdash}
      {\Theta;\Delta\vdash\mathcal{J}}
    \end{mathpar}
  \end{proposition}
  \begin{proof}
    We first show the first statement by induction on $\Delta$, using \cref{substitution-property}. Then, for the proof of the second statement we  instantiate the first with $\Theta;\Delta\vdash\id:\Delta$ to get $\Theta;\Delta\vdash\id:\Gamma$ and then conclude by applying \cref{substitution-property} and using the fact that $\mathcal{J}[\id] = \mathcal{J}$.
  \end{proof}

  \begin{proposition}[Sorts are well-typed]\label{types-are-well-typed}  The following rule is admissible when $\mathbb{T}$ is well-typed.
  \begin{mathpar}
    \inferrule
    {\Theta;\Gamma\vdash t : T}
    {\Theta;\Gamma\vdash T~\sort}
  \end{mathpar}
  \end{proposition}
  \begin{proof}
  By case analysis on $\Theta;\Gamma \vdash t : T$, and using \cref{substitution-property}. For rule \textsc{Dest} we use the fact that the theory is well-typed to deduce $\Xi_{pi}.(\texttt{x}:T).\Xi_{d}\vdash U~\sort$ from $d(\Xi_{pi}\sep\texttt{x}:T\sep\Xi_{d}):U~\in~\mathbb{T}$.
  \end{proof}

  Using \cref{types-are-well-typed}, the premise for typing the sort in rule \textsc{Cons} can now be dropped, as anticipated in \cref{verbose-cons-remark}. In the following we allow ourselves to use this economic version of the rule \textsc{Cons} without announcement.

  \begin{proposition}["Economic" \textsc{Cons}]\label{economic-cons} The following rule is admissible when $\mathbb{T}$ is well-typed.
    \begin{mathpar}
      c(\Xi_p\sep\Xi_c\sep\Bv_i/\Xi_i):T\in\mathbb{T}~
      \inferrule
      {\Theta;\Gamma\vdash \Bt, \Bu : \Xi_p.\Xi_c}
      {\Theta;\Gamma\vdash c(\Bu): T[\Bt, \Bv_i[\Bt, \Bu]]}
    \end{mathpar}
    \end{proposition}
    \begin{proof}
    By well-typedness of the theory we have $\Xi_p.\Xi_c\vdash\id,\Bv_i : \Xi_p.\Xi_i$ and $\Xi_p.\Xi_i\vdash T~\sort$, therefore by \cref{substitution-property} two times we get $\Theta;\Gamma\vdash \Bt, \Bv_i[\Bt,\Bu]:\Xi_p.\Xi_i$ and then $\Theta;\Gamma\vdash T[\Bt, \Bv_i[\Bt,\Bu]]~\sort$. Therefore, by rule \textsc{Cons} we conclude $\Theta;\Gamma\vdash c(\Bu): T[\Bt, \Bv_i[\Bt, \Bu]]$.
    \end{proof}

    Finally, we now show that asking rewrite rules to be type-preserving is sufficient to ensure subject reduction.\footnote{Note that this is actually an equivalence: subject reduction directly implies that all rules are type-preserving. }

    \begin{proposition}[Subject reduction]\label{subject-reduction}
      Suppose that the underlying theory is valid.
      \begin{itemize}
        \item If $\Gamma \vdash T~\sort$ and $T \red T'$ then $\Gamma\vdash T'~\sort$
        \item If $\Gamma \vdash t : T$ and $t \red t'$ then $\Gamma \vdash t' : T$
        \item If $\Gamma \vdash \Bt : \Xi$ and $\Xi\vdash$ and $\Bt \red \Bt'$ then $\Gamma\vdash\Bt' :\Xi$
      \end{itemize}
    \end{proposition}
    \begin{proof}
      By induction on the typing derivation, and by case analysis on the rewriting relation.

      \begin{itemize}
        \item Case $d(\Bv)[\Bt] \red r[\Bt]$. The result follows from the fact that all rules in $\mathbb{T}$ are type-preserving.
        \item Case $c(\Bt)\red c(\Bt')$ with $\Bt\red \Bt'$, and where $c(\Xi)~\sort\in\mathbb{T}$. By inversion of typing on $\Gamma\vdash c(\Bt)~\sort$ we have $\Gamma \vdash \Bt : \Xi$, and because $\mathbb{T}$ is well-typed we have $\Xi\vdash$, so by i.h. we get $\Gamma\vdash\Bt':\Xi$, allowing us to conclude $\Gamma\vdash c(\Bt')~\sort$.

        \item Case $c(\Bt)\red c(\Bt')$ with $\Bt\red \Bt'$, and where $c(\Xi_p\sep\Xi_c\sep\Bv_i/\Xi_i):U\in\mathbb{T}$. By inversion of typing on $\Gamma\vdash c(\Bt) : T$ we have $\Gamma \vdash \Bt_p,\Bt : \Xi_p.\Xi_c$ and $T \equiv U[\Bt_p,\Bv_i[\Bt_p,\Bt]]$. Because $\mathbb{T}$ is well-typed, we have $\Xi_p.\Xi_c\vdash$, therefore by i.h. we get $\Gamma \vdash \Bt_p, \Bt' : \Xi_p.\Xi_c$. We can thus derive $\Gamma\vdash c(\Bt'):U[\Bt_p,\Bv_i[\Bt_p,\Bt']]$, and because its sort is convertible to $T$, we conclude by applying conversion.

        \item Case $d(\Bt)\red d(\Bt')$ with $\Bt\red \Bt'$, and where $d(\Xi_{pi}\sep\texttt{x}:U\sep\Xi_d):V\in\mathbb{T}$. By inversion of typing on $\Gamma\vdash d(\Bt) : T$ we have $\Gamma \vdash \Bt_{pi},\Bt : \Xi_{pi}.(\texttt{x}:U).\Xi_d$ and $T \equiv V[\Bt_{pi},\Bt]$. Because $\mathbb{T}$ is well-typed, we have $\Xi_{pi}.(\texttt{x}:U).\Xi_d\vdash$, therefore by i.h. we get $\Gamma \vdash \Bt_{pi}, \Bt' : \Xi_{pi}.(\texttt{x}:U).\Xi_d$. We can thus derive $\Gamma\vdash d(\Bt'):V[\Bt_{pi},\Bt']$, and because its sort is convertible to $T$, we conclude by applying conversion.

        \item Case $\texttt{x}\{\vec{t}\}\red \texttt{x}\{\vec{t}'\}$ with $\vec{t}\red \vec{t}'$. Impossible because the metavariable context is empty.

        \item Case $\Bu, \vec{x}.t \red \Bt'$. By inversion on $\Gamma \vdash \Bu,\vec{x}.t : \Xi$ we get $\Xi=\Xi',\texttt{x}\{\Delta\}:U$ and $\Gamma\vdash \Bu : \Xi'$ and $\Gamma.\Delta[\Bu]\vdash t :U[\Bu]$. Moreover, from $\Xi\vdash$ we get $\Xi'; \Delta\vdash U~\sort$.
        \begin{itemize}
          \item Subcase $\Bt' = \Bu', \vec{x}.t$ with $\Bu \red \Bu'$. Then by i.h. we get $\Gamma \vdash \Bu' : \Xi'$. By \cref{substitution-property} with $\Xi'; \Delta\vdash U~\sort$ we get $\Gamma.\Delta[\Bu'] \vdash U[\Bu']~\sort$, so we can apply conversion and \cref{conversion-in-context} to $\Gamma.\Delta[\Bu]\vdash t: U[\Bu]$ to get $\Gamma.\Delta[\Bu']\vdash t:U[\Bu']$, allowing us to conclude $\Gamma \vdash \Bu', \vec{x}.t: (\Xi',\texttt{x}\{\Delta\}:U)$.
          \item Subcase $\Bt' = \Bu, \vec{x}.t'$ with $t\red t'$. By i.h. we get $\Gamma.\Delta[\Bu]\vdash t' : U[\Bu]$, and so we conclude $\Gamma \vdash \Bu, \vec{x}.t': (\Xi',\texttt{x}\{\Delta\}:U)$.\qedhere
        \end{itemize}
      \end{itemize}
    \end{proof}

\begin{remark}
    We have shown subject reduction only for terms without metavariables. It is nevertheless possible to extend \cref{subject-reduction} to terms with metavariable by strengthening the type-preservation condition to $\Theta;\Gamma \vdash l[\Bv]:T$ implies $\Theta;\Gamma\vdash r[\Bv]:T$ for all $\Theta,\Gamma,\Bv,T$. However, this generalization would be of no use for the proofs to come.
  \end{remark}

\section{Bidirectional type system}%
\label{sec:bidirectional-type-system}

Having seen the definition of the declarative system, we move now to the bidirectional type system. We start the section by discussing matching modulo rewriting, which is needed for recovering missing arguments. We then introduce the bidirectional syntax, over which the bidirectional type system is defined. We then give the bidirectional typing rules and prove they are correct with respect to the declarative system.
Finally, we conclude by showing the decidability of the bidirectional system for strongly-normalizing theories.

\subsection{Matching modulo rewriting}\label{matching-modulo}
\newcommand\tohead{!\textup{\textsf{h}}}
\newcommand\head{\textup{\textsf{h}}}
\newcommand\maxout{\textup{\textsf{m/o}}}

\newcommand\rpatt{\texttt{\textup{rpatt}}}

Suppose we want to type $\nApp(t,u)$ by first inferring the sort of $t$, yielding $T$. We know that the sort of the principal argument in the rule for $\nApp$ is the pattern $\nTm(\nPi(\texttt{A},x.\texttt{B}\{x\}))$, so we could hope to recover $A$ and $B$ by matching $T$ against this pattern. However, because of the conversion rule, in dependent type theories we cannot expect $T$ to be syntactically equal to an instance of this pattern, but only convertible to it. Therefore, our goal is instead to find $A$ and $B$ satisfying $\nTm(\nPi(\texttt{A},x.\texttt{B}\{x\}))[A, x.B] \equiv T$. This shows that the process of recovering missing arguments in bidirectional typing is actually an instance of \textit{matching modulo rewriting} --- a connection that apparently had not been discussed before in the literature. This also explains why we were careful in \cref{subsec:theories} to require the sort of a constructor rule and the sort of the principal argument of a destructor rule to be patterns, as they need to support decidable and unitary matching.

In order to solve matching modulo problems, we define in \cref{fig:matching} an inference system which, given a pattern $t$ and a term $u$, tries to compute a metavariable substitution $\Bv$ such that $u \redd t[\Bv]$. Here we write $t \red^\head u$ when $t \redd u$ and $u$ is \textit{head-normal}, meaning that for all $u'$ with $u\redd u'$, no rewrite rule can be applied to the head of $u'$. Note that, because all of our rewrite rules are headed by destructors, all terms not of the form $d(\Bt)$ are automatically head-normal.

\begin{figure}[h]\raggedright
  \fbox{$ t \prec u \leadsto \Bv \quad
    (t \in \Tm^{\Pat}~\theta~\gamma;~
    u \in \Tm~(\cdot)~\delta.\gamma;~
    \Bv \in \MSub~(\cdot)~\delta~\theta)$}
  \begin{mathpar}
    u \red^\head c(\Bu)
    \inferrule
    {\Bt \prec \Bu \leadsto \Bv}
    {c(\Bt) \prec u \leadsto \Bv}
    \and
    \inferrule
    { }
    {\texttt{x}\{\id_\gamma\} \prec u  \leadsto \vec{x}_\gamma.u}
  \end{mathpar}
  \fbox{$ \Bt \prec \Bu \leadsto \Bv\quad(
    \Bt \in \MSub^{\Pat}~\theta~\gamma~\xi;~
    \Bu \in \MSub~(\cdot)~\delta.\gamma~\xi;~
    \Bv \in \Sub~(\cdot)~\delta~\theta)$}
\begin{mathpar}
  \inferrule
  {  }
  {\varepsilon \prec \varepsilon   \leadsto \varepsilon}
  \and
  \inferrule
  {\Bt \prec \Bu \leadsto \Bv_{1} \\  t \prec u  \leadsto \Bv_{2}}
  {\Bt,\vec{x}.t \prec \Bu, \vec{x}.u \leadsto \Bv_1,\Bv_2}
\end{mathpar}
\caption{Inference system for matching modulo}
\label{fig:matching}
\end{figure}

\subsubsection{Correctness of matching modulo}

Let us now establish the correctness of this inference system in three steps, starting by  its soundness.
\begin{proposition}[Soundness of matching]\label{soundness-matching}~
  \begin{itemize}
    \item If $t \prec u \leadsto \Bv$ then $u \redd t[\Bv]$.
    \item If $\Bt \prec \Bu \leadsto \Bv$ then $\Bu \redd \Bt[\Bv]$.
  \end{itemize}
\end{proposition}
\begin{proof}
  By induction on the matching judgment.
\end{proof}

We then move to the converse of \cref{soundness-matching}, stating that we can always find a substitution if $u$ is convertible to an instance of $t$.

\begin{proposition}[Completeness of matching]\label{completeness-matching}Let $\Bv \in \MSub~(\cdot)~\delta~\theta$.~
  \begin{itemize}
    \item If $t\in\Tm^\Pat~\theta~\gamma$ and $t[\Bv] \equiv u$ then $t \prec u \leadsto \Bv'$ for some~$\Bv'\equiv \Bv$.
    \item If $\Bt\in\MSub^\Pat~\theta~\gamma~\xi$ and $\Bt[\Bv] \equiv \Bu$ then $\Bt \prec \Bu \leadsto \Bv'$ for some~$\Bv' \equiv \Bv$.
  \end{itemize}
\end{proposition}
\begin{proof}
  By induction on the pattern, the only interesting case being when $t = c(\Bt)$. In this case, by applying confluence to $c(\Bt[\Bv]) \equiv u$ we get $c(\Bt[\Bv])\redd u' \iredd u$ and because $c(\Bt[\Bv])$ is head normal we have $u' = c(\Bu)$ with $\Bt[\Bv]\redd \Bu$. We conclude by applying the i.h. to $\Bt$, which yields $\Bt \prec \Bu \leadsto \Bv'$ for some $\Bv' \equiv \Bv$, and allowing us to derive $t \prec u \leadsto \Bv'$.
\end{proof}

A direct consequence of soundness and completeness of matching is the following corollary, which will be useful in the proof of \cref{decidability-matching}.

\begin{corollary}[Matching respects conversion]\label{matching-invariant}~
  \begin{itemize}
    \item If $t \prec u \leadsto \Bv$ and $u \equiv u'$ then $t \prec u' \leadsto \Bv'$ for some $\Bv'\equiv \Bv$
    \item If $\Bt \prec \Bu \leadsto \Bv$ and $\Bu \equiv \Bu'$ then $\Bt \prec \Bu' \leadsto \Bv'$ for some $\Bv'\equiv \Bv$
  \end{itemize}
\end{corollary}

Finally, we establish that matching modulo is decidable when the expression $e$ from which we extract the substitution is \textit{strongly-normalizing} (often abbreviated as s.n.), meaning that all reduction sequences issuing from $e$ are finite.

\begin{proposition}[Decidability of matching]\label{decidability-matching}~
  \begin{itemize}
    \item If $u$ is strongly-normalizing  then $\exists \Bv.~t \prec u \leadsto \Bv$ is decidable for all $t$
    \item If $\Bu$ is strongly-normalizing  then $\exists \Bv.~\Bt \prec \Bu \leadsto \Bv$ is decidable for all $\Bt$
  \end{itemize}
\end{proposition}
\begin{proof}
  By induction on the pattern. We show the only interesting case, when $t = c(\Bt)$. Because $u$ is strongly-normalizing, we can use any reduction strategy to compute a head-normal form $u'$~for~$u$.

  If $u'$ is not headed by $c$ then $\exists \Bv.~t \prec u \leadsto \Bv$ cannot hold. Indeed, this would imply $u \redd c(\Bu)$ for some $\Bu$, so by confluence and the fact that $c(\Bu)$ and $u'$ are head-normal we could then show that $u'$ is headed by $c$, contradiction.

  If $u'$ is of the form $c(\Bu)$, then by i.h. we can decide $\exists \Bv.~\Bt \prec \Bu \leadsto \Bv$. If this holds, then it follows that $\exists \Bv.~t \prec u \leadsto \Bv$ holds. If $\exists \Bv.~\Bt \prec \Bu \leadsto \Bv$ does not hold, then $\exists \Bv.~t \prec u \leadsto \Bv$ also does not hold. Indeed, if $\exists \Bv.~t \prec u \leadsto \Bv$ holds then we have $u\red^\head c(\Bu')$ and $\Bt \prec \Bu' \leadsto \Bv'$ for some $\Bu'$ and $\Bv'$. But from $c(\Bu)\equiv c(\Bu')$ and confluence we  get $\Bu \equiv \Bu'$, but \cref{matching-invariant} would then imply $\exists \Bv.~\Bt \prec \Bu \leadsto \Bv$.
\end{proof}

\begin{remark}
  The condition of strong normalization in \cref{decidability-matching} could be slightly weaken to require only weak normalization, by instead employing in the proof a \textit{normalizing} strategy, such as the \textit{maximal-outermost} strategy~\cite{van1997outermost,van1999normalisation}.
  However, even if strategies derived from call-by-value have weaker theoretical guarantees, they are generally easier to implement efficiently --- for instance, by using \textit{normalization-by-evaluation}, as in the case of our implementation.
  Moreover, most theories used in practice are either strongly-normalizing or not normalizing at all, and so it is questionable whether asking only for weak- instead of strong-normalization would bring any benefit in practice. %

\end{remark}

\subsection{Bidirectional syntax}\label{bidi-syntax}

\newcommand\TmC{\Tm^\textsf{\textup{c}}}
\newcommand\TmI{\Tm^\textsf{\textup{i}}}
\newcommand\MSubC{\MSub^\textsf{\textup{c}}}

In order to define the bidirectional type system, we first have to address the problem that some terms without annotations cannot be algorithmically typed. Indeed, suppose for instance that we want to type the term $\nApp(\nLam(x.t),u)$ by inferring the sort of the principal argument of $\nApp$ to recover $A$ and $B$. But because $\nLam(x.t)$ is headed by a constructor, it can only be bidirectionally typed in mode check, so we are stuck. One could think that this limitation is specific to bidirectional typing, however a famous result by Dowek shows that, in a dependently-typed setting, the problem of typing unannotated terms is actually undecidable in its full generality~\cite{dowek1993undecidability}.
Therefore, instead of defining the bidirectional system over the regular syntax of terms, we will define it over the \textit{bidirectional syntax} which, given a signature $\Sigma$, is defined by the following grammar.
\begin{align*}
  \boxed{\TmC~\gamma} \ni & &t, u, v ::=
  &\mid c(\Bt\mkgray{\in\MSubC~\gamma~\xi})& &\stext{if $c(\xi)\in\Sigma$} \\
  &&&\mid \underline{t}\mkgray{\in\TmI~\gamma} \\
  \boxed{\TmI~\gamma} \ni & &t, u, v ::=& \mid x & &\stext{if $ x \in \gamma$}\\
  &&&\mid d(t \mkgray{\in \TmI~\gamma},\Bt\mkgray{\in\MSubC~\gamma~\xi})&& \stext{if $ d(\texttt{x},\xi)\in\Sigma$}\\ %
  &&&\mid t\mkgray{\in \TmC~\gamma} :: T \mkgray{\in \TmC~\gamma}\\
  \boxed{\MSubC~\gamma~\xi} \ni & &\Bt,\Bu,\Bv  ::= &\mid \epsilon && \stext{if $\xi = \cdot$}\\
  &&& \mid \Bt \mkgray{\in \MSubC~\gamma~\xi'}, \vec{x}_{\delta}.t \mkgray{\in \TmC~\gamma.\delta} &&\stext{if $\xi = \xi', \texttt{x} \{\delta\}$}
\end{align*}

By separating between \textit{checkable terms} $t \in \TmC~\gamma$ and \textit{inferrable terms} $t\in\TmI~\gamma$ we are now able to specify that the principal argument of a destructor can only be an inferable term, avoiding the situation described in the previous paragraph. The consequence of this is that terms of the form $d(c(\Bt), \Bu)$ are not directly part of the bidirectional syntax. Instead, we must first turn $c(\Bt)$ into an inferable term by adding a \textit{(sort) ascription} $c(\Bt)  :: T$, allowing us to then write $d(c(\Bt)::T, \Bu)$. The symmetric operation of \textit{embeding} then allows us to create a checkable term $\underline{t}$ from an~inferable~one~$t$.

\begin{example}
  The inferrable and checkable terms for the signature \ref{ex:sig} are given respectively by the following grammars, where we omit the scope information.%
  \begin{align*}
   t^\textsf{i}, u^\textsf{i} &::= x \mid  t^\textsf{c} :: T^\textsf{c}\mid \nApp(t^\textsf{i}, u^\textsf{c})\\
   t^\textsf{c}, u^\textsf{c}, A^\textsf{c},B^\textsf{c}, T^\textsf{c} &::= \nTy \mid \nTm(A^\textsf{c}) \mid \nPi(A^\textsf{c}, x.B^\textsf{c}) \mid \nLam(x.t^\textsf{c}) \mid \underline{t^\textsf{i}}
  \end{align*}
\end{example}

Because the bidirectional syntax requires us to add additional ascriptions when writing terms of the form $d(c(\Bt), \Bu)$, one can wonder if this requirement might be too inconvenient in practice.
If we remove sort ascriptions from the bidirectional syntax, then for most theories (like \ref{ex:theory}) the checkable terms coincide exactly with the normal forms, and the inferable terms with the neutrals. As argued in other works~\cite{norell:thesis,abel2011partial}, users of type theory almost never write redices, and because of this a large part of the bidirectional typing literature only supports the typing of normal forms~\cite{abel2005untyped,norell:thesis,abel2011partial,coquand1996algorithm,abel2011modular,abel2017normalization}, for which one needs no ascriptions. Our choice of \textit{also} supporting ascriptions is simply a matter of giving users an extra convenience for the few situations in which writing a redex is more convenient, yet we expect that in most cases they will not be needed.

\begin{remark}
  Note that we have omitted metavariables from the bidirectional syntax.
  Even if metavariables are needed in the core syntax for \textit{specifying} the theories in \cref{sec:theories} (and the well-typed theories in \cref{sec:declarative-type-sytem}), they are in general not needed for \textit{using} the theories, and this is why they are in general omitted from most presentations of type theories. It is therefore reasonable to leave them out of the bidirectional syntax, as they would be of no utility  for  users.
\end{remark}

\newcommand{\embed}[1]{ \ulcorner #1  \urcorner }

Given $t\in\TmC~\gamma$ or $t\in\TmI~\gamma$ we write $\embed{t}\in\Tm~(\cdot)~\gamma$ for its underlying term, obtained by forgetting the difference between checkable and inferable terms and by removing sort ascriptions. Similarly, if $\Bt\in\MSubC~\gamma~\xi$ we write $\embed{\Bt}\in\MSub~(\cdot)~\gamma~\xi$ for its underlying metavariable substitution.

\subsection{Bidirectional typing rules}

  Given a theory $\mathbb{T}$, we can now define its bidirectional type system by the rules in \cref{fig:bidirectional-typing}. The system is split in 4 judgments:

  \begin{itemize}
    \item $\Gamma \vdash T \Leftarrow \sort$ : Checking that a checkable term $T$ is a well-formed sort.
    \item $\Gamma\vdash t \Leftarrow T$ : Checking that a checkable term $t$ has sort $T$.
    \item $\Gamma\vdash t \Rightarrow T$ : Inferring a sort $T$ for an inferable term $t$.
    \item $\Gamma \mid \Bv : \Xi \vdash \Bt \Leftarrow \Theta$ : Checking that a checkable metavariable substitution $\Bt$ can be typed by $\Theta$, knowing that $\Bv:\Xi$.
  \end{itemize}

  \begin{remark}
    The reader could argue that, for the fourth judgment, the most natural choice would be something of the form $\Gamma \vdash \Bt \Leftarrow \Theta$. However, in some cases it will be necessary to check $\Bv.\embed{\Bt} : \Xi.\Theta$ knowing already that $\Bv : \Xi$ holds, so our more general judgment allows us to prevent reverifying that $\Bv$ is well-typed a second time.
  \end{remark}

\begin{figure}\raggedright

  \fbox{$ \Gamma  \vdash T \Leftarrow \sort \quad
    (\Gamma \in \Ctx;~
     T \in \TmC~|\Gamma|)$}
  \begin{mathpar}
    c(\Xi) ~\sort\in\mathbb{T}
    \inferrule[Sort]
    {\Gamma\mid \varepsilon : (\cdot) \vdash \Bt \Leftarrow \Xi}
    {\Gamma  \vdash c(\Bt) \Leftarrow \sort}
  \end{mathpar}

  \vspace{0.5em}

  \fbox{$\Gamma  \vdash t \Leftarrow T \quad (
    \Gamma \in \Ctx;~
    T \in \Tm~(\cdot)~|\Gamma|;~
    t \in \TmC~|\Gamma|)$}
  \begin{mathpar}
    c (\Xi_p\sep\Xi_c\sep \Bv_i/\Xi_i) : T \in \mathbb{T}~
    \inferrule[Cons]
      {T \prec T' \leadsto \Bt_p,\Bt_i \\\\ \Gamma\mid \Bt_p : \Xi_p \vdash \Bu \Leftarrow \Xi_c}
      {\Gamma  \vdash c(\Bu) \Leftarrow T'}
      \Bt_i \equiv \Bv_i[\Bt_p, \embed{\Bu}]
      \and
      \inferrule[Switch]
      {\Gamma \vdash t \Rightarrow T}
      {\Gamma \vdash \underline{t} \Leftarrow U}
      T \equiv U
  \end{mathpar}

  \vspace{0.5em}

  \fbox{$\Gamma  \vdash t \Rightarrow T \quad (
    \Gamma \in \Ctx;~
    T \in \Tm~(\cdot)~|\Gamma|;~
    t \in \TmI~|\Gamma|)$}
  \begin{mathpar}
    x :T \in \Gamma~
    \inferrule[Var]
    { }
    {\Gamma  \vdash x \Rightarrow T}
    \and
    \inferrule[Annot]
    {\Gamma \vdash T \Leftarrow \sort\\
     \Gamma \vdash t \Leftarrow \embed{T}}
    {\Gamma \vdash t :: T \Rightarrow \embed{T}}
    \and
    d (\Xi_{pi} \sep \texttt{x}:T \sep \Xi_d) : U \in \mathbb{T}~
    \inferrule[Dest]
    {\Gamma \vdash t \Rightarrow T'\\ T \prec T' \leadsto \Bt\\
     \Gamma\mid \Bt, \embed{t} : (\Xi_{pi},\texttt{x}:T) \vdash \Bu \Leftarrow \Xi_d}
    {\Gamma  \vdash d(t,\Bu) \Rightarrow U[\Bt,\embed{t},\embed{\Bu}]}
  \end{mathpar}

  \vspace{0.5em}

  \fbox{$ \Gamma\mid \Bv : \Theta \vdash \Bt \Leftarrow \Xi \quad
    (\Theta \in \MCtx;~
    \Gamma \in \Ctx;~
    \Bv \in \MSub~(\cdot)~|\Gamma|~|\Theta|;~
    \Xi \in \MCtx~|\Theta|;~
    \Bt \in \MSubC~|\Gamma|~|\Xi|)
    $}
  \begin{mathpar}
    \inferrule[EmptyMSub]
    { }
    {\Gamma\mid \Bv : \Theta  \vdash \varepsilon \Leftarrow (\cdot)}
    \and
    \inferrule[ExtMSub]
    {\Gamma\mid \Bv : \Theta \vdash \Bt \Leftarrow \Xi \\
     \Gamma.\Delta[\Bv,\embed{\Bt}] \vdash t \Leftarrow T[\Bv,\embed{\Bt}]}
    {\Gamma\mid \Bv : \Theta \vdash \Bt, \vec{x}_\Delta.  t \Leftarrow (\Xi , \texttt{x}\{\Delta\} : T)}
  \end{mathpar}

  \caption{Bidirectional typing rules}
  \label{fig:bidirectional-typing}
  \end{figure}

As in the declarative system, the most important rules are the one that instantiate the schematic typing rules: \textsc{Const}, \textsc{Dest} and \textsc{Sort}. However, differently from the declarative system, no more guessing is needed when building a type derivation. For instance, when using rule \textsc{Dest} with $d(t,\Bu)$ the omitted arguments are no longer guessed, but instead recovered by inferring the sort of the principal argument $t$ and then matching it against the associated pattern. %
\begin{example}
  Suppose we want to infer a sort for $\nApp(t,u)$ in the theory \ref{ex:theory}. To use rule \textsc{Dest}, we start by inferring a sort $T'$ for $t$, and then we try to match it against the pattern $\nTm(\nPi(\texttt{A},x.\texttt{B}\{x\}))$. If  matching succeeds, we recover the arguments $A$ and $B$, which together with $t$ are then used in
  \begin{align*}
    &\Gamma\mid (A,x.B,x.\embed{t}) : (\texttt{A}:\nTy,~ \texttt{B}\{x : \nTm(\texttt{A})\} : \nTy,~
    \texttt{t} \{ x : \nTm(\ttA)\} : \nTm(\ttB) )
    \vdash (u) \Leftarrow (\texttt{u}:\nTm(\texttt{A}))
  \end{align*}
  By applying the rules that define the judgment $\Gamma \mid \Bv : \Theta \vdash \Bt \Leftarrow \Xi$, we see that this amounts to showing just $\Gamma \vdash u \Leftarrow \nTm(A)$, and so the final shape of this derivation is the following, which corresponds to the usual bidirectional rule for application.
    \begin{mathpar}
      \inferrule
      {\Gamma\vdash t \Rightarrow T'\\  \nTm(\nPi(\texttt{A},x.\texttt{B}\{x\}))\prec T'\leadsto A, x.B\\ \Gamma \vdash u \Leftarrow \nTm(A)}
      {\Gamma\vdash\nApp(t,u)\Rightarrow \nTm(B[\id_\Gamma,\embed{u}])}
    \end{mathpar}
\end{example}

It is also instructive to see what happens when type-checking a constructor. In this case, we first use matching to recover both the parameters $\Bt_p$ and indices $\Bt_i$. With the parameters, we can then check the non-erased constructor arguments $\Bu$. To conclude, we however also need to verify that the expected indices $\Bt_i$ correspond to the actual ones, by checking if $\Bv_i[\Bt_p,\embed{\Bu}] \equiv \Bt_i$  holds.

\subsection{Correctness of bidirectional typing}

We now establish the correctness of the bidirectional type system with respect to the declarative typing rules of \cref{sec:declarative-type-sytem}.

\subsubsection{Soundness} In order to establish soundness, we will first need the following technical result for typing the arguments recovered through matching. It roughly states that, given a well-typed pattern $t$ such that the result of substituting $\Bv$ in $t$ is also well-typed, one can conclude that $\Bv$ is also well-typed --- an implication that generally does not hold when $t$ is not~a~pattern.

\begin{lemma}[Substitution typing inversion for patterns]\label{typing-a-substitution}  Let $\Bv \in \MSub~(\cdot)~|\Delta|~|\Theta|$.
  \begin{itemize}
    \item If $t \in \Tm^{\Pat}~|\Theta|~(\cdot)$ and $\Theta \vdash t : T$ and $\Delta \vdash t[\Bv] : T[\Bv]$ then $\Delta\vdash \Bv : \Theta$
    \item If $\Bt \in \MSub^{\Pat}~|\Theta|~(\cdot)~|\Xi|$ and $\Theta \vdash \Bt : \Xi$ and $\Delta \vdash \Bt[\Bv] : \Xi$ then $\Delta \vdash \Bv : \Theta$
    \item If $T \in \Tm^{\Pat}~|\Theta|~(\cdot)$ and $\Theta \vdash T~\sort$ and $\Delta \vdash T[\Bv]~\sort$ then $\Delta \vdash \Bv : \Theta$.
  \end{itemize}
\end{lemma}

\begin{proof}
  In order for the proof to go through, we show the following stronger statement. Let $\Bv = \Bv_1,\Bv_2,\Bv_3 $ and $\Theta=\Theta_1.\Theta_2.\Theta_3$ with $\Bv_{i} \in \MSub~(\cdot)~|\Delta|~|\Theta_{i}|$ for $i=1,2,3$, and suppose moreover that $\Delta \vdash \Bv_{1} : \Theta_{1}$ and that one of the following holds.
  \begin{itemize}
    \item $t \in \Tm^{\Pat}~|\Theta_{2}|~|\Gamma|$ and $\Theta; \Gamma \vdash t : T$ and $\Delta.\Gamma' \vdash t[\Bv_2] : T'$ with  $\Gamma'\equiv\Gamma[\Bv]$ and $T'\equiv T[\Bv]$.
    \item $\Bt_{2} \in \MSub^{\Pat}~|\Theta_{2}|~|\Gamma|~|\Xi_{2}|$ and $\Theta; \Gamma \vdash \Bt_{1},\Bt_{2} : \Xi_{1}.\Xi_{2}$ and $\Delta.\Gamma' \vdash \Bt'_{1},\Bt_{2}[\Bv_2] : \Xi_{1}.\Xi_{2}$ with $\Gamma'\equiv\Gamma[\Bv]$ and $\Bt_{1}'\equiv\Bt_{1}[\Bv]$.
    \item $T \in \Tm^{\Pat}~|\Theta_{2}|~|\Gamma|$ and $\Theta; \Gamma \vdash T~\sort$ and $\Delta.\Gamma' \vdash T'[\Bv_2]~\sort$ with  $\Gamma'\equiv\Gamma[\Bv]$.
  \end{itemize}
  Then we have $\Delta \vdash \Bv_{1}, \Bv_{2} : \Theta_{1}.\Theta_{2}$. The proof is by induction on the pattern.

\begin{itemize}
  \item Case $T = c(\Bt \mkgray{\in \MSub^{\Pat}~|\Theta_{2}|~|\Gamma|~|\Xi|})$ for $c (\Xi) ~\sort \in \mathbb{T}$. By inversion on $\Theta; \Gamma \vdash T~\sort$ and $\Delta.\Gamma' \vdash T'[\Bv_2]~\sort$ we obtain  $\Theta;\Gamma \vdash \Bt : \Xi$ and $\Delta.\Gamma' \vdash \Bt[\Bv_2] : \Xi$, so by i.h. we conclude.

  \item Case $t = c(\Bt_{c} \mkgray{\in \MSub^{\Pat}~|\Theta_{2}|~|\Gamma|~|\Xi_{2}|})$ for $c (\Xi_{p}\sep\Xi_c\sep\Bv_i/\Xi_i) : U \in \mathbb{T}$. By inversion on $\Theta; \Gamma \vdash c(\Bt_c) : T$ and $\Delta.\Gamma' \vdash c(\Bt_c)[\Bv_2] : T'$ we obtain $\Theta;\Gamma \vdash \Bt_p,\Bt_c : \Xi_p.\Xi_c$ with $T \equiv U[\Bt_p, \Bv_i[\Bt_p,\Bt_c]]$ and $\Delta.\Gamma' \vdash \Bt'_p,\Bt_c[\Bv_2] : \Xi_p.\Xi_c$ with $T' \equiv U[\Bt_p', \Bv_i[\Bt_p',\Bt_c[\Bv_2]]]$, for some $\Bt_p$ and $\Bt_p'$.
  We have
  \begin{align*}
    U[\Bt_p', \Bv_i[\Bt_p',\Bt_c[\Bv_2]]] \equiv T'
    \equiv T[\Bv]
    \equiv U[\Bt_p, \Bv_i[\Bt_p,\Bt_c]][\Bv]
    \equiv U[\Bt_p[\Bv], \Bv_i[\Bt_p,\Bt_c][\Bv]]
  \end{align*}
  so by \cref{pat-inj} we get $\Bt_p', \Bv_i[\Bt_p',\Bt_c[\Bv_2]]\equiv \Bt_p[\Bv], \Bv_i[\Bt_p,\Bt_c][\Bv]$. In particular we have $\Bt_p'\equiv\Bt_p[\Bv]$, allowing us to  apply the i.h. to conclude.

  \item Case $t = \texttt{x}\{\id_{\Gamma}\}$, in which case we must have $\Theta_{2} = \texttt{x}\{\Gamma_{\texttt{x}}\} : T_{\texttt{x}}$ for some $\Gamma_\ttx$ and $T_\ttx$. By inversion on $\Theta;\Gamma \vdash t : T$ we get $T \equiv T_{\texttt{x}}$ and $\Gamma \equiv \Gamma_{\texttt{x}}$, and therefore $T' \equiv T_{\texttt{x}}[\Bv]$ and $\Gamma' \equiv \Gamma_{\texttt{x}}[\Bv]$.  Moreover, as the only metavariables of $\Theta$ appearing in $\Gamma_{\texttt{x}}$ and $T_\texttt{x}$ are those of $\Theta_{1}$, we have $\Gamma_{\texttt{x}}[\Bv_{1}] = \Gamma_{\texttt{x}}[\Bv]$ and $T_{\texttt{x}}[\Bv_{1}]=T_\texttt{x}[\Bv]$, and therefore $T' \equiv T_{\texttt{x}}[\Bv_{1}]$ and $\Gamma' \equiv \Gamma_{\texttt{x}}[\Bv_{1}]$. Then, because $\Theta \vdash$, we have $\Theta_{1};\Gamma_{\texttt{x}} \vdash T_{\texttt{x}}~\sort$, so by applying \cref{substitution-property} with $\Delta \vdash \Bv_{1} : \Theta_{1}$ we get $\Delta.\Gamma_{\texttt{x}}[\Bv_{1}] \vdash T_{\texttt{x}}[\Bv_{1}]~\sort$. Now we can apply conversion and \cref{conversion-in-context} to $\Delta.\Gamma' \vdash t[\Bv_2] : T'$ to get $\Delta.\Gamma_{\texttt{x}}[\Bv_{1}] \vdash t[\Bv_2] : T_{\texttt{x}}[\Bv_{1}]$. Because $t[\Bv_2] = \Bv_{\texttt{x}}$ then together with $\Delta\vdash \Bv_{1}:\Theta_{1}$ we can conclude $\Delta \vdash \Bv_{1}, \vec{x}_{\Gamma}.\Bv_{\texttt{x}} : (\Theta_{1}, \texttt{x}\{\Gamma_{\texttt{x}}\}:T_{\texttt{x}})$.

  \item Case $\Bt_{2} = \varepsilon \mkgray{\in \MSub~(\cdot)~|\Gamma|~(\cdot)}$. Then the result follows by hypothesis.
  \item Case $\Bt_{2} = \Bu\mkgray{\in \MSub~|\Theta_{2l}|~|\Gamma|~|\Xi_{2}'|} , \vec{x}. u \mkgray{\in \Tm^{\Pat}~|\Theta_{2r}|~|\Gamma|.|\Delta_\texttt{x}|}$ for $\Theta_{2}=\Theta_{2l}.\Theta_{2r}$ and $\Xi_{2} = \Xi'_{2}, \texttt{x}\{\Delta_\texttt{x}\} : T_\texttt{x}$. Let $\Bv_{2}= \Bv_{2l},\Bv_{2l}$ be the splitting of $\Bv_{2}$ according to the decomposition $\Theta_{2}=\Theta_{2l}.\Theta_{2r}$.  By inversion on $\Theta; \Gamma \vdash \Bt_{1},\Bt_{2} : \Xi_1.\Xi_2$ and $\Delta. \Gamma' \vdash \Bt_{1}',\Bt_{2}[\Bv_2] : \Xi_1.\Xi_2$ we obtain the following.
  \begin{mathpar}
    \inferrule
    {\Theta;\Gamma \vdash \Bt_1,\Bu : \Xi_1.\Xi_2'\\\\ \Theta;\Gamma.\Delta_\texttt{x}[\Bt_1,\Bu] \vdash u : T_\texttt{x}[\Bt_1,\Bu]}
    {\Theta;\Gamma \vdash \Bt_1,\Bu, \vec{x}. u : (\Xi_1.\Xi_2', \texttt{x}\{\Delta_\texttt{x}\}:T_\texttt{x})}
    \and
    \inferrule
    {\Delta.\Gamma' \vdash \Bt_1',\Bu[\Bv_{2l}] : \Xi_1.\Xi_2'\\\\ \Delta.\Gamma'.\Delta_\texttt{x}[\Bt_1',\Bu[\Bv_{2l}]] \vdash u[\Bv_{2r}] : T_\texttt{x}[\Bt_1',\Bu[\Bv_{2l}]]}
    {\Delta.\Gamma' \vdash \Bt_1',\Bu[\Bv_{2l}], \vec{x}. u[\Bv_{2r}] : (\Xi_1.\Xi'_2, \texttt{x}\{\Delta_\texttt{x}\}:T_\texttt{x})}
  \end{mathpar}
  By the i.h. applied to the first premises we get $\Delta\vdash \Bv_{1},\Bv_{2l} : \Theta_{1}.\Theta_{2l}$.
  Then, note that we have $(\Gamma.\Delta_\texttt{x}[\Bt_{1},\Bu])[\Bv] = \Gamma[\Bv].\Delta_\texttt{x}[\Bt_{1}[\Bv],\Bu[\Bv_{2l}]] \equiv \Gamma'.\Delta_\texttt{x}[\Bt_1', \Bu[\Bv_{2l}]]$ and $T_\texttt{x}[\Bt_{1},\Bu][\Bv] \equiv T_\texttt{x}[\Bt_{1}[\Bv],\Bu[\Bv_{2l}]]\equiv T_\texttt{x}[\Bt_{1}',\Bu[\Bv_{2l}]]$, so by the i.h. applied to the second premises we get $\Delta\vdash \Bv_{1},\Bv_{2l},\Bv_{2r} : \Theta_{1}.\Theta_{2l}.\Theta_{2r}$, concluding the proof. \qedhere
\end{itemize}
\end{proof}

We can now show soundness:

\begin{theorem}[Soundness]\label{soundness-bidirectional} Suppose that $\mathbb{T}$ is valid.
  \begin{itemize}
    \item If $\Gamma \vdash $ and $\Gamma \vdash t \Rightarrow T$ then $\Gamma \vdash \embed{t} : T$
    \item If $\Gamma \vdash T~\sort$ and $\Gamma \vdash t \Leftarrow T$ then $\Gamma \vdash \embed{t} : T$
    \item If $\Gamma \vdash $ and $\Gamma \vdash T \Leftarrow \sort$ then $\Gamma \vdash \embed{T}~\sort$
    \item If $\Gamma \vdash \Bv : \Xi_{1}$  and $\Xi_{1}.\Xi_{2} \vdash$ and $\Gamma\mid \Bv : \Xi_{1} \vdash \Bt \Leftarrow \Xi_{2}$ then $\Gamma \vdash \Bv,\embed{\Bt} : \Xi_1.\Xi_{2}$.
  \end{itemize}
\end{theorem}
\begin{proof}
  By induction on the derivation.
  \begin{itemize}
    \item Case \textsc{Cons}.
    \begin{mathpar}
      c (\Xi_p\sep \Xi_c\sep \Bv_i/\Xi_i) : T \in \mathbb{T}~
      \inferrule
        {T \prec U \leadsto \Bt_p,\Bt_i \\ \Gamma\mid \Bt_p : \Xi_p \vdash \Bu \Leftarrow \Xi_c}
        {\Gamma  \vdash c(\Bu) \Leftarrow U}
        \Bt_i \equiv \Bv_i[\Bt_p, \embed{\Bu}]
    \end{mathpar}
    By \cref{soundness-matching} we have $U \redd T[\Bt_p,\Bt_i]$, so because we have $\Gamma\vdash U~\sort$ then by \cref{subject-reduction} we get $\Gamma\vdash T[\Bt_p,\Bt_i]~\sort$. We have $T\in\Tm^\Pat~|\Xi_p|.|\Xi_i|~(\cdot)$, and well-typedness of the theory also gives $\Xi_p.\Xi_i\vdash T~\sort$, therefore by \cref{typing-a-substitution} we get $\Gamma\vdash \Bt_p,\Bt_i : \Xi_p.\Xi_i$. By well-typedness of the theory we also have $\Xi_p.\Xi_c\vdash$, therefore by applying the i.h. to the second premise we get $\Gamma\vdash \Bt_p,\embed{\Bu} : \Xi_p.\Xi_c$. We can therefore derive $\Gamma \vdash c(\embed{\Bu}):T[\Bt_p,\Bv_i[\Bt_p,\embed{\Bu}]]$, and because $T[\Bt_p,\Bv_i[\Bt_p,\embed{u}]]\equiv T[\Bt_p,\Bt_i]\equiv U$ and $\Gamma\vdash U~\sort$ we can apply conversion to conclude $\Gamma\vdash c(\embed{\Bu}):U$.

    \item Case \textsc{Switch}.
    \begin{mathpar}
    T \equiv U
    \inferrule
    {\Gamma \vdash t \Rightarrow T}
    {\Gamma \vdash \underline{t} \Leftarrow U}
    \end{mathpar}
    By i.h. we have $\Gamma\vdash\embed{t}: T$, and because we have $\Gamma \vdash U~\sort$ and $T \equiv U$ we can apply the conversion rule to conclude $\Gamma\vdash\embed{\underline{t}}:U$.

    \item Case \textsc{Dest}.
    \begin{mathpar}
      d (\Xi_{pi} \sep \texttt{x}:T \sep \Xi_d) : U \in \mathbb{T}~
      \inferrule
      {\Gamma \vdash t \Rightarrow V \\ T \prec V \leadsto \Bt \\\\
        \Gamma\mid \Bt,\embed{t}:(\Xi_{pi},\texttt{x}:T) \vdash \Bu \Leftarrow \Xi_d}
      {\Gamma  \vdash d(t,\Bu) \Rightarrow U[\Bt,\embed{t},\embed{\Bu}]}
    \end{mathpar}
    By i.h. we have $\Gamma\vdash\embed{t}:V$. By \cref{soundness-matching} we have $V \redd T[\Bt]$, so by \cref{types-are-well-typed,subject-reduction} we get $\Gamma \vdash T[\Bt]~\sort$. By well-typedness of the theory, we have $\Xi_{pi}.(\texttt{x}:T).\Xi_d \vdash U~\sort$ and therefore $\Xi_{pi}\vdash T~\sort$, so because $T \in \Tm^\Pat~|\Xi_{pi}|~(\cdot)$ we can apply \cref{typing-a-substitution} to derive $\Gamma\vdash\Bt:\Xi_{pi}$. From $\Gamma\vdash\embed{t}:V$ and $V\equiv T[\Bt]$ and $\Gamma \vdash T[\Bt]~\sort$ we can derive $\Gamma\vdash\embed{t}:T[\Bt]$, and thus $\Gamma\vdash\Bt,\embed{t}:(\Xi_{pi},\texttt{x}:T)$.  We can now apply the i.h. to the third premise to derive $\Gamma\vdash \Bt, \embed{t}, \embed{\Bu}:\Xi_{pi}.(\texttt{x}:T).\Xi_d$, allowing us to conclude $\Gamma  \vdash d(\embed{t},\embed{\Bu}) : U[\Bt,\embed{t},\embed{\Bu}]$.

    \item Case \textsc{Var}. Trivial

    \item Case \textsc{Annot}.
    \begin{mathpar}
      \inferrule
      {\Gamma \vdash T \Leftarrow \sort\\ \Gamma \vdash t \Leftarrow \embed{T}}
      {\Gamma \vdash t :: T \Rightarrow \embed{T}}
    \end{mathpar}
    By the i.h. applied to the first premise we have $\Gamma \vdash \embed{T}~\sort$. Now we can apply the i.h. to the second premise and conclude $\Gamma\vdash \embed{t} : \embed{T}$, and because $\embed{t::T}=\embed{t}$ we are done.

    \item Case \textsc{Sort}.
    \begin{mathpar}
      c(\Xi) ~\sort\in\mathbb{T}~
      \inferrule
      { \Gamma\mid \varepsilon : (\cdot) \vdash \Bt \Leftarrow \Xi}
      {\Gamma  \vdash c(\Bt) \Leftarrow \sort}
    \end{mathpar}
    By well-typedness of the theory we have $\Xi\vdash$ and therefore we can apply the i.h. to show $\Gamma\vdash \embed{\Bt} : \Xi$, from which we conclude $\Gamma\vdash c(\embed{\Bt}) ~\sort$.

    \item Case \textsc{EmptyMSub}. Trivial.

    \item Case \textsc{ExtMSub}.
    \begin{mathpar}
      \inferrule
      {\Gamma\mid \Bv : \Theta \vdash \Bt \Leftarrow \Xi \\ \Gamma.\Delta[\Bv,\embed{\Bt}] \vdash t \Leftarrow T[\Bv,\embed{\Bt}]}
      {\Gamma\mid \Bv : \Theta \vdash \Bt, \vec{x}_\Delta.  t \Leftarrow (\Xi , \texttt{x}\{\Delta\} : T)}
    \end{mathpar}
    By hypothesis we have $\Theta.\Xi,\texttt{x}\{\Delta\}:T\vdash$, from which we get $\Theta.\Xi\vdash$ and $\Theta.\Xi;\Delta\vdash T~\sort$. By the i.h. applied to the first premise we get $\Gamma\vdash \Bv,\embed{\Bt} : \Theta.\Xi$, so by \cref{substitution-property} applied with $\Theta.\Xi;\Delta\vdash T~\sort$ we get $\Gamma.\Delta[\Bv,\embed{\Bt}]\vdash T[\Bv,\embed{\Bt}]~\sort$. Now we can apply the i.h. to the second premise and get $\Gamma.\Delta[\Bv,\embed{\Bt}] \vdash \embed{t} : T[\Bv,\embed{\Bt}]$, from which we can conclude $\Gamma\vdash \Bv,\embed{\Bt},\vec{x}.\embed{t} : \Theta.\Xi,\texttt{x}\{\Delta\}:T$. \qedhere
  \end{itemize}
\end{proof}

\subsubsection{Annotability}

We now want to show that the bidirectional system is complete with respect to the declarative typing rules, however what notion of completeness should we consider?
As argued by Dunfield and Krishnaswami~\cite{dunfield2021bidirectional}, completeness in bidirectional typing should correspond to \textit{annotability}: if $\Gamma \vdash t : T$ then for some $t'$ with $\embed{t'} = t$ we should have $\Gamma \vdash t' \Leftarrow T$. In other words, $t'$ should be equal to $t$ modulo the insertion of sort ascriptions for when a destructor meets a constructor. Our proof of annotability will need the following lemma, ensuring that the bidirectional system respects conversion.%

\begin{lemma}[Bidirectional system respects conversion]\label{invariance-bidi}~%
  \begin{itemize}
    \item If $\Gamma \vdash t \Rightarrow T$ and $\Gamma'\equiv \Gamma$ then $\Gamma'\vdash t \Rightarrow T'$ for some $T'\equiv T$.
    \item If $\Gamma \vdash t \Leftarrow T$ and $\Gamma'\equiv \Gamma$ and $T'\equiv T$ then $\Gamma'\vdash t \Leftarrow T'$.
    \item If $\Gamma \vdash T \Leftarrow \sort$ and $\Gamma'\equiv \Gamma$ then $\Gamma' \vdash T \Leftarrow \sort$
    \item If $\Gamma \mid \Bv : \Theta \vdash \Bt \Leftarrow \Xi$ and $\Gamma' \equiv \Gamma$ and $\Bv'\equiv \Bv$ then $\Gamma'\mid\Bv':\Theta\vdash\Bt\Leftarrow\Xi$.
  \end{itemize}
  \end{lemma}
  \begin{proof}
    By straightfoward induction, using \cref{matching-invariant} for cases \textsc{Cons} and \textsc{Dest}.
  \end{proof}

Our actual statement for annotability will be slightlier stronger than what we anticipated in the previous paragraphs. Let us call a bidirectional expression \textit{minimal} if it contains no occurrences of $\underline{t::T}$ or $\underline{t}::T$. Our theorem will not only ensure that a regular term can be annotated into a bidirectional one, but also that the resulting term is minimal. In the end of the subsection, this will allow us to derive an alternative completeness result  as a corollary  of \cref{annotability}.

\begin{theorem}[Annotability]\label{annotability}~%
  \begin{enumerate}
    \item If $\Gamma\vdash t : T$ then $\Gamma \vdash t' \Leftarrow T$ for some $t'\in \TmC~|\Gamma|$ minimal with $\embed{t'}=t$.
    \item If $\Gamma\vdash t : T$ then $\Gamma \vdash t' \Rightarrow T'$ for some $T'\equiv T$ and $t'\in \TmI~|\Gamma|$ minimal with $\embed{t'}=t$.
    \item If $\Gamma \vdash T~\sort$ then $\Gamma\vdash T' \Leftarrow\sort$ for some $T'\in \TmC~|\Gamma|$ minimal with $\embed{T'}=T$.
    \item If $\Gamma \vdash \Bv, \Bt : \Theta.\Xi$ then $\Gamma\mid \Bv :\Theta \vdash \Bt' \Leftarrow \Xi$ for some $\Bt' \in \MSubC~|\Gamma|~|\Xi|$ minimal with $\embed{\Bt'}=\Bt$.
  \end{enumerate}
\end{theorem}

\begin{proof}
  The proof is by induction on the derivation. For the rules \textsc{Var}, \textsc{Cons}, \textsc{Dest} and \textsc{Conv} we need to show both points 1 and 2, and so we will organize these cases accordingly.
  \begin{itemize}
    \item Case \textsc{Var}. Trivial.

    \item Case \textsc{Cons}.
    \begin{mathpar}
      c (\Xi_p\sep \Xi_c\sep \Bv_i/\Xi_i) : T \in \mathbb{T}~
      \inferrule
      { \Gamma \vdash \Bt_p, \Bu:\Xi_p.\Xi_c\\
        \Gamma \vdash T[\Bt_p, \Bv_i[\Bt_p,\Bu]]~\sort}
      {\Gamma  \vdash c(\Bu) : T[\Bt_p, \Bv_i[\Bt_p,\Bu]]}
    \end{mathpar}
    \begin{enumerate}
      \item[1] By \cref{completeness-matching} we have $T \prec T[\Bt_p, \Bv_i[\Bt_p,\Bu]] \leadsto \Bt_p', \Bt_i'$ with $\Bt_p',\Bt_i'\equiv \Bt_p, \Bv_i[\Bt_p,\Bu]$. By the i.h. applied to the first premise, for some $\Bu'\in\MSubC~|\Gamma|~|\Xi_c|$ minimal we have $\Gamma \mid \Bt_p : \Xi_p \vdash \Bu' \Leftarrow \Xi_c$ and $\embed{\Bu'}=\Bu$, and by \cref{invariance-bidi} with $\Bt_p\equiv\Bt_p'$ we have  $\Gamma \mid \Bt_p' : \Xi_p \vdash \Bu' \Leftarrow \Xi_c$. Finally, we also have $\Bt_i'\equiv \Bv_i[\Bt_p',\embed{\Bu'}]$, so we conclude $\Gamma \vdash c(\Bu')\Leftarrow T[\Bt_p, \Bv_i[\Bt_p,\Bu]]$ with $\embed{c(\Bu')}=c(\Bu)$ and $c(\Bu')$ minimal.%
      \item[2] By the previous paragraph,  we have $\Gamma\vdash t' \Leftarrow T[\Bt_p, \Bv_i[\Bt_p,\Bu]]$ for some $t'\in\TmC~|\Gamma|$ minimal with $\embed{t'}=c(\Bu)$. Moreover, by the i.h. applied to the second premise we get $T' \in\TmC~|\Gamma|$ such that $\Gamma\vdash T'\Leftarrow\sort$ and $\embed{T'}=T[\Bt_p, \Bv_i[\Bt_p,\Bu]]$, so we have $\Gamma \vdash t'::T'\Rightarrow T[\Bt_p, \Bv_i[\Bt_p,\Bu]]$ with $\embed{t'::T'}=c(\Bu)$. Finally, by inspection on the previous paragraph, $t'$ is not of the form $\underline{t''}$, and so $t'::T'$ is indeed minimal
    \end{enumerate}

    \item Case \textsc{Dest}.
    \begin{mathpar}
      d (\Xi_{pi} \sep\texttt{x}:T \sep \Xi_d) : U \in \mathbb{T}~
      \inferrule
      {\Gamma \vdash \Bt, \Bv : \Xi_{pi}.(\texttt{x}:T).\Xi_d
      }
      {\Gamma  \vdash d(\Bv) : U[\Bt,\Bv]}
    \end{mathpar}
    \begin{enumerate}
      \item[2] Because $\Bv \in \MSub~(\cdot)~|\Gamma|~(\texttt{x}, |\Xi_d|)$ we can split $\Bv$ into $t,\Bu$. We can then extract a strictly smaller derivation of $\Gamma\vdash t : T[\Bt]$, so by i.h. we get $t'\in\TmI~|\Gamma|$ minimal with $\Gamma\vdash t' \Rightarrow T'$ and $T'\equiv T[\Bt]$ and $\embed{t'}=t$, and by \cref{completeness-matching} we get $T \prec T' \leadsto\Bt'$ with $\Bt'\equiv \Bt$. By the i.h. again we also get $\Bu'\in\MSubC~|\Gamma|~|\Xi_d|$ minimal with $\Gamma \mid \Bt,\embed{t'} : (\Xi_{pi},\texttt{x}:T) \vdash \Bu' \Leftarrow \Xi_d$ and $\embed{\Bu'}=\Bu$, so by \cref{invariance-bidi} with $\Bt, \embed{t'} \equiv \Bt',\embed{t'}$ we have $\Gamma \mid \Bt',\embed{t} : (\Xi_{pi},\texttt{x}:T) \vdash \Bu' \Leftarrow \Xi_d$. Finally, we conclude~$\Gamma \vdash d(t',\Bu')\Rightarrow U[\Bt', \embed{t'},\embed{\Bu'}]$, and we indeed have $U[\Bt',  \embed{t'},\embed{\Bu'}]\equiv U[\Bt,  \Bv]$ and $\embed{d(t',\Bu')}=d(t,\Bu)$ and $d(t',\Bu')$ minimal as required.
      \item[1]By the previous paragraph, for some $t'\in\TmI~|\Gamma|$ minimal we have  $\Gamma \vdash t' \Rightarrow U'$ with $U'\equiv U[\Bt,\Bv]$ and $\embed{t'} = d(\Bv)$.  Therefore, by rule \textsc{Switch} we get $\Gamma \vdash \underline{t'} \Leftarrow U[\Bt,\Bv]$. Finally, by inspection on the previous paragraph, $t'$ is not of the form $t'':: T$ and so $\underline{t'}$ is indeed minimal.
    \end{enumerate}

    \item Case \textsc{Conv}.
    \begin{mathpar}
      T\equiv U
      \inferrule
      {\Gamma \vdash t : T\\ \Gamma \vdash U~\sort}
      {\Gamma \vdash t : U}
    \end{mathpar}
    \begin{enumerate}
      \item[1] By the i.h. applied to the first premise we get $t' \in \TmC~|\Gamma|$ minimal with $\Gamma \vdash t' \Leftarrow T$ and $\embed{t'}=t$. By \cref{invariance-bidi} with $T \equiv U$ we get $\Gamma\vdash t'\Leftarrow U$, so we are done.
      \item[2] By the i.h. applied to the first premise we get $t' \in \TmI~|\Gamma|$  minimal with  $\embed{t'}=t$ and $\Gamma\vdash t' \Rightarrow T'$ for some $T'\equiv T$. Because $T'\equiv T \equiv U$, we are done.
    \end{enumerate}

    \item Case \textsc{Sort}.
    \begin{mathpar}
      c (\Xi)~\sort\in\mathbb{T}
      \inferrule
      {\Gamma \vdash \Bt : \Xi}
      {\Gamma  \vdash c(\Bt)~\sort}
    \end{mathpar}
    By i.h. we have $\Bt'\in\MSubC~|\Gamma|~|\Xi|$ minimal such that $\Gamma\mid \varepsilon :(\cdot)\vdash \Bt' \Leftarrow\Xi$ and $\embed{\Bt'}=\Bt$, and thus $\Gamma\vdash c(\Bt')\Leftarrow\sort$ with $\embed{c(\Bt')}=c(\Bt)$ and $c(\Bt')$ minimal.

    \item Cases \textsc{EmptyMCtx} or \textsc{ExtMSub} where $\Bt = \varepsilon$. Trivial.

    \item Case \textsc{ExtMCtx} with $\Bt=\Bu,\vec{x}.u$.
    \begin{mathpar}
      \inferrule
      {\Gamma \vdash \Bv,\Bu : \Theta.\Xi \\ \Gamma.\Delta[\Bv,\Bu] \vdash u : T[\Bv,\Bu]}
      {\Gamma \vdash \Bv,\Bu, \vec{x}_\Delta.  u : (\Theta.\Xi , \texttt{x}\{\Delta\} : T)}
    \end{mathpar}
    By the i.h. applied to both premises we get $\Bu'\in\MSubC~|\Gamma|~|\Xi|$ minimal such that $\Gamma\mid \Bv : \Theta\vdash \Bu' \Leftarrow\Xi$ and $\embed{\Bu'}=\Bu$, and we also get $u' \in \TmC~|\Gamma.\Delta[\Bv,\Bu]|$ minimal such that $\Gamma.\Delta[\Bv,\Bu] \vdash u' \Leftarrow T[\Bv,\Bu]$ and $\embed{u'}=u$. Therefore, we conclude $\Gamma\mid \Bv :\Theta\vdash \Bu', \vec{x}.u' : (\Xi,\texttt{x}\{\Delta\}:T)$ with $\embed{\Bu',\vec{x}.u'}=\Bu,\vec{x}.u$ and $\Bu',\vec{x}.u'$ minimal.\qedhere
  \end{itemize}
\end{proof}

\subsubsection{Ascription-free completeness}

When considering a bidirectional system with ascriptions, completeness is nicely expressed by the notion of annotability. However, as mentioned in \cref{bidi-syntax}, some authors prefer to leave ascriptions out of the bidirectional syntax, given that in most practical cases they are not much used. In this setting, completeness instead ensures that, if a  bidirectional term (seen as a regular one) is typable by the regular type system, then it is also typable by the bidirectional one\footnote{This is actually the notion of completeness we employed in our preliminary work~\cite{felicissimo-esop}.}. We now show that, when considering the subset of the bidirectional syntax which removes ascriptions, this form of \textit{ascription-free completeness} can be deduced almost for free from \cref{annotability}. More precisely, let us say that a bidirectional expression is \textit{ascription-free} if it contains no occurrence of $t::T$. We will then show that, if $\Gamma \vdash \embed{t}: T$ for $t$ a checkable ascription-free term, then $\Gamma \vdash t \Leftarrow T$. Our main lemma for proving this will be the following one, stating that $\embed{-}$ satisfies a restricted form of injectivity.

\begin{lemma}[Restricted injectivity of $\embed{-}$]\label{lem:inj}~
  \begin{itemize}
    \item If $t\in\TmC~\gamma$ is ascription-free and $t'\in\TmC~\gamma$ is minimal and $\embed{t}=\embed{t'}$ then $t = t'$.
    \item If $t\in\TmI~\gamma$ is ascription-free and $t'\in\TmI~\gamma$ is minimal and $\embed{t}=\embed{t'}$ then $t = t'$.
    \item If $\Bt\in\MSubC~\gamma~\xi$ is ascription-free and $\Bt'\in\MSubC~\gamma~\xi$ is minimal and $\embed{\Bt}=\embed{\Bt'}$ then $\Bt = \Bt'$.
  \end{itemize}
\end{lemma}
\begin{proof}
  By straightfoward induction on $t$ (or $\Bt$) and case analysis on $t'$ (or $\Bt'$).
\end{proof}

Ascription-free completeness now follows directly, by composing \cref{annotability} with \cref{lem:inj}.

\begin{corollary}[Ascription-free completeness] \label{thm:completeness-bidirectional}
  ~%
  \begin{itemize}
    \item If $t\in\TmI~|\Gamma|$ is ascription-free and $\Gamma \vdash \embed{t} : T$ then $\Gamma \vdash t \Rightarrow U$ with $T \equiv U$
    \item If $t\in\TmC~|\Gamma|$ is ascription-free and $\Gamma \vdash \embed{t} : T$ then $\Gamma \vdash t \Leftarrow T$
    \item If $T\in\TmC~|\Gamma|$ is ascription-free and $\Gamma \vdash \embed{T}~\sort$ then $\Gamma \vdash T \Leftarrow \sort$
    \item If $\Bt\in\MSubC~|\Gamma|~|\Xi|$ is ascription-free and $\Gamma \vdash \Bv,\embed{\Bt} : \Theta.\Xi$ then $\Gamma\mid \Bv : \Theta \vdash \Bt \Leftarrow \Xi$
  \end{itemize}
\end{corollary}

\subsection{Decidability of bidirectional typing}

We now come to the main property of interest of the bidirectional typing system: its decidability, allowing it to be used when implementing a type-checker for our theories. Our proof will need the following two lemmas, ensuring that matching and type-inference are \textit{functional}, in the sense that when starting from convertible inputs we can only deduce convertible outputs.

\begin{lemma}[Functionality of matching]\label{matching-functional}~
  \begin{itemize}
    \item If $t \prec u \leadsto \Bv$ and $t \prec u' \leadsto \Bv'$ and $u \equiv u'$ then $\Bv \equiv \Bv'$
    \item If $\Bt \prec \Bu \leadsto \Bv$ and $\Bt \prec \Bu' \leadsto \Bv'$ and $\Bu \equiv \Bu'$ then $\Bv \equiv \Bv'$
  \end{itemize}
\end{lemma}
\begin{proof}
By induction on $t$ or $\Bt$. The only interesting case is when $t = c(\Bt)$, in which case we have $u \red^\head c(\Bu)$ and $\Bt \prec \Bu\leadsto \Bv$ and $u' \red^\head c(\Bu')$ and $\Bt \prec \Bu'\leadsto \Bv'$. By applying confluence to $c(\Bu)\equiv c(\Bu')$ we get $\Bu \equiv \Bu'$, allowing us to apply the i.h. to conclude $\Bv\equiv\Bv'$.
\end{proof}

\begin{lemma}[Functionality of inference]\label{inferance-functional} If $\Gamma\vdash t\Rightarrow T$ and $\Gamma \vdash t \Rightarrow T'$ then $T \equiv T'$
\end{lemma}
\begin{proof}
  By straightfoward induction, using \cref{matching-functional} for the case \textsc{Dest}.
\end{proof}

A final hypothesis we will ask for ensuring the decidability of bidirectional typing is for the theory to be \textit{strongly-normalizing}, meaning that $\Gamma \vdash t : T$ should imply that $t$ is s.n. --- a direct consequence of this is that all $T$ with $\Gamma \vdash T~\sort$ and $\Bt$ with $\Gamma\vdash\Bt:\Xi$ must also be s.n.. Indeed, type-checking with dependent types requires checking the conversion of terms (in rules \textsc{Conv} and \textsc{Cons}), whose decidability requires normalization. Moreover, strong normalization is also a requirement for the decidability of matching (\cref{decidability-matching}).

\begin{theorem}[Decidability of bidirectional typing]\label{decidability-bidi} Suppose that $\mathbb{T}$ is valid~and~s.n.
  \begin{itemize}
    \item If $t$ is inferable and $\Gamma \vdash$ then the statement $ \exists T.~(\Gamma \vdash t \Rightarrow T)$ is decidable.

    \item If $t$ is checkable and $\Gamma \vdash T~\sort$ then the statement  $\Gamma \vdash t \Leftarrow T$ is decidable.

    \item If $T$ is checkable and $\Gamma \vdash $ then the statement $\Gamma \vdash T \Leftarrow\sort$ is decidable.

    \item If $\Bt$ is checkable and $\Theta.\Xi\vdash$ and $\Gamma \vdash \Bv : \Theta$ then the statement $\Gamma\mid \Bv : \Theta \vdash\Bt \Leftarrow\Xi$ is decidable.
    \end{itemize}
\end{theorem}
\begin{proof}
  We now proceed with the proof, which is by induction on the bidirectional expression.
    \begin{itemize}
      \item Case $t = c(\Bu)$ with $c(\Xi_p\sep\Xi_c\sep\Bv_i/\Xi_i):U\in\mathbb{T}$. Given $\Gamma,T$ with $\Gamma\vdash T~\sort$, we are to decide if $\Gamma\vdash c(\Bu) \Leftarrow T$ holds. Because $\Gamma \vdash T~\sort$ then $T$ is s.n., so by \cref{decidability-matching} the statement $\exists \Bt_p,\Bt_i.~ U \prec T \leadsto \Bt_p,\Bt_i$ is decidable. %

      \begin{itemize}
        \item If $U \prec T \leadsto \Bt_p,\Bt_i$ does not hold for any $\Bt_p,\Bt_i$, then $\Gamma\vdash c(\Bu) \Leftarrow T$ is not derivable.

        \item If $U \prec T \leadsto \Bt_p,\Bt_i$ holds, we apply \cref{soundness-matching} to derive $T \redd U[\Bt_p,\Bt_i]$. Then, by \cref{subject-reduction} applied to $\Gamma\vdash T~\sort$ we get $\Gamma\vdash U[\Bt_p,\Bt_i]~\sort$, and by well-typedness of the theory we have  $\Xi_p.\Xi_i\vdash U~\sort$, so by \cref{typing-a-substitution} we derive $\Gamma\vdash\Bt_p,\Bt_i:\Xi_p.\Xi_i$. By well-typedness of the theory once again we have $\Xi_p.\Xi_c\vdash$, so by i.h. we get that $\Gamma\mid \Bt_p:\Xi_p\vdash \Bu \Leftarrow\Xi_c$ is decidable. %
        \begin{itemize}
          \item If $\Gamma\mid \Bt_p:\Xi_p\vdash \Bu \Leftarrow\Xi_c$ does not hold, if follows that $\Gamma \vdash c(\Bu)\Leftarrow T$ does not hold. Indeed, if $\Gamma \vdash c(\Bu)\Leftarrow T$ holds then we must have $U \prec T \leadsto \Bt_p',\Bt_i'$ and $\Gamma \mid \Bt_p' : \Xi_p \vdash \Bu \Leftarrow \Xi_c$ for some $\Bt_p',\Bt_i'$, so by \cref{matching-functional} we get $\Bt_p,\Bt_i\equiv\Bt_p',\Bt_i'$. But then \cref{invariance-bidi} gives $\Gamma\mid \Bt_p:\Xi_p\vdash \Bu \Leftarrow\Xi_c$, contradiction.

          \item If $\Gamma\mid \Bt_p:\Xi_p\vdash \Bu \Leftarrow\Xi_c$ holds then by \cref{soundness-bidirectional} we get  $\Gamma \vdash \Bt_p, \embed{\Bu} : \Xi_p.\Xi_c$.  By well-typedness of the theory we have $\Xi_p.\Xi_c\vdash \id,\Bv_i:\Xi_p.\Xi_i$, so by \cref{substitution-property} we get $\Gamma \vdash \Bt_p, \Bv_i[\Bt_p,\embed{\Bu}]:\Xi_p.\Xi_c$. Because the theory is s.n. and confluent, we can decide $\Bt_i\equiv\Bv_i[\Bt_p,\embed{\Bu}]$. If this holds, then it follows that $\Gamma\vdash c(\Bu) \Leftarrow T$ is derivable, otherwise it cannot be derivable. Indeed, if $\Gamma\vdash c(\Bu) \Leftarrow T$ holds then for some $\Bt_p',\Bt_i'$ we have $U \prec T \leadsto \Bt_p',\Bt_i'$ with $\Bt_i'\equiv\Bv_i[\Bt_p',\embed{\Bu}]$, but \cref{matching-functional} implies $\Bt_p',\Bt_i'\equiv \Bt_p,\Bt_i$ and thus $\Bt_i\equiv\Bv_i[\Bt_p,\embed{\Bu}]$, contradiction.
        \end{itemize}
      \end{itemize}

      \item Case $T= c(\Bt)$ with $c(\Xi)~\sort \in \mathbb{T}$. Given $\Gamma$ with $\Gamma\vdash$ we are to decide if $\Gamma\vdash c(\Bu)\Leftarrow\sort$ holds. By well-typedness of the theory we have $\Xi\vdash$, so by i.h. we get that $\Gamma\mid \varepsilon:(\cdot)\vdash \Bu\Leftarrow\Xi$ is decidable. Because this holds iff $\Gamma\vdash c(\Bu)\Leftarrow\sort$, it follows that the latter is also decidable.

      \item Case $t = x$. Trivial, as $\exists T.~(\Gamma \vdash x \Rightarrow T)$ holds iff $x:T \in \Gamma$ for some $T$.

      \item Case $t = u :: U$. Given $\Gamma$ with $\Gamma \vdash$ we are to decide if $\exists T.~(\Gamma \vdash u :: U \Rightarrow T)$ holds. By i.h., we can decide $\Gamma \vdash U \Leftarrow \sort$. If this does not hold, it follows that $\exists T.~(\Gamma\vdash u :: U \Rightarrow T)$ also does not hold. If $\Gamma \vdash U \Leftarrow \sort$ holds, by \cref{soundness-bidirectional} we get $\Gamma\vdash \embed{U}~\sort$, hence by i.h. again we can decide $\Gamma \vdash u \Leftarrow\embed{U}$. If this is the case then we get $\Gamma\vdash u :: U \Rightarrow \embed{U}$, otherwise $\exists T.~(\Gamma\vdash u :: U \Rightarrow T)$ does not hold.

      \item Case $t = d(u, \Bu)$ with $d(\Xi_{pi}\sep\texttt{x}:U\sep\Xi_d):V\in\mathbb{T}$. Given $\Gamma$ with $\Gamma \vdash$ we are to decide if $\exists T.~(\Gamma \vdash t \Rightarrow T)$ holds. By i.h. it follows that $\exists U'.~\Gamma \vdash u \Rightarrow U'$ is decidable.%
      \begin{itemize}
        \item If $\exists U'~.(\Gamma \vdash u \Rightarrow U')$ does not hold, it is clear that $\exists T.~\Gamma\vdash d(u,\Bu)\Rightarrow T$ does not hold.
        \item If $\Gamma \vdash u \Rightarrow U'$ is derivable, then by \cref{soundness-bidirectional} it follows that $\Gamma\vdash \embed{u} : U'$ holds. Therefore, $U'$ is s.n. so by \cref{decidability-matching} it  follows that $\exists\Bt.~U \prec U'\leadsto\Bt$ is decidable. %
        \begin{itemize}
          \item If $U \prec U'\leadsto\Bt$ does not hold for no $\Bt$, it follows that $\exists T.~\Gamma\vdash d(u,\Bu)\Rightarrow T$ does not hold neither. Indeed, if $\Gamma\vdash d(u,\Bu)\Rightarrow T'$ holds for some $T'$, then we have $\Gamma \vdash u \Rightarrow U''$ and $U \prec U'' \leadsto \Bt'$ for some $U''$ and $\Bt'$. But by \cref{inferance-functional,matching-invariant} we get $U \prec U' \leadsto \Bt$ for some $\Bt$, contradiction.

          \item If  $U \prec U'\leadsto\Bt$ is derivable, then by \cref{soundness-matching} we get $U'\redd U[\Bt]$, so by \cref{subject-reduction} applied to $\Gamma\vdash U'~\sort$ we get $\Gamma\vdash U[\Bt] ~\sort$. By well-typedness of the theory we have $\Xi_{pi}.(\texttt{x}:U).\Xi_d\vdash V~\sort$ and thus $\Xi_{pi}\vdash U~\sort$. Therefore by \cref{typing-a-substitution} we get $\Gamma\vdash\Bt :\Xi_{pi}$ and so $\Gamma\vdash \Bt,\embed{u}: (\Xi_{pi},\texttt{x}:U)$. By i.h., the statement $\Gamma\mid \Bt,\embed{u} : (\Xi_{pi},\texttt{x}:U)\vdash \Bu \Leftarrow \Xi_d$ is decidable. If it holds, we conclude that $\Gamma\vdash d(u,\Bu) \Rightarrow U[\Bt,\embed{u},\embed{\Bu}]$ also holds. Otherwise $\Gamma\vdash d(u,\Bu)\Rightarrow T$ cannot hold for no $T$. Indeed, this would imply $\Gamma\vdash u \Rightarrow U''$ and $U\prec U'' \leadsto \Bt'$ and  $\Gamma \mid \Bt',\embed{u} : (\Xi_{pi},\texttt{x}:U)\vdash \Bu \Leftarrow\Xi_d$ for some $U''$ and $\Bt'$,
          so  \cref{inferance-functional,matching-invariant,invariance-bidi} would give $\Gamma \mid \Bt,\embed{u}:(\Xi_{pi},\texttt{x}:U)\vdash \Bu\Leftarrow\Xi_d$, a contradiction.
        \end{itemize}
      \end{itemize}

      \item Case $t = \underline{u}$.  Given $\Gamma,T$ with $\Gamma \vdash T~\sort$, we are to decide if $\Gamma\vdash t\Leftarrow T$ holds. By i.h. we have that $\exists U. ~\Gamma\vdash u \Rightarrow U$ is decidable. If this statement does not hold, it follows that $\Gamma\vdash \underline{u} \Leftarrow T$ does not hold. If $\exists U. ~\Gamma\vdash u \Rightarrow U$ holds, then by \cref{soundness-bidirectional} we get $\Gamma\vdash \embed{u} : U$, which by \cref{types-are-well-typed} implies $\Gamma\vdash U~\sort$. We also have $\Gamma\vdash T~\sort$ so it follows that both $U$ and $T$ are s.n., allowing us to decide $T\equiv U$. If this is the case, then it follows that $\Gamma\vdash \underline{u} \Leftarrow T$ holds. Otherwise $\Gamma\vdash \underline{u} \Leftarrow T$ cannot hold, as this would imply $\Gamma\vdash u\Rightarrow U'$ for some $U'$, but then \cref{inferance-functional} implies $U \equiv U'$ and so  $T\equiv U$, contradiction.

      \item Case $\Bt = \varepsilon$. Trivial, as $\Gamma \mid \Bv :\Theta \vdash \Bt \Leftarrow \Xi$ holds iff $\Xi=\cdot$.

      \item Case $\Bt = \Bu, \vec{x}.t$. Given $\Gamma, \Theta, \Xi, \Bv$ with $\Theta.\Xi\vdash$ and $\Gamma \vdash \Bv:\Theta$, we are to decide if $\Gamma \mid \Bv : \Theta \vdash \Bt \Leftarrow \Xi$ holds.
      If $\Xi =\cdot$ then this clearly does not hold, so let us now suppose that  $\Xi = \Xi', \texttt{x}\{\Delta\} : T$. From $\Theta.\Xi',\texttt{x}\{\Delta\}:T\vdash$  we then get $\Theta.\Xi'\vdash$ and $\Theta.\Xi';\Delta\vdash T~\sort$. By i.h. we then get that $\Gamma\mid\Bv:\Theta\vdash \Bu \Leftarrow \Xi'$ is decidable. If this does not hold, then it is clear that $\Gamma\mid \Bv : \Theta\vdash \Bt \Leftarrow\Xi$ is not derivable, so in the following we assume that we have $\Gamma\mid\Bv:\Theta\vdash \Bu \Leftarrow \Xi'$. Then, by \cref{soundness-bidirectional} we get $\Gamma\vdash\Bv,\embed{\Bu}:\Theta.\Xi'$, so by applying \cref{substitution-property} with $\Theta.\Xi';\Delta\vdash T~\sort$ we get $\Gamma.\Delta[\Bv,\embed{\Bu}]\vdash T[\Bv,\embed{\Bu}]~\sort$. By i.h. we therefore get that $\Gamma.\Delta[\Bv,\embed{\Bu}]\vdash t\Leftarrow T[\Bv,\embed{\Bu}]$ is decidable, hence by testing this statement we can decide $\Gamma\mid \Bv : \Theta\vdash \Bu,\vec{x}.t \Leftarrow\Xi',\texttt{x}\{\Delta\}:T$. \qedhere
    \end{itemize}
\end{proof}

\section{More bidirectional type theories}\label{sec:more-examples}

In the previous sections we have illustrated our framework with the theory~\ref{ex:theory}, defining a basic Martin-L\"of Type Theory with dependent products. We now show other examples of theories covered by our framework. All of the theories we present here are valid: our implementation can automatically check that all schematic rules are well-typed, and can also check that all but a few\footnote{The ones it cannot check automatically are exactly the rewrite rules declared in the implementation using the keyword \texttt{skipcheck}, which skips the implemented check for type-preservation.} rewrite rules are type-preserving (the remaining rewrite rules can be shown to be type-preserving manually, using the same technique as in \cref{ex:well-typed-theory}). Throughout this section, we use the informal notation for schematic typing rules discussed in \cref{subsec:theories} for readability purposes. We refer to the files of the implementation for more details about the examples.

\subsection{Inductive types}

Our framework supports the definition of arbitrary inductive types. For instance, starting from~\ref{ex:theory}, dependent sums can be defined by the following declarations. Note that, as one would wish, the parameters $\ttA$ and $\ttB$ are completely omitted in the constructor and the projections.

\begin{smalldisplay}
\begin{mathpar}
  \inferrule
  { \ttA:\nTy\\ x:\nTm(\ttA) \vdash \ttB:\nTy }
  { \nSigma(\ttA,x.\ttB\{x\}):\nTy}
  \and
  \inferrule
  { \ttA:\nTy\\ x:\nTm(\ttA) \vdash \ttB:\nTy  \\\\
     \ttt : \nTm(\ttA)\\  \ttu :\nTm(\ttB\{\ttt\})}
  {  \nPair(\ttt,\ttu) : \nTm(\nSigma(\ttA,x.\ttB\{x\})) }
  \\
  \inferrule
  { \ttA:\nTy\\ x:\nTm(\ttA) \vdash \ttB:\nTy  \\\\
     \ttt : \nTm(\nSigma(\ttA,x.\ttB\{x\}))}
  { \nProjLeft(\ttt) : \nTm(\ttA)}
  \and
  \inferrule
  { \ttA:\nTy\\ x:\nTm(\ttA) \vdash \ttB:\nTy  \\\\
     \ttt : \nTm(\nSigma(\ttA,x.\ttB\{x\}))}
  { \nProjRight(\ttt) : \nTm(\ttB\{\nProjLeft(\ttt)\})}
  \\
  \nProjLeft(\nPair(\ttt,\ttu)) \longmapsto \ttt
  \and
  \nProjRight(\nPair(\ttt,\ttu)) \longmapsto \ttu
\end{mathpar}\end{smalldisplay}

Dependent sums are an example of \textit{negative inductive types}, which are types that are eliminated by means of projections. We can also define \textit{positive inductive types}, which feature instead a dependent eliminator. The main example of positive inductive type is the W-type, which can be used to define any other positive inductive type~\cite{hugunin2021not}. Once again, note how the parameters $\ttA$ and $\ttB$ are omitted from both the constructor $\nSup$ and the eliminator $\nWRec$. %

\begin{smalldisplay}
\begin{mathpar}
  \inferrule
  {\texttt{A} : \nTy\\
  x:\nTm(\texttt{A}) \vdash \texttt{B} : \nTy}
  {\nW(\texttt{A},x.\texttt{B}\{x\}) : \nTy}
  \and
  \inferrule
  {\texttt{A} : \nTy\\ x:\nTm(\texttt{A}) \vdash \texttt{B} : \nTy\\\\
    \texttt{a}:\nTm(\texttt{A})\\
    \texttt{f}:\nTm(\nPi(\texttt{B}\{\texttt{a}\}, \_.\nW(\texttt{A},x.\texttt{B}\{x\})))}
  {\nSup(\texttt{a}, \texttt{f}) : \nTm(\nW(\texttt{A},x.\texttt{B}\{x\}))}
  \and
  \inferrule
  {\texttt{A} : \nTy\\ x:\nTm(\texttt{A}) \vdash \texttt{B} : \nTy\\
    \texttt{t} : \nTm(\nW(\texttt{A},x.\texttt{B}\{x\}))\\
    x : \nTm(\nW(\texttt{A},x.\texttt{B}\{x\})) \vdash \texttt{P}  : \nTy\\
    x : \nTm(\texttt{A}),
    y : \nTm(\nPi(\texttt{B}\{x\}, \_. \nW(\texttt{A},x.\texttt{B}\{x\}))),
    z : \nTm(\nPi(\texttt{B}\{x\}, x'. \texttt{P}\{\nApp(y, x')\})) \vdash \texttt{p} : \nTm(\texttt{P}\{\nSup(x, y)\})
  }
  { \nWRec(\texttt{t}, x.\texttt{P}\{x\}, x y z.\texttt{p}\{x,y,z\}) : \nTm(\texttt{P}\{\texttt{t}\})}
  \and
  \nWRec(\nSup(\texttt{a},\texttt{f}), x.\texttt{P}\{x\}, x y z. \texttt{p}\{x,y,z\})
  \longmapsto
  \texttt{p}\{\texttt{a}, \texttt{f}, \nLam(x'. \nWRec(\nApp(\texttt{f}, x'), x.\texttt{P}\{x\}, x y z. \texttt{p}\{x,y,z\}))\}
\end{mathpar}\end{smalldisplay}

\subsection{Indexed inductive types}

The types of the previous subsection are non-indexed, in the sense that they are specified uniformly in its parameters (in the examples, $\ttA$ and $\ttB$). As anticipated in \cref{subsec:theories}, our framework also supports the definition of \textit{indexed} inductive types~\cite{dybjer1994inductive}, which are also specified by indices that can vary along the definition. An example of such a type is the one of vectors, for which the length $\ttn$ takes the value $\nZero$ in the constructor $\nNil$ but $\nSucc(\ttm)$ in the constructor $\nCons$.

\begin{smalldisplay}
  \begin{mathpar}
      \inferrule
      {\ttA:\nTy\\ \ttn : \nTm(\nNat)}
      {\nVec(\ttA, \ttn):\nTy}
      \and
      \inferrule
      {\ttA:\nTy\\ \ttn \mapsto \nZero : \nTm(\nNat)}
      {\nNil : \nTm(\nVec(\ttA, \ttn))}
      \and
      \inferrule
      {\ttA:\nTy\\ \ttm:\nTm(\nNat)\\
      \ttt:\nTm(\ttA)\\\\
      \ttl:\nTm(\nVec(\ttA,\ttm))\\
      \ttn\mapsto \nSucc(\ttm) : \nTm(\nNat)}
      {\nCons(\ttm,\ttt, \ttl) :
      \nTm(\nVec(\ttA,\ttn))}
    \end{mathpar}
  \end{smalldisplay}
  \begin{smalldisplay}
    \begin{mathpar}
      \inferrule
      {\ttA:\nTy \\ \ttn:\nTm(\nNat)\\ \ttl :\nTm(\nVec(\ttA, \ttn)) \\\\ x : \nTm(\nNat), y : \nTm(\nVec(\ttA,x)) \vdash \ttP:\nTy\\ \texttt{pnil} : \nTm(\ttP\{\nZero,\nNil\})\\
      x : \nTm(\nNat), y : \nTm(\ttA), z : \nTm(\nVec(\ttA, x)), w : \nTm(\ttP\{x, z\}) \vdash \texttt{pcons} : \nTm(\ttP\{\nSucc(x), \nCons(x, y, z)\})
      }
      {\nVecRec(\ttl, x.\ttP\{x\}, \texttt{pnil}, xyzw.\texttt{pcons}\{x,y,z,w\}) : \nTm(\ttP\{\ttn, \ttl\})}
    \end{mathpar}
\end{smalldisplay}

As also mentioned in \cref{subsec:theories}, note that the argument $\ttm$ of $\nCons$ cannot be omitted because it needs to be passed to $\texttt{pcons}$ in the second rewrite rule for $\nVecRec$.

\begin{smalldisplay}
\begin{align*}
  &\nVecRec(\nNil, xy.\ttP\{x,y\}, \texttt{pnil}, xyzw.\texttt{pcons}\{x,y,z,w\}) \longmapsto \texttt{pnil}\\
  &\nVecRec(\nCons(\ttm,\ttt,\ttl), xy.\ttP\{x,y\}, \texttt{pnil}, xyzw.\texttt{pcons}\{x,y,z,w\}) \longmapsto\\ &\quad\quad\texttt{pcons}\{\ttm,\ttt,\ttl,\nVecRec(\ttl, xy.\ttP\{x,y\}, \texttt{pnil}, xyzw.\texttt{pcons}\{x,y,z,w\})\}
\end{align*}\end{smalldisplay}

Another main example of indexed type is equality, also known as Martin-L\"of's \textit{identity type}.

\begin{smalldisplay}
  \begin{mathpar}
  \inferrule
  {\ttA:\nTy\\ \tta:\nTm(\ttA)\\ \ttb:\nTm(\ttA)}
  {\nEq(\ttA,\tta, \ttb):\nTy}
  \and
  \inferrule
  { \ttA:\nTy\\ \tta : \nTm(\ttA)\\
  \ttb\mapsto \tta : \nTm(\ttA)
  }
  {\nrefl : \nTm(\nEq(\ttA, \tta, \ttb))}
  \and
  \inferrule
  {\ttA:\nTy\\ \tta:\nTm(\ttA)\\ \ttb:\nTm(\ttA)\\ \ttt:\nTm(\nEq(\ttA,\tta, \ttb))\\\\
  x : \nTm(\ttA), y : \nTm(\nEq(\ttA,\tta, x))\vdash \ttP : \nTy \\
  \ttp : \nTm(\ttP\{\tta, \nrefl\})}
  {\nJ(\ttt,x y.\ttP\{x,y\} ,\ttp) : \nTm(\ttP\{\ttb, \ttt\})}
  \and
  \nJ(\nrefl, x y. \ttP\{x, y\}, \ttp) \longmapsto \ttp
\end{mathpar}\end{smalldisplay}

\subsection{Higher-order logic}

It is well-known that the Curry-Howard correspondence allows to embed many kinds of logic into type theories, enabling us to use the previously introduced types as propositions. Nevertheless, it can be useful sometimes to explicitly separate types from propositions, as done for instance in the proof assistant Coq. Let us illustrate how this can be done in our framework by defining a variant of Higher-Order Logic. We start by extending \ref{ex:theory} with a type of propositions and a sort $\nPrf(\ttP)$ for representing the judgment form "$P$ is provable".

\begin{smalldisplay}
\begin{mathpar}
  \inferrule
  { }
  { \nProp:\nTy}
  \and
  \inferrule
  { \ttP :\nTm(\nProp)}
  { \nPrf(\ttP)~\textsf{sort}}
\end{mathpar}\end{smalldisplay}

We can then add arbitrary connectives or quantifiers. For instance, we can add universe quantification with the following declarations. We refer to the file \texttt{hol.bitts} of the implementation in which we also add implication and define conjunction using the impredicative encoding~\cite{girard1989proofs}.

\begin{smalldisplay}
\begin{mathpar}
  \inferrule
  { \ttA : \nTy\\ x : \nTm(\ttA)\vdash \ttP : \nTm(\nProp)}
  { \nForall(\ttA,x.\ttP\{x\}) : \nTm(\nProp)}
  \and
  \inferrule
  { \ttA : \nTy\\ x : \nTm(\ttA)\vdash \ttP : \nTm(\nProp)\\\\
   x : \nTm(\ttA)\vdash \ttp : \nPrf(\ttP) }
  { \nForallIntro(x.\ttp\{x\}) : \nPrf(\nForall(\ttA,x.\ttP\{x\}))}
  \and
  \inferrule
  { \ttA : \nTy\\ x : \nTm(\ttA)\vdash \ttP : \nTm(\nProp)\\\\
    \ttq : \nPrf(\nForall(\ttA,x.\ttP\{x\}))\\ \ttt:\nTm(\ttA) }
  { \nForallElim(\ttq,\ttt) : \nPrf(\ttP\{\ttt\})}
  \and
  \nForallElim(\nForallIntro(x.\ttp\{x\}), \ttt)\longmapsto \ttp\{\ttt\}
\end{mathpar}\end{smalldisplay}

\subsection{Universes}

In dependent type theories, types can be reified as terms by adding \textit{universes}.  Starting from \ref{ex:theory}, we define a \textit{Tarski-style universe} by adding add a type $\nU$ of \textit{codes} and a decoding function $\nEl$ mapping each code to an associated type. We then must close $\nU$ under the type formers of our theory, by adding the codes $\nuu$ for $\nU$ and $\npi$ for $\nPi$, and stating that $\nEl$ decodes them to the expected types.

\begin{smalldisplay}
  \begin{mathpar}
    \inferrule
    { }
    {\nU: \nTy}
    \and
    \inferrule
    { \tta : \nTm(\nU) }
    { \nEl(\tta) : \nTy}
    \and
    \inferrule
    { }
    { \nuu : \nTm(\nU)}
    \and
    \nEl(\nuu) \longmapsto \nU
    \\
    \inferrule
    {\tta : \nTm(\nU)\\ x:\nTm(\nEl(\tta))\vdash \ttb : \nTm(\nU)}
    {\npi(\tta, x.\ttb\{x\}) : \nTm(\nU)}
    \and
    \nEl(\npi(\tta, x.\ttb\{x\})) \longmapsto \nPi(\nEl(\tta), x. \nEl(\ttb\{x\}))
  \end{mathpar}\end{smalldisplay}

  For illustrative purposes, in the above we  have defined a \textit{type-in-type} universe, which is known to be inconsistent~\cite{coquand:inria-00076023}. This can however be easily solved if we stratify universes into an hierarchy, by instead introducing a family of symbols $\nU_l, \nEl_l,\dots$ indexed by some set $l\in\mathcal{L}$ of \textit{universe levels}. In particular, this allows us to define a Tarski-style variant of \textit{Pure Type Systems}~\cite{barendregt2013lambda}, which are usually presented using \textit{Russell-style} universes.

\paragraph*{Internal universe levels} By indexing the above symbols externally, they become annotated with levels $l$ in their names. A better approach is instead to index them \textit{internally}, which then allows us to leverage our support for erased arguments. Let us illustrate this by taking~$\mathcal{L}:= \mathbb{N}$. We start by declaring a sort of levels along with constructors for zero and successor.

\begin{smalldisplay}\begin{mathpar}
  \inferrule
  { }
  {\nLvl~\sort}
  \and
  \inferrule
  { }
  {\nZero : \nLvl}
  \and
  \inferrule
  {\ttl : \nLvl}
  {\nSucc(\ttl) : \nLvl}
\end{mathpar}\end{smalldisplay}

We then update the definitions of $\nU, \nEl, \nuu, \npi$ in the following manner, so that they now take a level as an argument. With these definitions, we can omit the level annotations in $\nEl$ and $\npi$. Note that in the case of $\nuu$ we cannot omit $\ttl$ from the syntax, otherwise we would not be able to define the rewrite rule $\nEl(\nuu) \longmapsto \nU(?)$.

\begin{smalldisplay}\begin{mathpar}
  \inferrule
  {\ttl:\nLvl }
  {\nU(\ttl) : \nTy}
  \and
  \inferrule
  {\ttl:\nLvl\\ \ttA : \nTm(\nU(\ttl)) }
  { \nEl(\ttA) : \nTy}
  \and
  \inferrule
  {\ttl : \nLvl \\ \tti \mapsto\nSucc(\ttl): \nLvl}
  {\nuu(\ttl) : \nTm(\nU(\tti))}
  \and
  \nEl(\nuu(\ttl)) \longmapsto \nU(\ttl)
  \\
  \inferrule
  {\ttl : \nLvl\\\tta : \nTm(\nU(\ttl))\\\\
   x:\nTm(\nEl(\tta))\vdash \ttb : \nTm(\nU(\ttl))}
  {\npi(\tta, x.\ttb\{x\}) : \nTm(\nU(\ttl))}
  \and
  \nEl(\npi(\tta, x.\ttb\{x\})) \longmapsto \nPi(\nEl(\tta), x. \nEl(\ttb\{x\}))
\end{mathpar}\end{smalldisplay}

\paragraph*{Cumulativity}
For now, in the above theory we can only find a code decoding to $\nPi(\nEl(a), x.\nEl(b))$ if $a$ and $b$ live in the same universe. In order to fix this, we could define an heterogeneous version of $\npi$ allowing for $a$ and $b$ to be in different universes --- see the file \texttt{mltt-tarski-heterogeneous.bitts} in the implementation where this solution is developed. We instead prefer to add \textit{cumulativity} to the theory. This is done by adding a \textit{lift} $\nUp$ mapping a code $a$ from $\nU(l)$ to $\nU(\nSucc(l))$, and a rewrite rule identifying its elements with the ones for $a$. Now we can form a code for $\nPi(\nEl(a), x.\nEl(b))$ by simply lifting the smaller code to the universe of the bigger one.

\begin{smalldisplay}\begin{mathpar}
  \inferrule
  {\ttl:\nLvl\\ \ttA:\nTm(\nU(\ttl))}
  {\nUp(\tta) : \nTm(\nU(\nSucc(\ttl))) }
  \and
  \nEl(\nUp(\tta)) \longmapsto \nEl(\tta)
\end{mathpar}
\end{smalldisplay}

\begin{remark}
  Some authors consider a stronger version of cumulativity in which the lift operator~$\nUp$ commutes with all codes~\cite{assaf2014calculus,sterling2019algebraic,kovacs:LIPIcs.CSL.2022.28}. This would require us to add the rule $\nUp(\npi(\tta,x.\ttb\{x\}))\longmapsto \npi(\nUp(\tta),x.\nUp(\ttb\{x\}))$, yet because its left-hand side is headed by a constructor this would not be a valid rewrite rule in our setting.
\end{remark}

\paragraph*{Universe polymorphism} With our current theory we can write the polymorphic identity function at the base universe $\nLam(A. \nLam(x.x)) : \nTm(\nPi(\nU(\nZero), a. \nPi(\nEl(a),\_.\nEl(a))))$, but there is no way of writing an identity function that can be used with types in all universes. This can be achieved by extending our theory with \textit{universe polymorphism}~\cite{kovacs:LIPIcs.CSL.2022.28,bezem_et_al:LIPIcs.TYPES.2022.13,coq}, in which we add a type former for quantifying over levels. We can then write the universe-polymorphic identity function $\nBigLam(i.\nLam(A. \nLam(x.x))) : \nTm(\nForallUP(i.\nPi(\nU(i), a. \nPi(\nEl(a),\_.\nEl(a)))))$

\begin{smalldisplay}\begin{mathpar}
  \inferrule
  {i : \nLvl \vdash \ttA : \nTy}
  {\nForallUP(i.\ttA\{i\}) : \nTy}
  \and
  \inferrule
  {i : \nLvl \vdash \ttA : \nTy\\
   i : \nLvl \vdash \ttt : \nTm(\ttA\{i\})}
  {\nBigLam(i.\ttt\{i\}) : \nTm(\nForallUP(i.\ttA\{i\}))}
  \\
  \inferrule
  {\ttt: \nTm(\nForallUP(i.\ttA\{i\}))\\
   \ttl : \nLvl}
  {\nInst(\ttt,\ttl):\nTm(\ttA\{\ttl\})}
  \and
  \nInst(\nBigLam(i.\ttt\{i\}),\ttl) \longmapsto \ttt\{\ttl\}
\end{mathpar}\end{smalldisplay}

\paragraph*{Coquand-style universes} Up until now we have only seen examples showing the use of Tarski-style universes, however in our framework we can also define \textit{(weak) Coquand-style universes}~\cite{coquand-style,kaposi2019gluing,kovacs:LIPIcs.CSL.2022.28}. In this approach, we start by redefining the sorts $\nTy$ and $\nTm$ so that they become themselves stratified --- note therefore that this example is the first which is not an extension of \ref{ex:theory}.

 \begin{smalldisplay}\begin{mathpar}
  \inferrule
  { }
  {\nLvl~\sort}
  \and
  \inferrule
  { }
  {\nZero : \nLvl}
  \and
  \inferrule
  {\ttl : \nLvl}
  {\nSucc(\ttl) : \nLvl}
  \and
  \inferrule
  {\ttl : \nLvl}
  {\nTy(\ttl)~\sort}
  \and
  \inferrule
  {\ttl:\nLvl\\ \ttA : \nTy(\ttl) }
  { \nTm(\ttl, \ttA)~\sort}
  \end{mathpar}
   \end{smalldisplay}

  We then postulate once again a type for the universe $\nU$ and a decoding function $\nEl$, but instead of adding codes for each type former manually, we add a code constructor $\ncode$. A rewrite rule then states that decoding $\ncode(A)$ yields precisely the type $A$.\footnote{We call this variant of Coquand-style universes \textit{weak} because the rule $\nEl(\ncode(\ttA))\longmapsto \ttA$ only amounts to asking $\nTy(l)$ to be a retract of $\nTm(\nSucc(l),\nU)$. In contrast, in regular Coquand-style universes one has a definitional isomorphism $\nTm(\nSucc(l),\nU) \simeq \nTy(l)$. This would require also the equation $\ncode(\nEl(\tta)) = \tta$, but because $\ncode$ is not a destructor this is not a valid rewrite rule.} We refer to the file \texttt{mltt-coquand.bitts} of the implementation in which this approach is further developed.

  \begin{smalldisplay}
  \begin{mathpar}
  \inferrule
  {\ttl:\nLvl }
  {\nU : \nTy(\nSucc(\ttl))}
  \and
  \inferrule
  {\ttl:\nLvl\\ \ttA : \nTy(\ttl) }
  {\ncode(\ttA) : \nTm(\nSucc(\ttl),\nU)}
  \and
  \inferrule
  {\ttl:\nLvl\\ \tta : \nTm(\nSucc(\ttl),\nU) }
  {\nEl(\tta) : \nTm(\nU(\ttl))}
  \and
  \nEl(\ncode(\ttA)) \longmapsto \ttA
\end{mathpar}
\end{smalldisplay}

\subsection{Exceptional type theory}

We have seen that our framework supports the definition of many features commonly found in dependent type theories. However, we can also define theories with features that are more unusual. Let us now see how to define in our framework a variant of Pédrot and Tabareau's Exceptional Type Theory~\cite{10.1007/978-3-319-89884-1_9}, that extends MLTT with exceptions. Starting from \ref{ex:theory}, we add a new sort $\nEx$ for exceptions, a default exception $\nErr$ and a constructor $\nRaise$ that allows us to raise an exception at any type. Note that $\nRaise$ renders the theory inconsistent, howeover in their paper Pédrot and Tabareau show how to define a parametricity layer to isolate a consistent subset of the language, in which exceptions have to be caught locally. Nevertheless, if one wishes to use the theory for programming instead of proving theorems, it is reasonable to drop this extra layer and work in a language where all types are inhabited (like almost all commonly used programming languages).

\begin{smalldisplay}\begin{mathpar}
  \inferrule
  { }
  {\nEx~\sort}
  \and
  \inferrule
  { }
  {\nErr : \nEx}
  \and
  \inferrule
  {\ttA : \nTy\\ \tte : \nEx }
  {\nRaise(\tte) : \nTm(\ttA)}
\end{mathpar}\end{smalldisplay}

We then must add rules for ensuring that destructors properly propagate the exceptions, such as the following one.

\begin{smalldisplay}\begin{mathpar}
  \nApp (\nRaise(\tte), \ttu) \longmapsto \nRaise(\tte)
\end{mathpar}\end{smalldisplay}

We can then extend the theory with generalized eliminators for the positive types in order to allow for exception-catching. For instance, supposing we also have extended the theory with booleans --- and added the associated rule $\nBoolRec(\nRaise(\tte), x.\ttP\{x\}, \texttt{pt}, \texttt{pf}) \longmapsto \nRaise(\tte)$ for propagating exceptions ---, we can add a new eliminator $\nCatchBool$ in which $\nRaise$ is treated like a constructor of $\nBool$, allowing us to handle the raised exception.

\begin{smalldisplay}
\hspace{-1em}\begin{minipage}{0.54\linewidth}
  \begin{mathpar}
    \inferrule
    {\ttt:\nTm(\nBool)\\
     x : \nTm(\nBool)\vdash \ttP : \nTy \\
     \texttt{pt} : \nTm(\ttP\{\nTrue\})\\\\
     \texttt{pf} : \nTm(\ttP\{\nFalse\})\\
     x : \nEx \vdash \texttt{pe} : \nTm(\ttP\{\nRaise(x)\})}
    {\nCatchBool(\ttt,x.\ttP\{x\}, \texttt{pt}, \texttt{pf}, x. \texttt{pe}\{x\}) : \nTm(\ttP\{\ttt\})}
  \end{mathpar}
\end{minipage}\hspace{0.7em}
\begin{minipage}{0.45\linewidth}
\begin{align*}
  &\nCatchBool(\nTrue,x.\ttP\{x\}, \texttt{pt}, \texttt{pf}, x. \texttt{pe}\{x\}) \longmapsto \texttt{pt}
    \\
  &\nCatchBool(\nFalse,x.\ttP\{x\}, \texttt{pt}, \texttt{pf}, x. \texttt{pe}\{x\}) \longmapsto \texttt{pf}
    \\
  &\nCatchBool(\nRaise(\tte),x.\ttP\{x\}, \texttt{pt}, \texttt{pf}, x. \texttt{pe}\{x\}) \longmapsto \texttt{pe}\{\tte\}
\end{align*}
\end{minipage}
\end{smalldisplay}

This theory can then be extended with various types, and we refer to the file \texttt{exceptional.bitts} of the implementation for more details.

\newcommand\nMode{{\color{objectblue}\mathfrak{M}}}
\newcommand\nExe{{\color{objectblue}\mathbb{E}}}
\newcommand\nPure{{\color{objectblue}\mathbb{P}}}

We can also interface the exceptional theory with a pure one by restricting the eliminators, in the spirit of the Multiverse Type Theory~(MuTT)~\cite{DBLP:journals/corr/abs-2108-10259}.
For this, we redefine the theory by parametrizing $\nTy$ and $\nTm$ by a \textit{mode} $\nMode$, which can either be $\nPure$ (for pure) or $\nExe$ (for exceptional). The $\nRaise$ constructor is then restricted to only allow for raising exceptions when in the exceptional mode, and the only way for eliminating from types in the exceptional mode to the pure world is by using catching eliminators, which then ensure that exceptions cannot be propagated. We refer to the file \texttt{exceptional-multiverse.bitts} of the implementation, where this approach is sketched.

\subsection{Observational type theory}

As our last example, we show how to define a variant of McBride et al's \textit{observational type theory~(OTT)}~\cite{altenkirch2007observational,pujet2022observational} in our framework. Starting from \ref{ex:theory}, we begin by extending the theory with an \textit{heterogeneous} equality type.

\begin{smalldisplay}
\begin{mathpar}
  \inferrule
  {\ttA:\nTy\\ \tta:\nTm(\ttA)\\ \ttB:\nTy\\ \ttb:\nTm(\ttB)}
  {\nEq(\ttA,\tta,\ttB, \ttb):\nTy}
\end{mathpar}\end{smalldisplay}

The defining characteristic of OTT is that its equality $\nEq(A, t, A', t')$ is defined inductively on the structure of the types $A$ and $A'$. For instance,  a proof of $\nEq(\nPi(A,x.B),f, \nPi(A',x'.B'\{x'\}), f')$ should correspond exactly to a function mapping proofs of equality between $t$ and $t'$ to proofs of equality between $\nApp(f,t)$ and $\nApp(f',t')$. In the original formulation of OTT~\cite{altenkirch2007observational} this correspondence was made by adding a rewrite rule explaining how to reduce $\nEq(\nPi(A,x.B),f, \nPi(A',x'.B'\{x'\}), f')$. Here we instead adopt the approach taken by Atkey~\cite{sott} and more recently by Pujet et al~\cite{obsind}, in which one instead postulates symbols $\nEqPii$ and $\nEqPie$ for constructing and eliminating equality proofs between $f$ and $f'$. Note that our support for omitting arguments is vital here to ensure that the numerous arguments do not get all recorded in the syntax.

\begin{smalldisplay}
  \begin{mathpar}
  \inferrule
  {\ttA:\nTy\\ x : \nTm(\ttA) \vdash \ttB : \nTy\\
  \ttf : \nTm(\nPi(\ttA,x.\ttB\{x\}))\\\\
   \ttA':\nTy\\ x' : \nTm(\ttA') \vdash \ttB' : \nTy\\
   \ttf' : \nTm(\nPi(\ttA',x'.\ttB'\{x'\}))\\
   x :\nTm(\ttA), x' :\nTm(\ttA'), y : \nTm(\nEq(\ttA,x,\ttA',x')) \vdash \ttp : \nTm(\nEq(\ttB\{x\}, \nApp(\ttf, x), \ttB'\{x'\}, \nApp(\ttf',x')))}
  { \nEqPii(x x'y. \ttp\{x,x',y\}) : \nTm(\nEq(\nPi(\ttA,x.\ttB\{x\}), \ttf, \nPi(\ttA',x'.\ttB'\{x'\}),\ttf'))}
  \and
  \inferrule
  {\ttA:\nTy\\ x : \nTm(\ttA) \vdash \ttB : \nTy\\
  \ttf : \nTm(\nPi(\ttA,x.\ttB\{x\}))\\\\
   \ttA':\nTy\\ x' : \nTm(\ttA') \vdash \ttB' : \nTy\\
   \ttf' : \nTm(\nPi(\ttA',x'.\ttB'\{x'\}))\\
    \ttp : \nTm(\nEq(\nPi(\ttA,x.\ttB\{x\}), \ttf, \nPi(\ttA',x'.\ttB'\{x'\}),\ttf'))\\
    \ttx : \nTm(\ttA)\\ \ttx' : \nTm(\ttA')\\ \tte : \nTm(\nEq(\ttA,\ttx,\ttA',\ttx'))}
  {\nEqPie(\ttp,\ttx,\ttx',\tte) : \nTm(\nEq(\ttB\{\ttx\}, \nApp(\ttf, \ttx), \ttB'\{\ttx'\}, \nApp(\ttf',\ttx')))}
\end{mathpar}
\end{smalldisplay}

\definecolor{gray3}{RGB}{130, 130, 130}
\newcommand\mLet{{\color{gray3} \textsf{let}}}
\newcommand\mIn{{\color{gray3} \textsf{in}}}

If we add a Tarski-style universe $\nU$ to the theory, we would then like to allow for transporting a term from $\nEl(a)$ to $\nEl(b)$ when we have a proof of $\nEq(\nU,a,\nU,b)$. In OTT, this is achieved by adding a cast operator to the theory. Note that the arguments  $\tta$ and $\ttb$ could be recovered from $\ttp$, but they cannot be omitted because they are needed to know how to reduce casts.%

\begin{smalldisplay}
\begin{mathpar}
  \inferrule
  {\tta : \nTm(\nU) \\
   \ttb : \nTm(\nU) \\
   \ttp : \nTm(\nEq(\tta,\nU,\ttb,\nU))\\
   \ttt : \nTm(\nEl(\tta))
  }
  {\nCast(\tta,\ttb,\ttp,\ttt) : \nTm(\nEl(\ttb))}
  \and
  \nCast(\npi(\tta, x.\ttb\{x\}), \npi(\tta', x'.\ttb'\{x'\}), \ttp, \ttt )
  \longmapsto
  \begin{matrix}
    \nLam(x'.\hspace{-1em} &\mLet~ e := \nEqUpiel(\ttp) ~\mIn \hfill \\
    &\mLet~ x := \nCast(\tta',\tta, e, x') ~\mIn \hfill \\
    &\nCast(\ttb\{x\}, \ttb\{x'\}, \nEqUpier(\ttp, x, x', e), \nApp(\ttt, x)))
  \end{matrix}
\end{mathpar}\end{smalldisplay}

We refer to the file \texttt{ott.bitts} of the implementation in which this construction is worked out in detail. We also provide a second variant of OTT in the file \texttt{ott-2.bitts}, this time using an homogeneous equality, in the style of Pujet et al~\cite{pujet2022observational}.

\section{BiTTs: An implementation of our framework}\label{implementation}

The bidirectional type system of \cref{sec:bidirectional-type-system} has been implemented in the tool BiTTs (for \underline{Bi}directional \underline{T}ype \underline{T}heorie\underline{s}), which we present in this section. We start by illustrating how the tool can be used in practice, and then briefly discuss some aspects of the implementation.

\subsection{A quick introduction to the tool}

As we have seen, our framework is not designed to target a specific type theory, but instead a whole class of theories. Therefore, in order to use our tool, the first step is to specify the theory we want to work in. This is done with the commands \texttt{sort}, \texttt{constructor}, \texttt{destructor} and \texttt{equation} which specify respectively sort, constructor, destructor and rewrite rules. For instance, the theory~\ref{ex:theory} can be defined by the following declarations --- note the support for unicode characters.

\begin{verbatim}
sort Ty ()
sort Tm (A : Ty)
constructor Π () (A : Ty, B{x : Tm(A)} : Ty) : Ty
constructor λ (A : Ty, B{x : Tm(A)} : Ty) (t{x : Tm(A)} : Tm(B{x})) : Tm(Π(A, x. B{x}))
destructor ﹫ (A : Ty, B{x : Tm(A)} : Ty) [t : Tm(Π(A, x. B{x}))] (u : Tm(A)) : Tm(B{u})
equation ﹫(λ(x. t{x}), u) --> t{u}
\end{verbatim}

The implementation then tries to check automatically that the specified theory is valid. For rewrite rules, it implements an incomplete criterion for verifying type preservation, and in case a rule falls out of the covered fragment, the user can disable the checker with the keywork \texttt{skipcheck} and verify that the rewrite rule preserves typing manually (for instance, by using the same technique as in \cref{ex:well-typed-theory}). By \cref{soundness-bidirectional}, the validity of the theory then ensures that the type-checker implemented is sound with respect to the declarative type system of the theory.

Once the theory is specified, we can start writing and type-checking terms inside it. For instance, supposing we have also added a Tarski-style universe \texttt{U}, we can check the following definition of the polymorphic identity function.

\begin{verbatim}
let idU : Tm(Π(U, a. Π(El(a), _. El(a)))) := λ(a. λ(x. x))
\end{verbatim}

To type-check this definition, the tool first verifies that the sort given in the annotation is well-typed, and then type-checks the body of the definition against the sort. If all the steps succeed, the identifier is added to a global scope of top-level definitions and becomes available to be used in the rest of the file.

The implementation also supports local let definitions, as illustrated in the following example, which also shows how sort ascriptions can be used.

\begin{verbatim}
let redex' : Tm(ℕ) :=
    let ty : Ty := Π(ℕ, _. ℕ) in
    ﹫(λ(x. x) :: Tm(ty), 0)
\end{verbatim}

Finally, we also provide commands for evaluating terms to normal form and asserting that two terms are definitionally equal. For instance, assuming we have defined factorial, we can use these commands to compute the factorial of 3 and check that it is equal to 6.

\begin{verbatim}
let fact3 : Tm(ℕ) := ﹫(fact, S(S(S(0))))
evaluate fact3
let 6 : Tm(ℕ) := S(S(S(S(S(S(0))))))
assert fact3 = 6
\end{verbatim}

The implementation also comes with many theories that can be defined in the framework, along with some examples of terms written in these theories. These can be found in the directory \texttt{examples/}, and most of them are discussed in \cref{sec:more-examples}.

\subsection{The implementation}

The core of the implementation can be separated into two main parts: the type-checking and the normalization algorithms. The type-checking algorithm follows very closely the bidirectional system described in \cref{sec:bidirectional-type-system} with only minor differences, so we will not discuss it here.

Regarding normalization, we have implemented it by employing an untyped variant of \textit{Normalization by Evaluation (NbE)}, inspired by the works of Coquand~\cite{coquand1996algorithm}, Abel~\cite{abel2013normalization} and Kovacs~\cite{elaborationzoo}. In NbE, terms are evaluated into a separate syntax of runtime values, in which binders are represented by closures and free variables by unknowns. This evaluation roughly corresponds to a call-by-value reduction to weak-head normal form.  Values can then be compared for equality by entering closures and recursively evaluating and comparing their bodies. One of the benefits of this approach is that, by using de Bruijn indices in the syntax of regular terms but de Bruijn \textit{levels} in the syntax of values, we completely avoid the need of implementing substitution or index-shifting functions, which are often complicated and inefficient.

Preliminary tests suggest that our implementation of normalization has good performances: we can compute factorial of 8 with unary natural numbers in around half a second, whereas the same test leads Coq to crash with a stack overflow and Agda to run for more than a minute before we terminate the process. In the future, we plan to further test the general performances of our tool with larger and more realistic examples. In particular, we would like to compare it with type-checkers for Dedukti~\cite{dedukti,thU}. Because Dedukti has no support for erased arguments, its terms are highly-annotated, which can have an important impact on performance~\cite{felicissimo:LIPIcs.FSCD.2022.25}, whereas our support for non-annotated syntaxes should allow for shorter type-checking times.

\section{Related work}\label{sec:related-work}

Our general definition of dependent type theories draws much inspiration from other frameworks for type theory, such as GATs/QIITs~\cite{CARTMELL1986209,thorsten-ambrus,10.1145/3290315}, SOGATs~\cite{uemura2021abstract}, FTTs~\cite{haselwarter2021finitary}, and logical frameworks such as Dedukti~\cite{dedukti,thU} and Harper's Equational LF~\cite{harper2021equational}. However, we differ from these works by supporting non-annotated syntaxes and enforcing a constructor/destructor separation of symbols and rules, both of which seem to be important ingredients for bidirectional typing.

Another point of divergence from these frameworks is that most of them allow the use of arbitrary equations when defining the definitional equality of theories. However, it then becomes hard to give an implementation, as it would require deciding arbitrary equational theories. We instead take the approach of Dedukti of supporting only rewrite rules, which allows us to decide the definitional equality of theories in a uniform manner, and made it possible to implement our framework. A different approach is taken in Andromeda, an implementation of FTTs, where they also allow for extensionality rules~\cite{DBLP:journals/lmcs/BauerK22}. They however provide no proof of completeness for their equality-checking algorithm.%

Our proposal also draws inspiration from the works of McBride, a main advocate of dependent bidirectional typing. His ongoing work on a framework for bidirectional typing~\cite{mcbride,typeswhosayni} shares many similarities with ours, for instance by adopting a constructor/destructor separation of rules. However, an important difference with our framework is that he takes the bidirectional type system as the definition of the theory. Therefore, there is no discussion on how to show soundness and completeness with respect to a declarative system, as the bidirectional one is the only type system defined in his setting. This approach differs from most of the literature on dependent bidirectional typing~\cite{dunfield2021bidirectional,abel2005untyped,lennonbertrand:LIPIcs.ITP.2021.24,abel2011partial,abel2011modular,abel2017normalization}, in which one first defines the type theory by a "platonic" declarative type system and then shows it equivalent to a bidirectional system which can be implemented.
Finally, this choice also makes the metatheoretic study of theories quite different from what we have done here: for instance, even to be able to state subject reduction for the bidirectional system, the notion of reduction has to be updated to take ascriptions into account.

Another work from which ours drew inspiration is the one of Reed~\cite{reed2008redundancy}, where he proposes a variant of the Edinburgh Logical Framework in which arguments can be omitted. Crucially, these arguments are not elaborated through global unification, but instead locally recovered by annotating each declaration with modes to guide a bidirectional algorithm. However, his framework does not allow for extending the definitional equality, meaning that one cannot define dependent type theories with non-trivial equalities directly, but instead has to encode its derivations trees (as in~\cite{harper2007mechanizing}). This also means that his system does not need to deal with some complications that arise in our more general setting, such as matching modulo. %

Finally, concurrently to our work, Chen and Ko~\cite{bisig} have proposed a framework for simply-typed bidirectional typing. They also define declarative and bidirectional systems and establish a correspondence between them. Compared to our work, their restriction to simple types removes many of the complexities that appear with dependent type theories. For instance, while their types are first-order terms with no notion of computation or typing, our sorts are higher-order terms  considered modulo a set of rewrite rules and subject to typing judgments, making the process  of recovering missing arguments much more intricate. The restriction to simple types also rules out examples like the ones presented in \cref{sec:more-examples}, given that they are all dependent type theories. They however provide an impressive formalization of all their results in Agda.%

\section{Conclusion and future work}\label{sec:conclusion}

In this work we have given a generic account of bidirectional typing for a large class of dependent type theories. Our main results, \cref{soundness-bidirectional,annotability}, establish an equivalence between declarative and bidirectional type systems for a whole class of theories. The decidability of the bidirectional type system (\cref{decidability-bidi}) allowed for its implementation in the tool BiTTs, which has been used in practice with multiple theories. Compared to other theory-independent type-checkers, such as Dedukti and Andromeda, BiTTs' support for unannotated syntaxes can allow for better performances, making it  a good candidate for cross-checking real proof~libraries.

Regarding future work, the most important omission that we would like to address is that of type-directed equality rules, which are needed for handling $\eta$-laws and definitional proof irrelevance.
As mentioned in \cref{sec:related-work}, our choice of supporting only rewrite rules was motivated by the fact that they allow for deciding the definitional equality in a uniform way, which made it possible to implement our framework. Indeed, as long as a rewrite system $\mathcal{R}$ is both confluent and strongly-normalizing, computing and comparing normal forms is a complete equality-checking algorithm, regardless of any other specificity of $\mathcal{R}$. In contrast, even if it is well known how to design complete equality checking algorithms for specific theories with type-directed equalities~\cite{abelDecidabilityConversionType2018}, doing so in a general setting like ours  seems to be an important challenge. We could take inspiration from the customizable equality-checking algorithm implemented in Andromeda~\cite{DBLP:journals/lmcs/BauerK22}. However, as mentioned in the previous section, their algorithm is not proven complete, so further research in this direction seems to be required.

Moreover, even if our system builds heavily on the constructor/destructor distinction in type theory, some theories employ equations that do not respect this separation.
For instance, to define Russell universes we need the rewrite~rule $\nTm(\nU) \longmapsto \nTy$~\cite{sterling2019algebraic}, which is not valid as $\nTm$ is~not~a~destructor.
Whether there is a way of accommodating these constructions without fully abandoning the constructor/destructor separation is something we would like to investigate in future work.

Finally, some could argue that our choice of declarative type system is not "declarative" enough, as some authors often prefer more abstract definitions for when specifying what is a type theory.
For instance, the point of view that syntax should correspond to the initial model (for some notion of semantics) often leads one to consider fully-anotated terms with typed equality (or even quontiented terms~\cite{thorsten-ambrus}), whereas our declarative type system uses non-annotated terms and untyped equality. In the future, we would like to investigate if our results could be adapted to such a setting, for instance by considering a variant of Uemura's SOGATs~\cite{uemura2021abstract} for the declarative type system. In this setting, our bidirectional system would be adapted into an elaboration algorithm, producing core fully-annotated syntax from the user-friendly bidirectional one --- similarly to \cite{gratzer2022controlling}.

\paragraph*{Acknowledgements} The author thanks Gilles Dowek, Vincent Moreau, Théo Winterhalter for helpful remarks on a preliminary version of this paper, Meven Lennon-Bertrand, Jesper Cockx, András Kovács for informative discussions about bidirectional typing, Kenji Maillard for suggesting the example of Exceptional Type Theory, and Frédéric Blanqui for valuable remarks about this~work.

\bibliographystyle{splncs04}
\bibliography{ref}
\end{document}